\documentstyle[aps,prl,epsf,floats,multicol,amssymb,preprint,tighten]{revtex}

\begin{document}

\newcommand{\fig}[2]{\epsfxsize=#1\epsfbox{#2}} \reversemarginpar 
\newcommand{\mnote}[1]{$^*$\marginpar{$^*$ {\footnotesize #1}}}

\bibliographystyle{prsty}

\title{Random Walkers in 1-D Random Environments: \\
Exact Renormalization Group Analysis} 

\author{Daniel S. Fisher}
\address{Lyman Laboratory of Physics, Harvard University, Cambridge MA 02138, USA}
\author{Pierre Le Doussal}
\address{CNRS-Laboratoire de Physique Th\'eorique de l'Ecole\\
Normale Sup\'erieure, 24 rue Lhomond, F-75231
Paris}
\author{C\'ecile Monthus}
\address{Laboratoire de Physique Th\'eorique et Mod\`eles Statistiques \\
CNRS-Universit\'e Paris XI, B\^at. 100, 91405 Orsay, France}
\maketitle
\begin{abstract}
Sinai's model of diffusion in one-dimension with random local bias is studied by 
a real space renormalization group which yields exact results at long times.
The effects of an additional small uniform bias force
are also studied. We obtain analytically
the scaling form of the distribution of the position $x(t)$ of a particle,
the probability of it not returning to the origin and the distributions
of first passage times, in an infinite sample as well as in the presence
of a boundary and in a finite but large sample. We compute
the distribution of meeting time of two particles in the same
environment. We also obtain a
detailed analytic description of the thermally averaged trajectories 
by computing quantities such as the joint distribution of the
number of returns and of the number of jumps forward. These quantities
obey multifractal scaling, characterized by generalized persistence
exponents $\theta(g)$ which we compute. In the presence of
a small bias, the number of returns to the origin becomes
finite, characterized by a universal scaling function which we obtain.
The full statistics of the distribution of successive 
times of return of thermally averaged trajectories is obtained,
as well as detailed analytical information about correlations
between directions and times of successive jumps.
The two time distribution of the positions of a particle, $x(t)$ and
$x(t')$ with $t>t'$, is also computed exactly.
It is found  
to exhibit ``aging'' with several 
time regimes characterized by different behaviors.
In the unbiased case, for $t-t' \sim t'^\alpha$ with
$\alpha>1$, it exhibits a $\frac{\ln t}{\ln t'}$  
scaling, with a singularity at coinciding rescaled positions
$x(t)=x(t')$. This singularity is a novel feature, 
and corresponds to particles which remain
in a renormalized valley. For closer times $\alpha<1$,
the two time diffusion front exhibits a quasi-equilibrium regime
with a $\ln(t-t')/\ln t'$ behavior which we compute.
The crossover to a $t/t'$ aging form in the presence
of a small bias is also obtained analytically. Rare events
corresponding to intermittent splitting of the thermal
packet between separated wells which dominate
some averaged observables, are also characterized in detail.
Connections with the Green's
function of a one-dimensional Schr\"odinger problem
and quantum spin chains are discussed.

\end{abstract}


\widetext

\newpage

\tableofcontents

\section{Introduction} \label{secint}

Studying non-equilibrium dynamics provides
a useful route to elucidate the properties of systems
with quenched disorder. In addition it is 
very relevant for experiments, since most such 
systems form glassy states with ultraslow 
dynamics and usually do not reach full thermal equilibrium 
within the accessible time scales. This is the case for a 
variety of experimental systems such spin glasses
\cite{Vihaoc}, random field systems
\cite{anderson,fisher_huse},
vortex lines in superconductors 
\cite{blatter_vortex_review,Legi},
and domain growth in presence of quenched disorder.
Despite decades of extensive work,
there are still a number of unresolved issues
in the theoretical description of the dynamics of
systems with quenched disorder. This uncertainty is 
due, to a large extent, to the lack of
physically relevant models for which  analytical solutions
can be obtained, providing clear cut answers to 
well posed questions. The need for such models 
is all the more acute since it is prohibitively difficult 
to obtain unambiguous answers from numerical simulations 
when the dynamics is ultraslow, especially since the
interpretation is often blurred by the absence of
precise theoretical predictions. Solvable models, where the
answers are known, should also provide useful testing ground
for numerical methods by giving clues on the 
necessary simulation time scales and averaging procedures
in disordered systems which are often dominated by rare events.

Some progress has been made in obtaining analytical solutions for
the large time behavior of mean field type models \cite{Cuku}.
Although it is still extremely unclear how much these mean field results
will carry through to short range, finite dimensional systems, one 
outcome of these works \cite{Cuku,horner,Cugledou}
has been to demonstrate the existence 
of several possible large time regimes and to attempt to classify them. 
This provides further motivation to study aging dynamics
in a larger class of models, in particular to study
the possible ways of taking the large time $t, t' \to \infty$ limits
for correlations between configurations of the system at
a waiting time $t'=t_w$ after a quench at $t=0$, and
a later observation time $t$. 

Other types of approaches, such as 
droplet descriptions of the statics and  the non-equilibrium
dynamics of disordered systems \cite{fisher_huse}, 
make use of domain growth arguments.
These approaches emphasize the leading role of thermally activated 
processes, which should play an important role in short range models,
while mean field dynamics may be dominated by other type of collective
processes \cite{kurchan}. The ``coarsening'' of domain structures evolving towards
equilibrium has been studied extensively in  pure models
\cite{bray} but little is known rigorously for domain growth 
with quenched disorder. Thus these approaches are still to a large extent 
phenomenological and one would like to find models where solid results 
about aging in the presence of activated dynamics can be obtained 
analytically. A natural hope for that would be to study 1D models
which could be used as testing grounds for more complex $D>1$ cases 
which have resisted analytic attack.

A celebrated 1D toy model for glassy activated dynamics
is the Sinai model, which describes the diffusion of a random walker
in a 1D random static force field---equivalent to 
a random potential which itself has the statistics 
of a 1D random walk \cite{sinai}. Although this model
(with or without a bias) has been much studied, the
known analytical results 
\cite{sinai,kesten,derrida_pomeau,ledou_1d,monthus_diffusion,laloux_pld_sinai}
usually concern single time and single particle quantities
and are technically hard to obtain. It is known that
this model without a bias exhibits non trivial ultraslow 
logarithmic behavior, as the walker typically moves as $x \sim (\ln t)^2$,
as well as several dynamical phases with anomalous diffusion
as the bias is increased from zero. By contrast,
there was until now no exact results about two time
aging dynamics, despite several mostly qualitative and 
numerical studies \cite{feigelman_aging_1d,laloux_pld_sinai}
which found interesting aging behavior in this model.
In addition, the Sinai model
has interesting extensions to many interacting particles,
and via domain walls, to the Glauber
dynamics of 1D random field Ising ferromagnets and spin glasses in a 
magnetic field.

Recently we have proposed an approach, based on
a real space renormalization group (RSRG) method, which 
allows us to obtain many exact results for the non equilibrium
dynamics of several 1D disordered systems \cite{us_prl}. We have
shown that it applies to the Sinai model as well as to
1D disordered spin models and diffusion-reaction processes 
in Sinai's type of energy landscapes. This RSRG method is closely
related to that used to study disordered quantum spin chains 
\cite{ma_dasgupta,danfisher_rg1,danfisher_rg2,danfisher_rg3,hyman_yang,%
monthus_golinelli}. The crucial feature of the 
RG is coarse-graining the energy landscape 
in a way that  preserves the long time dynamics.
In Sinai's model the way to implement the RSRG
is very direct: one decimates iteratively the {\it smallest
energy barrier} in the system stopping when the time to surmount 
the smallest remaining barrier is of order the time scale of interest.
Despite its approximate character, the RSRG yields for many quantities
asymptotically exact results. As in \cite{danfisher_rg2} it works because
the iterated distribution of barriers grows infinitely wide,
consistent with \cite{sinai}.

The aim of the present paper is to show in detail
how the RSRG method applies to the Sinai model,
allowing one to obtain in a simple way a large 
number of exact results. We obtain a 
host of quantities such as return and first passage probabilities,
single time correlations as well as two time correlations
of the type that are probed in aging experiments.
Given the long history of 
Sinai's model, some of the results obtained here
have been derived previously, by completely 
different methods. These methods include
those from probability theory
\cite{kesten1,sinai,golosov,golosov2,tanaka}, as well as
conventional methods of the physics of disordered
systems, such as the Dyson Schmidt method, replica methods,
supersymmetry, transfer matrix etc.
\cite{footnote3}. Despite that,
a large number of our results are, to our knowledge, novel.
Indeed, as we aim to illustrate in this paper,
the most interesting feature of the RSRG, besides
being simple to apply, is that it allows one to obtain all 
these results (new and old) from a {\it single} method,
while other methods usually allow  
access to only specific types of exact results.
As we will explain, the only limitations 
are the ones usually associated with any RG 
method. First, almost by definition,
it only addresses and obtains exactly the universal 
quantities, i.e., the ones which are independent of the
short scale details of the model. Second, it does rely on 
the global assumption that the starting model is within the
basin of attraction of the zero temperature fixed point 
studied here, and is thus not ``exact from first principles''.
This is not a restriction in the case of the 
Sinai model:  because rigorous results
already exist from probability theory, this
last assumption can be considered established.

In two companion papers \cite{us_rfim,us_rd}, we will detail the
applications of the RSRG, given as a short account in 
\cite{us_prl}, to the Glauber dynamics
of disordered spin models and to 
diffusion-reaction processes in presence of quenched
disorder. These works  rely heavily
on the Sinai model and  the present work.
Thus we here 
give a  detailed presentation of the
results for Sinai model.

An interesting feature of the RSRG is that it demonstrates
in a simple and operational way how the Sinai model is related to other
one-dimensional disordered models. More formal derivations 
of such mappings can also be made in some cases via
free fermion models. For instance, the quantum XX spin chains
with disorder and the random transverse field
Ising chain (RTFIC) are related via Jordan Wigner transformations
to free fermion models near half filling with disorder 
in the hopping term. This problem is in turn related,
via its expression as a random Dirac problem, to a supersymmetric
random Schr\"odinger operator \cite{balents_fisher}
identical to the Fokker Planck diffusion operator associated with
the  Sinai model 
\cite{ledou_1d,fokker_dirac}. Most of these
relations have been detailed previously in various contexts
(see, e.g. for a review \cite{comtet_texier,mckenzie,rieger_freefermion}).
These disordered fermion models have been much reinvestigated
recently as they provide examples of quantum delocalization transitions.
It may sometimes be useful to recast them in terms of the
Sinai model where some quantities have a 
straightforward physical interpretation (e.g., the logarithmic 
Arrhenius diffusion over barriers growing as a random
walk gives the logarithmic energy dependence of
the local density of states).
The RSRG demonstrates such mappings for the 
low energy (large time) properties in a very direct way,
as we will illustrate. Zero drift in Sinai's model corresponds to the
self dual critical case in the RTFIC
\cite{danfisher_rg2} 
and to the antiferromagnetic XX chain \cite{danfisher_rg1,ma_dasgupta},
while the zero velocity biased phase \cite{derrida_pomeau,ledou_1d}
corresponds to the Griffiths phase of the RTFIC \cite{danfisher_rg2}
and  a dimerized XX chain. As we show here, magnetization properties
correspond to persistence properties in Sinai's model.

Although the idea of studying random diffusion problem via 
real space decimation techniques has been used previously,
it has been mostly applied to fractal or hierarchical 
landscapes (see e.g. \cite{maritan}) which are designed for
such methods. By contrast, here the RSRG emerges from the structure of
the zero temperature fixed point itself, as 
the natural way to treat diffusion in a
statistically translationally invariant disordered system,
with no ad-hoc assumptions. Interestingly, a similar property arises
in the problem of the coarsening of the pure 1D $\Phi^4$ model
at zero temperature, which can be treated exactly by successive elimination of
the smallest domains in the system \cite{derrida_coarsening_phi4},
a method reminiscent of the RSRG studied here.
Finally note that since \cite{us_prl} appeared, several new
papers have been devoted to the Sinai model 
\cite{rieger_anomalous,comtet_dean,chave_guitter,stephen_sinai}.

\subsection{Outline}

\label{subsecoutline}

The outline of this paper is as follows.
Section \ref{secmodels} contains a
pedagogical introduction to
the real space renormalization (RSRG) approach for the
Sinai model. It terminates with the explicit
expressions for the fixed points of the RSRG
(\ref{secsolutionrg}). In Section \ref{secsinglediffusion} 
we compute the averaged single time diffusion front 
for the symmetric Sinai model in (\ref{secsinglediffusionsymm})
and with a bias in (\ref{secsinglediffusionbias}).
In Section \ref{thermal averages} we study the
returns to the origin (persistence properties)
of the thermally averaged motion, as well as
the statistics of the jumps, in the symmetric
and biased case. In Section \ref{truedynamics},
we study returns to the origin of a single walker,
distributions of first passage time and of the maximum
position as well as the probability of
meeeting of two walkers. Section \ref{sectwotime} is
devoted to the aging properties of the Sinai model and contains
a general discussion (\ref{subsecgeneral}), calculations of singular parts of
the diffusion front (\ref{sub:singular}), the full two-time probability
distribution
(\ref{sub:full}), and  the analysis
of a simpler case (\ref{sub:resultsymmetric}). The section 
terminates with the
analysis of rare events and calculation of the 
front in the quasi-equilibrium regime
(\ref{sub:inwell}) and fluctuations in the single time
diffusion front. In Section \ref{finitesizerg}
the RSRG is studied in a finite size system;
equilibration properties, first passage times
with boundaries and finite size diffusion 
fronts are computed. Finally in section
\ref{shrofokk} we obtain the Green's function of the
associated Schr\"odinger operators in 
\ref{greenfunction}. Section \ref{secconclusion}
contains the conclusions.
Further technicalities are relegated to various
Appendices.

\section{Models and real space renormalization procedure}

\label{secmodels}

\subsection{Diffusion models}

\label{subsecmodels}

Diffusion in one-dimensional random media has been 
modeled in three ways, which usually lead to equivalent classes
of behavior in the large time limit. Probabilists have often 
studied models discrete in time {\it and} space; for instance,
a particle on points of a one-dimensional lattice, $n$, which jumps 
to the right ($n+1$) with probability $p_n$ and to the
left with probability $1-p_{n}$. Physicists on the
other hand have often considered  random hopping models,
continuous in time but discrete in space, described by
the master equation:

\begin{eqnarray}  \label{defhopping}
\frac{dP_n(t)}{dt} = - (J_{n+1,n} - J_{n,n-1}) 
\qquad J_{n+1,n} = W_{n+1,n} P_n - W_{n,n+1} P_{n+1}
\end{eqnarray}
$W_{n+1,n}$ and $J_{n+1,n}$ are respectively
the transition rate and the current from $n$ to $n+1$,
and $W_{n,n+1}$ and $J_{n,n+1}=-J_{n+1,n}$ from $n+1$ to $n$.
Finally fully continuum models, with Fokker Planck equation:

\begin{eqnarray}  \label{deffplanck}
\partial_t P(x,t) = H_{FP} P = \partial_x D(x) (T \partial_x P - F(x) P)
\end{eqnarray}
have also been studied (with $D(x) >0$).

It is useful to distinguish three classes of disorder (within
each description)  leading to different types of generic 1D
large time behavior (for uncorrelated disorder).

(i) {\it detailed balance, random diffusion coefficient}

This corresponds to $W_{n,n+1}=W_{n+1,n} \equiv D_{n,n+1}$
in (\ref{defhopping}) or to $F(x)=0$ and $D(x)$ a random positive function.
It is well known that the large time diffusion 
coefficient is $D_{{\rm eff}} = \langle 1/W
\rangle^{-1}$ for uncorrelated disorder and thus that
this model exhibits asymptotic ``normal diffusion'' unless the $D_{n,n+1}$ have 
a broad distribution, with a tail near the origin, $P(D) \sim D^{-\alpha}$
($0 \le \alpha <1$).

(ii) {\it random traps}

This corresponds to $W_{n,n+1}=1/\tau_{n+1}$ and
$W_{n+1,n}=1/\tau_{n}$. Each site is characterized by a release time,
but the exit is with the same probability $1/2$ to the left or to
the right (the jump probability depends only on the starting point). Again this
model exhibits asymptotic ``normal diffusion'' unless the release
times have an algebraically broad distribution.

(iii) {\it generic case: Sinai model}

In the generic case one can always parametrize the hopping rates as:

\begin{eqnarray}  \label{paramhopping}
W_{n,n+1} = \tau^{-1}_0 e^{\beta E_{n,n+1}} 
e^{ - \frac{1}{2} \beta (U_n - U_{n+1}) }
\qquad 
W_{n+1,n} = \tau^{-1}_0 e^{\beta E_{n,n+1}} 
e^{ - \frac{1}{2} \beta (U_{n+1} - U_{n}) }
\end{eqnarray}
where $\beta=1/T$ and $T$ is the temperature.
This can be illustrated as in Fig. 1: there is a symmetric barrier $E_{n,n+1}$
between sites $n$ and $n+1$, plus an additional potential difference. The
barrier $E_{n,n+1}$ gives the average diffusion coefficient (or attempt frequency)
on the bond. The  ``forces'' on the bonds are
$f_{n+1,n} = - (U_{n+1} - U_{n})$
which represent a local bias. In a finite size system
(periodic in the $U_n$) the expressions (\ref{paramhopping})
correctly lead to the Gibbs zero current equilibrium measure,
$e^{-\beta U_n}/Z$.

The main case of interest here and studied by Sinai
is that of {\it independent random forces}.
The generic case, for uncorrelated disorder and
for distributions of $f_n$ and  $E_{n,n+1}$ with
fast enough decay (e.g. faster than exponential) all
belong to the class of Sinai's model, which is a discrete
time model. A similar potential can be introduced for this model:
$U_n - U_0 = T \sum_{i=1}^{n-1}
\log ( p_i/(1-p_i))$. One can most easily visualize these
as  Arrhenius motion in a random potential $U_n$ which itself
performs a random walk,  either symmetric
or biased. This motion has been studied extensively
and it is known that diffusion is logarithmic $x \sim \ln^2 t$
in the symmetric case, sublinear $x \sim t^\mu$ for a small bias
($\mu <1$) and with a finite velocity $x \sim V t$ for
$\mu >1$, where $\mu$ is related to the asymmetry of the 
force distributions as defined later. 

\begin{figure}[thb] 
\centerline{\fig{8cm}{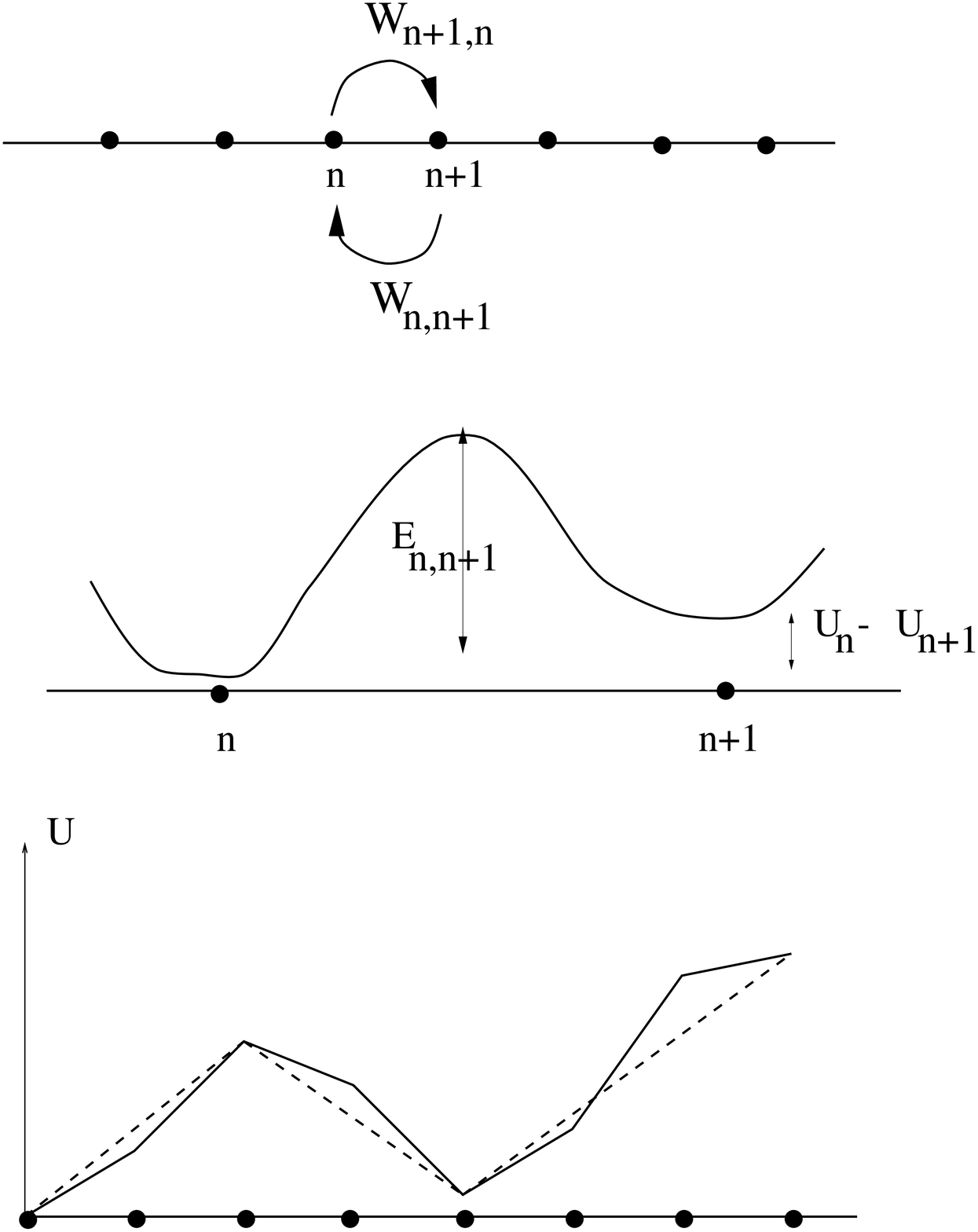}}
\caption{\label{fig1} \narrowtext (a) General hopping model; (b) 
interpretation of (a) with a barrier between each site and
a potential difference (local bias); (c) model studied
here. }
\end{figure}

For convenience and to be specific in what follows 
we will mostly study, as our basic model,
the random hopping model with the choice $E_{n,n+1}=0$. We will also 
compare with the discrete time model originally studied by Sinai.
However, our results 
are much more general and apply to any model 
within the locally random force class (with short 
range correlations).

\subsection{Renormalization method}

\label{sub:rgmethod}

\subsubsection{Definitions and RG equations}

\label{subsub:rgmethod}

As described above, we  consider models of diffusion in 1D landscapes
in which walkers perform Arrhenius diffusion in a
potential ${U_n}$ ($n$ is a site index). 
A ``force'' variable $f_n = U_n - U_{n+1}$
is defined on each bond $(n,n+1)$ (indexed as bond $n$)
and as in the  Sinai model,  the $f_n$ are independent
random variables with distribution $Q(f)df$.
The long-time dynamics in such landscapes are primarily determined
 by the large barriers and deep valleys. 
Thus we need to be able to focus on these aspects 
of the landscape while eliminating as much as possible the effects of the finer 
scale structure.

We therefore introduce a renormalization procedure, for a
given landscape, which will allow us, in this way,
 to study the asymptotic dynamics.
We should emphasize that we will apply it mainly to the case
of  forces  independent from
bond to bond, but  it can in principle be applied to
any 1D landscape. The crucial feature which is needed
for the RG to yield asymptotically exact results, is that the 
landscapes have extremal values of the potential which 
grow with length scale. This will make the distributions
of the renormalized barriers broader and broader.
In the case of the Sinai model, it is possible to follow
exactly the RG flow (because the forces remain uncorrelated under the RG)
and thus to check {\it a posteriori}  that at large scales
the distributions of renormalized
barriers are indeed very broad. However, the procedure
is much more general and would also lead to asymptotically exact
results for correlated landscapes in which the  renormalized barriers become
higher and higher. The difficulty in such correlated cases 
is to follow the distributions. Of course there are
1D landscapes for which the RSRG  would not give exact
results for the diffusion behavior: in particular, bounded potentials which 
have normal diffusive behavior.

\begin{figure}[thb] 
\centerline{\fig{8cm}{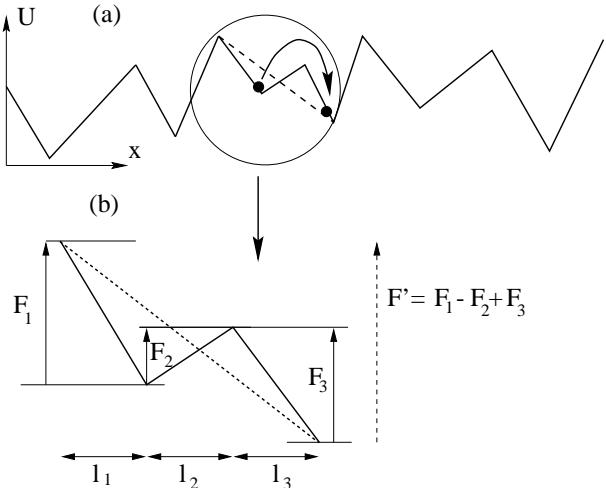}}
\caption{\label{fig2} \narrowtext (a) Energy landscape in Sinai model; (b) 
decimation method: the bond with the smallest barrier
$F_{\min}=F_2$ is eliminated as three bonds are grouped into
one (see text).}
\end{figure}

The RSRG procedure on a given landscape
is implemented as follows. One can first group the bonds with the
same sign of the force (see Fig. \ref{fig1} c), and then can start,
with no loss of generality, 
from an ``antiferromagnetic'' landscape (see Fig. \ref{fig2})
with the $f'_n$ alternatively positive
and negative) but with a distribution of bond lengths $l_i$.
Our starting model is thus defined by 
$f'_n= (-1)^n F_n$ where the $F_n=|U_n - U_{n+1}|$ are
the useful variables---called here ``barriers''---and
the two bond variables $F,l$ are chosen independently
from bond to bond with an initial distribution $P(F,l)$.
In the presence of a bias one needs two distinct
distributions
$P^{+}(F,l)$ for ``descending bonds''
and $P^{-}(F,l)$ for ``ascending bonds''
(opposing the mean force), both 
normalized to unity. Note that the combining together of consecutive
descending bonds in this way naturally leads to an exponential tail in
the distribution $P^-$ and likewise in $P^+$. Such exponential tails in
barrier distributions will play an important role in 
the physics and in our analysis.

We are interested in long times when the behavior will be dominated by 
large barriers
and it is on these that we must focus. Our RG procedure is conceptually  
simple:
in a given energy
landscape it consists of iterative decimation of 
the bond with the {\it smallest barrier} $\Gamma=F_{\min}$, 
say $F_2=U_3 - U_2=\Gamma$
as illustrated in Fig. \ref{fig2}. At time scales much longer than $\exp 
(F_2/T)$, 
local equilibrium will be established between sites 2 and 3 and the rate for the 
walker to get from 4 to 1 will be essentially the same as it would be if sites 2 
and 3 did not exist but 1 and 4 were instead connected by a bond with barrier 
\begin{eqnarray}
F'=F_1 - F_2 + F_3
\end{eqnarray}
 and length 
\begin{eqnarray}
l'=l_1 + l_2 + l_3.
\end{eqnarray}
We thus carry out exactly this replacement. This  preserves 
the zigzag structure (the model remains alternating ``up-down'')
and the larger scale extrema of the potential since the total length 
and the extrema of $U$ in the segments are exactly preserved.
Furthermore if the starting distribution is independent forces 
from bond to bond, this 
remains so under the RG. One then keeps on iteratively eliminating barriers
$\Gamma<F<\Gamma +d\Gamma$ thereby gradually decreasing the 
minimum remaining barrier height, $\Gamma$. Note that there is 
no ambiguity in the case
of continuous distributions as considered here,
as one can always neglect 
the unlikely events when two neighbors or next nearest neighbors
are within $d \Gamma$.

The above rules for $F$ and $l$
define the RSRG transformation for arbitrary landscapes.
In the case of the Sinai landscape where bonds remain statistically
independent one can define 
\begin{eqnarray*}
\zeta_n\equiv F_n-\Gamma 
\end{eqnarray*}
and
introduce $P_\Gamma^{+}(\zeta=F-\Gamma,l)$ 
and $P_\Gamma^{-}(\zeta=F-\Gamma,l)$ which denote the probabilities
that a $\pm$ renormalized bond at scale $\Gamma$ has a barrier
$F=\Gamma + \zeta > \Gamma$ and a length $l$, each normalized by
$\int_0^{\infty} d\zeta \int_0^{\infty} dl P_{\Gamma}^{\pm}(\zeta,l)=1$.
One can then explicitly write closed RG equations for these
two distributions describing their evolution under the
decimation represented in Fig. \ref{fig2}:
\begin{eqnarray}   \label{rg1}
 (\partial_\Gamma - \partial_\zeta) P_\Gamma^{\pm} (\zeta,l)  =
&& P_\Gamma^{\mp}(0,.)*_l P_\Gamma^{\pm}(.,.) *_{\zeta,l} P_\Gamma^{\pm}(.,.) 
-2 P_\Gamma^\pm(\zeta , l) \int_0^{\infty} dl' P_\Gamma^{\mp}(0,l') \nonumber \\
 && + P_\Gamma^{\pm} (\zeta,l) \int_0^{\infty} dl' \left(P_\Gamma^{\pm}(0,l') 
+ P_\Gamma^{\mp}(0,l')\right)
\end{eqnarray}
where $*_\zeta$ denotes a convolution with respect
to $\zeta$ only and $*_{\zeta,l}$ with respect to both $\zeta$ and $l$
with the variables to be convoluted denoted by dots.
The first term on the r.h.s. represents the new renormalized bonds,
the second the bonds which are decimated as neighbors of
the smallest barrier and the last  comes from keeping 
the distribution normalized. The total number
$n_\Gamma$ of bonds in the system evolves as
\begin{eqnarray}   \label{rgng}
\partial_\Gamma n_{\Gamma} =- n_{\Gamma} \int_0^{\infty} 
dl' \left(P_\Gamma^{+}(0,l') + P_\Gamma^{-}(0,l')\right)
\end{eqnarray}
We need also to introduce the average lengths 
\[
\overline{l}_\Gamma^{\pm}
=\int_0^{\infty} d\zeta \int_0^{\infty} dl\; l P_{\Gamma}^{\pm}(\zeta,l)
\]
of a $\pm$ bond,
and the total average length $\overline{l}_\Gamma
=\overline{l}_\Gamma^{+}+\overline{l}_\Gamma^{-}$ of a valley
that evolves as
\begin{eqnarray}   \label{rglength}
\partial_\Gamma \overline{l}_\Gamma =\overline{l}_\Gamma
\int_0^{\infty} 
dl' \left(P_\Gamma^{+}(0,l') + P_\Gamma^{-}(0,l')\right)
\end{eqnarray}
We have of course that $n_{\Gamma} \sim 1/\overline{l}_\Gamma$.

The RG equations (\ref{rg1}) derived here for Sinai's model
are identical to those derived to study the low energy
properties of the random transverse field Ising chain (RTFIC) 
in \cite{danfisher_rg2} (we choose notations
and conventions as in \cite{danfisher_rg2}) using a  perturbative
analysis of the effects of the   strongest bonds and fields. 
The reason for this is that the two models
are in fact formally related, 
as mentionned in the Introduction.
At the level of the RSRG equations, the mapping appears in
a very simple way: the local random fields $h_k$ and the random exchanges $J_k$ in the 
RTFIC correspond to the ascending and descending barriers
respectively, through the relations $F_{2 k}/T=-\ln h_k$ 
and $F_{2k+1}/T= -\ln J_k$. We can also identify the renormalization
scale $\Gamma$ in both models. For the diffusion model 
it corresponds to an Arrhenius time scale $t = t_0 \exp(\Gamma/T)$
to go over a barrier $F=\Gamma$,
whereas in the quantum model it corresponds to the minimal energy scale of
the levels which have been eliminated $\Omega = \Omega_0 e^{-\Gamma}$.
The duality between $J$ and $h$ 
in the RTFIC simply corresponds
to reversing the average force (i.e., $x \to -x$) in
Sinai's model. As will be discussed below, 
the deviation from criticality parameter $2 \delta$ in the RTFIC
corresponds to the parameter $\mu/T$ in Sinai's 
model (see \cite{derrida_pomeau,ledou_1d})
which controls the long time properties and the various phases 
and is defined for the original model with unit length bonds by 
\[
\langle\exp (-\mu f_n/T)\rangle =1.
\]
Zero drift corresponds to criticality in the RTFIC \cite{danfisher_rg2},
while the biased phase with zero velocity \cite{derrida_pomeau,ledou_1d}
corresponds to the Griffiths phase of the RTFIC \cite{danfisher_rg2}
as will be discussed below. Note however that the physical quantities
of interest in the two models can be different.

\subsubsection{Effective dynamics and validity of the method}

\label{validity}

Throughout the paper, we define the ``effective dynamics'' as
the dynamics which consists in putting the particle at time $t$
at the bottom of the renormalized valley
at scale $\Gamma = T \ln (t/t_0)$
which contains the starting point at $t=0$ (see Fig. \ref{fig2}).
Thus in the effective dynamics the particle does not move
unless one of the bonds  which are the sides 
of the renormalized valley to which it
belongs is decimated, in which case  it jumps to the bottom of the
new renormalized bond as in Fig. \ref{fig2}.
Here,  $t_0$ is 
a non-universal microscopic time scale, which 
throughout the paper we set to unity by appropriate 
redefinition of time units; we can then use interchangably
$\Gamma$ and $T \ln t$.

{\it Symmetric case} 

This effective dynamics is an approximation of the true dynamics.
But within the RG approach it can be seen that this 
approximation becomes better and better as $\Gamma= T \ln t$ increases
since the distribution of barriers $P_\Gamma(F)$ becomes broader and
broader, as  is detailed below.
Thus the renormalized landscape consists entirely of deep valleys 
separated by high barriers and with high probability the particle
will be near the bottom of the valley in which it began. 
Upon rescaling of space as $X=x/\Gamma^2$
the effective dynamics of the diffusion front becomes exact as $\Gamma$ tends 
to $\infty$ as was proven in ref. \cite{sinai}.
Indeed, the probability that the walker is close---in a 
precise sense that we discuss later---to its position given by the 
effective dynamics,  approaches one at long times. This 
stronger result has also been rigorously established. 
\cite{sinai,kesten,golosov}.

There is clearly a source of error in the approximation of the true dynamics
by the effective dynamics when two neighboring bonds have barriers $F$'s that are 
within order $T$ of each other. However, the error introduced by assigning 
a particle to one 
of two almost-equal-depth neighboring valleys rather than splitting its 
distribution 
between the two valleys will occur more and more rarely at long scales. 
Furthermore, any such error is wiped out by a later decimation which eliminates 
the two valleys in favor of a deeper valley. 
These errors thus lead only to subdominant
contributions to the quantities that we will compute---with the exception 
of tails of certain distributions 
which are dominated by rare configurations of the lansdscape. These
subdominant corrections can themselves be estimated via the RG. For instance
the rescaled mean square thermal width of a packet $\frac{1}{\Gamma^4} 
\overline{\langle x(t)^2\rangle -\langle x(t)\rangle^2}$ (with overbars
denoting
averaging over landscapes) tends to $0$ for large $\Gamma$ which is
its value in the effective dynamics, but has $1/\Gamma$ corrections 
coming from barrier degeneracies, estimated in Section (\ref{sub:rareevents}).

Strong differences between the real dynamics and the effective dynamics can
appear in some quantities, such as the persistence properties studied
in Sections (\ref{thermal averages}, \ref{truedynamics}). These quantities are 
usually in some way 
nonlocal in time and depend on 
the behavior of the system over time. Even in these cases though, as 
we show in Section (\ref{thermal averages}, \ref{truedynamics}),
it is possible to compute some of these quantities
by a proper interpretation and examination of the RG procedure. 

{\it Biased case}

In the biased case with a bias $\delta > 0$ one finds within the RG 
that the distribution of
barriers against the drift is no longer infinitely broad.
However if $\delta$ is small the barriers remain large enough 
so that the RSRG remains a good approximation. Again, this
approximation remains exact, in the same sense as above,
in the appropriate scaling limit
fixed $\delta \Gamma$ and $x/\Gamma^2$ (corresponding to the critical region 
of the RTFIC). For a fixed $\delta$ one expects that
the thermal packet is spread over several deep wells, but
when $\delta \to 0$ the contributions of these few additional wells
becomes subdominant. 

To conclude this Section, we stress that
despite its approximate character, our RSRG method allows us to
obtain exact results for many   quantities
both for the symmetric and the weakly biased Sinai model.

\subsubsection{RG with one boundary: reflecting or absorbing}

\label{boundaryrg}

We now consider the problem of diffusion in
a semi-infinite one-dimensional medium defined
as $x>0$. In practice there are two main types
of boundary conditions for the diffusing particle:
(i) reflecting (the current
at the boundary is zero) and 
(ii) absorbing (the probability is zero at the boundary).
We show in this section how both boundary conditions can
be treated by adding to the bulk RSRG specific rules 
near the boundary, which we call boundary
RSRG.

\begin{figure}[thb]
\centerline{\fig{12cm}{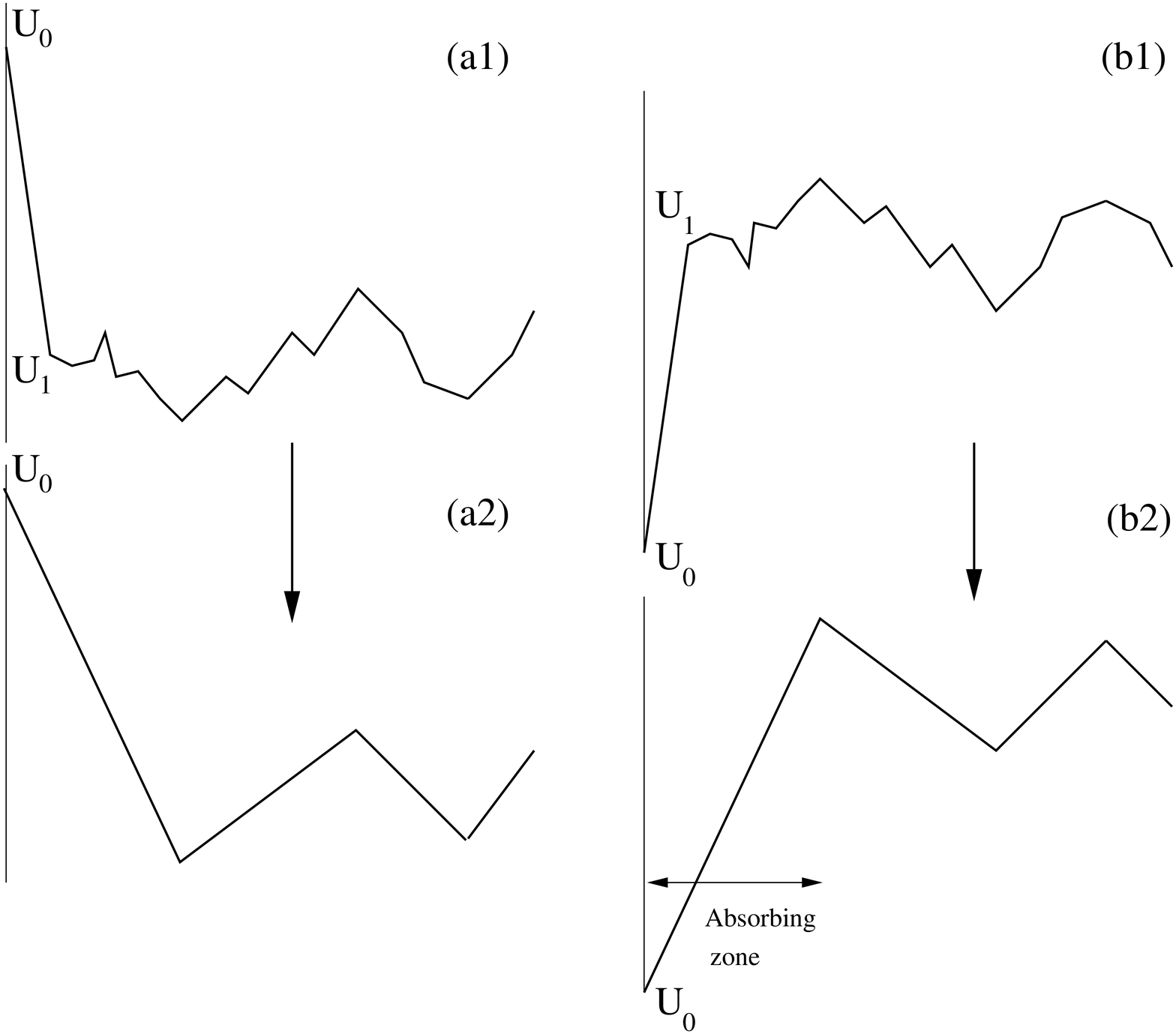}}
\caption{\label{figboundary} 
\narrowtext Illustration of the RG in presence of
a boundary. (a1) Reflecting boundary conditions: the boundary
at site $x=0$ can be represented by setting $U_0= +\infty$. 
(a2)~Renormalized landscape. (b1) Absorbing boundary conditions:
the boundary at site $x=0$ can be represented by setting $U_0= -\infty$
(b2) Renormalized landscape, with the absorbing zone (see text).}
\end{figure}

Let us start with the zero bias case and a reflecting boundary.
This condition
can be represented by placing a barrier with infinite
potential $U_0 = +\infty$ at $x=0$ with $U_1$  finite,
as is illustrated in Fig. \ref{figboundary}.

When grouping bonds with the same sign as in the previous
section, the first bond will always be descending with an infinite barrier
$F_1= +\infty$ and a length $l_1$. The decimation of the landscape
then proceeds as in the bulk case except that now the
first bond is never decimated and when the second bond 
gets decimated (at $\Gamma=F_2$) it simply increases the
length of the first bond $l_1' = l_1 + l_2 + l_3$.
One can easily see that starting
from a landscape where bonds are statistically uncorrelated---with
a distribution $E_\Gamma(l)$ for the first bond
and $P_\Gamma(F,l)$ for all the other bonds---they remain 
so under the boundary RSRG. Upon increase in $\Gamma$, the bulk
distribution $P$ obeys the same RG equation (\ref{rg1}) 
while $E$ satisfies:

\begin{eqnarray}  \label{brsrg}
\partial_\Gamma E_\Gamma (l) = - P_{\Gamma}(\Gamma) E_\Gamma(l)
+ E_\Gamma(.)*_l P_{\Gamma}(\Gamma,.)*_l P_{\Gamma}(.) 
\end{eqnarray}

The case of an absorbing boundary can be treated in the same
way since it amounts to setting the potential of the site $x=0$ to
$U_0=-\infty$. This is illustrated in Fig. \ref{figboundary}.
Thus the first bond will always be ascending
with an infinite barrier $F_1= +\infty$ and a length $l_1$
(and thus cannot be decimated). The rules are thus the same
as above with the same RG equation (\ref{brsrg}) for the distribution
$E_\Gamma(l)$ of the length of the first bond. The interpretation
is however different: the first bond represent an ``absorbing zone''
such that any particle starting from a 
point within this zone will be absorbed by the boundary before
time $\Gamma = T \ln t$, while the particles starting outside this
zone are still ``alive'' (and outside
this zone) at $\Gamma = T \ln t$ (with probability
asymptotically close to one).

We note at this stage that this equation coincides with
the RG equation for the endpoint magnetization in the
RTFIC; i.e., with the first exchange being $J_0=0$. Conversely, a 
reflecting boundary corresponds to the first transverse field being $h_1=0$.
The equivalence is reversed on the other end of the chain 
\cite{danfisher_rg2}. 

In the case of a bias, the probability distribution of
the first renormalized bond $E_{\Gamma}^{\pm}(l)$ 
($+$ when the bond is along the bias, and 
$-$ when it is against) satisfies:

\begin{eqnarray} \label{brsrgdrift}
\partial_\Gamma  E_{\Gamma}^{\pm}(l) = P_{\Gamma}^{\mp}(0,.) *_l E_{\Gamma}^{\pm} (.) *_l \int_0^{\infty} d\zeta' P_{\Gamma}^{\pm}(\zeta' ,.) -
E_{\Gamma}^{\pm}(l) \int_0^{\infty} dl' P_{\Gamma}^{\mp}(0,l')
 \end{eqnarray}
which generalizes  equation (\ref{brsrg}) of the zero bias case.

\subsection{General analysis of the RSRG equations}

\label{secsolutionrg}

In this Section we recall some results from references 
\cite{danfisher_rg1,danfisher_rg2,us_prl,danfisher_rg3} 
which will be used
extensively in this paper, about the large
$\Gamma$ behavior of the solutions of the RSRG equations 
(\ref{rg1}, \ref{brsrgdrift}) and 
discuss them in the context of the Sinai model.

\subsubsection{Symmetric model}

\label{subsecsolutionrgsymm}

We start with the symmetric Sinai model (zero bias, self-dual)
and thus the RG of a single distribution $P^+=P^-$. One first defines
the large-scale variance $\sigma$ of the potential as 
\[
\overline{(U_i-U_j)^2} 
\approx 2 \sigma |l_{i-j}|,
\]
with $l_{i-j}$ the distance from $i$ to $j$.
Since both $l_{i-j}$ and $U_i-U_j$ are preserved by the RG,
$\sigma$ is also preserved and determined by the initial model
as $2 \sigma = \int df f^2 Q(f)$. In the remainder of the
paper we will absorb $\sigma$ in $l$ and simply study
the case $\sigma=1$. The units of length are then $1/\sigma$.
To obtain the full results one must
change $l \to \sigma l$ (and $dl \to \sigma dl$) 
in the following formulae. The rescaled probability
$P_\Gamma(\eta,\lambda) \equiv \Gamma^3 P_\Gamma(\eta \Gamma, \lambda \Gamma^2)$
in terms of the rescaled variables 
\[
\eta=\zeta/\Gamma, \quad
\lambda=l/\Gamma^2
\]
 satisfies, when Laplace transformed 
in $\lambda\to p$:
\begin{eqnarray}
&& (\Gamma \partial_\Gamma - (1+\eta) \partial_\eta + 2 p \partial_p -1) P_\Gamma(\eta,p)
= P_\Gamma(0,p) P_\Gamma(.,p)*_\eta P_\Gamma(.,p) \nonumber
\end{eqnarray}
The fixed point solution \cite{danfisher_rg2} is found to be
\begin{eqnarray}  \label{solu} 
&& \tilde{P}(\eta,p) =  a(p) e^{- \eta b(p)} \\
&& {\rm with} \ \  a(p)=\frac{\sqrt{p}}{\sinh(\sqrt{p})} \qquad
b(p)=\sqrt{p}~\coth(\sqrt{p}). \nonumber
\end{eqnarray}
Thus, taking $p=0$, one finds the physically natural 
result that, due to the occurrence of long regions which are predominantly up
or predominantly down, the coarse-grained
probability distribution of barriers in Sinai's model
is {\it exponential}
\begin{eqnarray}   
P_{\Gamma}(F) \simeq \frac{\theta(F-\Gamma)}{\Gamma} e^{- \frac{(F-\Gamma)}{\Gamma}}
\end{eqnarray}
with a width which grows as $\langle F \rangle \sim \Gamma \sim T \ln t$.
The total number of bonds satisfies (\ref{rgng})  
and thus
decays asymptotically as $n_\Gamma \sim \Gamma^{-2}$ and
the average bond length (\ref{rglength}) grows as 
$\sim \Gamma^2$. Since
$\Gamma \sim \ln t$ one recovers Sinai's scaling \cite{sinai}
\begin{eqnarray}
x \sim \ln^2 t.
\end{eqnarray}

In the following we will need the explicit form of the distribution
$P(\lambda)=\int_0^{\infty} d \eta P(\eta,\lambda)$

\begin{eqnarray}  \label{pdel}
&& P(\lambda)  = LT^{-1}_{p \to \lambda } \left( \frac{a(p)}{b(p)}\right)
= LT^{-1}_{p \to \lambda } \left( \frac{1}{ \cosh(\sqrt{p})}\right) \\
&&= \sum_{n = -\infty}^{\infty} \left(n+\frac{1}{2}\right)
\pi (-1)^n e^{- \pi^2 \lambda \left(n+\frac{1}{2}\right)^2}
= \frac{1}{\sqrt \pi \lambda^{3/2}}
\sum_{m = -\infty}^{\infty} (-1)^m (m+\frac{1}{2})
 e^{-  \frac{1}{\lambda} (m+\frac{1}{2})^2}
\end{eqnarray}

It was shown in \cite{danfisher_rg2} that the convergence towards
the fixed point solution $P^*(\eta)=e^{-\eta}$ on the critical manifold
(i.e., symmetric perturbations) is like $\frac{1}{\Gamma}$
with eigenvector $P^{(1)}_{\Gamma}(\eta)=(\eta-1) e^{-\eta}$---corresponding
simply to a shift in $\Gamma$---plus other
parts that decay exponentially in $\Gamma$ and 
depend on tails in the initial distributions.

\subsubsection{Biased model}

\label{subbiasedrg}

In the case of the biased model one must follow the
descending bond distribution $P_{+}$ and the ascending bond 
distribution $P_{-}$ which are different. Contrary to the previous
section, it is more convenient in this case to use the unrescaled
distributions and variables. In terms of the Laplace transforms
$P_\Gamma^{\pm}(p,l) = \int_0^{+ \infty} dl e^{-p l} P_\Gamma^{\pm}(\zeta,l)$
the equation (\ref{rg1}) reads:

\begin{eqnarray}  \label{rg1laplace}
(\partial_\Gamma - \partial_\zeta) P^{\pm}(\zeta,p) =
P^{\mp}(0,p) P^{\pm}(., p) *_{\zeta} P^{\pm}(.,p)
+ (P^{\pm}(0,0) - P^{\mp}(0,0)) P^{\pm}(\zeta,p)
\end{eqnarray}

As was shown in \cite{danfisher_rg2}, for large $\Gamma$ the distributions
$P_{\pm}$ take the following form, in the scaling  regime of small $\delta$ and
small $p$ with $\delta \Gamma$ fixed and $p \Gamma^2$ fixed:

\begin{eqnarray} 
\label{solu-biased}
P_{\Gamma}^{\pm}(\zeta,p) & = & U_{\Gamma}^{\pm}(p) e^{- \zeta u_{\Gamma}^{\pm}(p)} \\
u_{\Gamma}^{\pm}(p) & = & \sqrt{p + \delta^2} \coth{[\Gamma \sqrt{p + \delta^2}]} 
\mp \delta  \\
U_{\Gamma}^{\pm}(p)  & = & \frac{\sqrt{p 
+ \delta^2}}{\sinh{[\Gamma \sqrt{p + \delta^2}]}} e^{\mp \delta \Gamma}
\label{solu-biased-u}
\end{eqnarray}
[Note that here we use $U$ rather than the $\Upsilon$ of 
\cite{danfisher_rg2}.] 
We will also use the evolution equations 
obtained by substituting (\ref{solu-biased}) in the RG equation
(\ref{rg1laplace}):

\begin{eqnarray} \label{equpm}
&& \partial_\Gamma u^{\pm}_\Gamma (p) 
= - U^{+}_\Gamma (p) U^{-}_\Gamma (p) \\
&& \partial_\Gamma U^{\pm}_\Gamma (p) 
= - u^{\mp}_\Gamma (p) U^{\pm}_\Gamma (p) 
\end{eqnarray}

The distributions of barriers alone are:

\begin{eqnarray} \label{solu-biased}
P^{-}(\zeta) = \frac{2 \delta}{1 - e^{-2 \Gamma \delta}} 
\exp\left(- \zeta \frac{2 \delta}{1 - e^{-2 \Gamma \delta}}\right) \\
P^{+}(\zeta) = \frac{2 \delta}{e^{2 \Gamma \delta} - 1} 
\exp\left(- \zeta  \frac{2 \delta}{e^{2 \Gamma \delta} - 1}\right)
\end{eqnarray}
and the average lengths of the ($\pm$)-bonds are respectively given by:

\begin{eqnarray} \label{lengthpm}
&&\overline{l}^{+} = \frac{1}{2 \delta^2} (e^{\delta \Gamma} \sinh(\delta \Gamma)- \delta \Gamma) \\
&&\overline{l}^{-} = \frac{1}{2 \delta^2} ( \delta \Gamma-e^{-\delta \Gamma} \sinh(\delta \Gamma) )
\end{eqnarray}
When $\delta \to 0$ one has $\overline{l}^{\pm} \to \frac{1}{2} \Gamma^2$.
and thus the total average length of a valley is:

\begin{eqnarray}
\overline{l}_\Gamma = \overline{l}^{+} + \overline{l}^{-} =
\left(\frac{\sinh(\Gamma \delta)}{\delta}\right)^2 \sim n_\Gamma^{-1}
\end{eqnarray}
where $n_\Gamma$ is the total number of bonds.

The convergence towards the solution (\ref{solu-biased})
has been discussed in \cite{danfisher_rg2}. 
The above solution (\ref{solu-biased})
thus depends on an ``integration constant'' 
$2 \delta = u_\Gamma^{-}(p=0) - u_\Gamma^{+}(p=0)$
which is determined by the initial condition, and is proportional
to the drift. In \cite{danfisher_rg2} it was identified for small $\delta$ as
the ratio of the mean to the variance of the {\it original} distribution 
$Q(f)$ of the initial independent bonds of unit length 1; i.e. before
grouping the bonds together:
\begin{eqnarray}
\delta = \frac{\overline{f}}{\overline{f^2}-
\overline{f}^2}. 
\label{delta}
\end{eqnarray}

It is  useful to introduce a parameter
$\mu$ defined by the unique non zero solution of the
equation
\begin{eqnarray} \label{defmu}
\overline{e^{- \mu f/T}} = \int_{-\infty}^{+\infty} df Q(f) e^{- \mu f/T} = 1
\end{eqnarray}
This parameter $\mu$ has been introduced previously
in \cite{kesten1} and is  known to determine
exactly the various phases of the dynamics of the Sinai
model with a bias ($x \sim t^{\mu}$ for $\mu>1$,
$x \sim t$ for $\mu>1 $). Indeed it is also
known in random walk theory \cite{randomwalk}
to control the probability of large excursions against
the bias. We now show that we can interpret these
properties within the RG as associated with
the exact decimation of the landscape (\ref{rg1}). 

The distributions $P_0^{\pm}(F)$ of the barriers $F>0$ of
the zig-zag landscape (see Fig. \ref{fig2})
obtained by grouping together the consecutive ascending
or descending bonds of
the original discrete model, are related to the original $Q(f)$
distribution through
 \begin{eqnarray}
\int_0^{\infty} dF e^{-sF} P_0^{\pm}(F) =
 \frac{Q^{\mp}(0)}{Q^{\pm}(0)} ~~ \frac{Q^{\pm}(s)}{1-Q^{\pm}(s)}
\end{eqnarray}
where $Q^{+}(s)=\int_{0}^{+\infty} df e^{- s f} Q(f)$
 and $Q^{-}(s)=\int_{-\infty}^{0} df e^{ s f} Q(f)$.
The difference of potential $(F_+-F_-)$
of the boundaries of
a valley of the initial zig-zag potential has  Laplace transform
 \begin{eqnarray}
\int_0^{\infty} dF_+ e^{-s F_+} P_0^{+}(F_+)
\int_0^{\infty} dF_- e^{s F_-} P_0^{-}(F_-) =
\frac{Q^{+}(s) Q^{-}(-s)}{1- Q(s) + Q^{+}(s) Q^{-}(-s)}
 \end{eqnarray}
where $Q(s)=Q^{+}(s) + Q^{-}(-s)=\int_{-\infty}^{+\infty} df Q(f) e^{- s f}$.
The definition of $\mu$, Eq. (\ref{defmu}), for the original model
is thus equivalent for the initial zig-zag landscape to  defining  $\mu$
as the unique non-zero solution of the equation
\[
\langle e^{- \mu F_+ /T}\rangle _{P_0^+} 
\langle e^{ \mu F_-/T}\rangle_{P_0^-}= 1.
\]
But since the renormalized valleys at scale $\Gamma$ are 
constructed from the valleys of the initial zig-zag potential
and are statistically uncorrelated, this implies that for the probability
distributions $P_{\Gamma}^{\pm}(F)$ of the renormalized
barriers at any scale $\Gamma$,

 \begin{eqnarray}
\langle e^{- \mu F_+/T}\rangle_{P_{\Gamma}^+} \langle e^{ \mu F_-/T}
\rangle_{P_{\Gamma}^-}
\equiv \int_{\Gamma}^{\infty} dF_+ e^{- \mu F_+/T} P_{\Gamma}^{+}(F_+)
\int_{\Gamma}^{\infty} dF_- e^{ \mu F_-/T} P_{\Gamma}^{-}(F_-) = 1
 \end{eqnarray}

Using the explicit solutions (\ref{solu-biased}) for the
  distributions of barriers, we obtain in terms of 
 $u_{\Gamma}^{\pm}=u_{\Gamma}^{\pm}(p=0)$ the following equation for $\mu$
\begin{eqnarray}
1= \left( \frac{u_{\Gamma}^{+}}{u_{\Gamma}^{+}+ \frac{\mu}{T}} \right)
\left( \frac{u_{\Gamma}^{-}}{u_{\Gamma}^{-}- \frac{\mu}{T}} \right)
=\frac{1}{1+\frac{\mu}{T u_{\Gamma}^{+}u_{\Gamma}^{-}} 
\left( (u_{\Gamma}^{-} -u_{\Gamma}^{+}) - \frac{\mu}{T} \right)} 
\end{eqnarray}
and we thus obtain that the parameter 
$ 2 \delta =u_{\Gamma}^{-} -u_{\Gamma}^{+}$ parametrizing the RG solutions
(\ref{solu-biased}) indeed corresponds to the parameter $\mu/T$.
Note, however, that the expression (\ref{delta}) is only 
valid for small $\delta$.

Thus even away from small $\mu$, 
the RSRG allows one to obtain exact information on the structure
of the landscape, in particular the behavior of the probability of large 
barriers impeding the drift. For large $\Gamma$ we have, from (\ref{solu-biased})
that (for positive $\delta$) $P^-(\zeta)\approx 2\delta e^{-2\delta\zeta}$
implying that the probability of a large barrier, $F$, is
$ \sim \exp(- \mu F/T)$.
As we will see shortly,
this controls the anomalous drift exponent $\mu$.

\subsubsection{Boundary fixed point solutions}

\label{finitesolu}

The RG equation for the distribution of lengths of the boundary bond
$E^{\mp}(l)$ defined in section (\ref{boundaryrg})
was given in (\ref{brsrgdrift}).
In the Laplace variable with respect to  length the RG equation
(\ref{brsrgdrift}) reads:

\begin{eqnarray}
\partial_\Gamma  E_\Gamma^{\pm}(p) = 
E_\Gamma^{\pm}(p) ( P_\Gamma^{\mp}(0,p) P_\Gamma^{\pm}(p) -
P_\Gamma^{\mp}(0,p=0) )
\end{eqnarray}

For large $\Gamma$ using the properties of the fixed point
solution (\ref{solu-biased}) for $P^\pm$ and
the properties (\ref{equpm}) of the functions $U$ and $u$,
this can be rewritten as:

\begin{eqnarray}
&& \partial_\Gamma  \ln E_\Gamma^{\pm}(p) 
= \frac{U_\Gamma^{+}(p) U_\Gamma^{-}(p)}{u_\Gamma^{\pm}(p)} - u_\Gamma^{\mp}(0)
= \partial_\Gamma ( \ln U_\Gamma^{\pm}(0) - \ln u_\Gamma^{\pm}(p) )
\end{eqnarray}
Finally we find:

\begin{eqnarray}
E_\Gamma^{\pm}(p) = \frac{u_\Gamma^{\pm}(0)}{u_\Gamma^{\pm}(p)}
=\frac{\delta e^{\mp \delta \Gamma}}{
\sinh(\delta \Gamma) ( \sqrt{p + \delta^2}
 \coth{[\Gamma \sqrt{p + \delta^2}]} \mp \delta )} 
\label{finitex}
\end{eqnarray}

In the symmetric case $\delta=0$ we get

\begin{eqnarray}
E_{\Gamma}(p) =  \frac{\sinh{\Gamma\sqrt{p}}}{ \Gamma \sqrt{p} \cosh{\Gamma \sqrt{p}}}
\end{eqnarray}
whose inverse Laplace transform reads in the rescaled variable
\[
\lambda\equiv \frac{l}{\Gamma^2}
\]
is
\begin{eqnarray} \label{elambda}
E(\lambda)= \sum_{n=-\infty}^{+\infty} e^{-\lambda \pi^2 (n+\frac12)^2} 
=  \frac1{\sqrt{\pi \lambda}}  \sum_{m=-\infty}^{+\infty} (-1)^m 
e^{- \frac{m^2}{\lambda}} 
\end{eqnarray}

\section{Single time diffusion properties in Sinai model}

\label{secsinglediffusion}

In this Section we study the diffusion front using the
effective dynamics introduced in  Section (\ref{validity}).

\subsection{Single time diffusion front for the symmetric model}

\label{secsinglediffusionsymm}

We now consider the (single time) diffusion front, i.e.,  the
probability $\mbox{Prob}(x,t|x_0,0)$ that a particle starting
at $x_0$ at $t=0$ be located at $x$ at time $t$, for the symmetric,
zero bias case. At large time $t$ the effective (renormalized)
dynamics corresponds to moving the particle from its starting point 
$x_0$ to the lower-potential end of the renormalized bond at
scale $\Gamma = T \ln t$ that contains $x_0$. This is illustrated in
Fig \ref{fig2}. In a {\it single} environment $\mbox{Prob}(x,t|x_0,0)$ is thus
localized near the bottom of the bond---i.e., the bottom of a 
valley---and the rescaled position
$x/\ln^2 t$ has a delta function shape at large time.

One can compute averages over environments, or equivalently over
initial conditions $x_0$ (with a spatially uniform measure)
in a single environment. The average diffusion front
$\overline{\mbox{Prob}(x_0+x,t|x_0,0)}$ is obtained as follows. 
The probability that a given
bond has length $l$ is $P_\Gamma(l)=\int d\eta P_{\Gamma}(\eta,l)$ and the 
probability density that $x_0$ belongs to a renormalized bond of length
$l$ at scale $\Gamma$ is $l P_\Gamma(l)/\int_l l P_\Gamma(l)$.
Taking into account that the 
distance $|x|$ between the starting point $x_0$ and the
bottom of the bond is uniformly distributed on $[0,l]$,
one finds after averaging over $l$:
\begin{eqnarray}
\overline{\mbox{Prob}(x,t|0,0)} = \frac{1}{2 \int_l l P_\Gamma(l)} 
\int_{|x|}^{\infty} dl P_\Gamma(l)
\end{eqnarray}

Using the fixed point solution (\ref{solu},\ref{pdel}) with
$\Gamma= T \ln t$, we find that the diffusion front takes
the scaling form:

\begin{eqnarray} \label{kesten}
&& \overline{\mbox{Prob}(x,t|0,0)}=\frac{\sigma}{T^2 \ln^2 t} 
q\left(\frac{\sigma x}{T^2 \ln^2 t}\right) \\
\mbox{with}\qquad&& q(X) = \frac{4}{\pi} 
\sum_{n = 0}^{\infty} \frac{(-1)^{n}}{2 n+ 1}
e^{- \frac{1}{4} \pi^2 |X| (2 n + 1)^2}
\end{eqnarray}
where we have reinserted $\sigma$. This 
coincides with the Kesten-Golosov
rigorous result \cite{kesten,golosov2}
for a Brownian potential, as it should \cite{sinai}
since our method gives  exact results for properties of the 
rescaled walks $x(t)/\ln^2t$.

\subsection{Single time diffusion front for the biased model}

\label{secsinglediffusionbias}

The case of a global bias $\langle f \rangle_Q >0$ is described
by the RG equations (\ref{rg1}) with $P^{+} \neq P^{-}$.
The fixed point (\ref{solu-biased})
was analyzed in the previous section. It shows that at
large scales $\Gamma$ the barriers impeding the drift 
have an exponential distribution that does not continue to broaden:
\begin{eqnarray}
P^{-}_\Gamma(F) \sim 2 \delta e^{- 2 \delta (F-\Gamma) } \theta(F-\Gamma)
\label{pminus}
\end{eqnarray}
 On the other hand, the
bonds along the drift become very long with large barriers:
\begin{eqnarray}
P_\Gamma^+(F) \sim \frac{1}{F_\Gamma}  e^{- \frac{(F-\Gamma)}{F_\Gamma} } 
\theta(F-\Gamma)
\end{eqnarray}
where $F_\Gamma \sim \frac{1}{2\delta} e^{2 \delta \Gamma} \sim 
\frac{1}{2\delta} t^{\mu}$. Asymptotically
in the RG only barriers impeding the drift are decimated, since the 
barriers to go against the bias are very large. The distribution 
Eq. (\ref{pminus}) is then simply that of potential drops between the impeding 
barriers. 
One thus recovers
the physical picture \cite{ledou_1d} that 
Sinai's biased diffusion renormalizes onto a directed model
with traps (ascending bonds) of release times $\tau$ with 
distribution  $\rho(\tau) \sim
\tau^{-(1+\mu)}$. The average length $\overline{l}_{\Gamma}^+$
of the descending bond distribution Eq. (\ref{lengthpm})
yields the anomalous diffusion scaling $x \sim t^{\mu}$.

We now compute the average diffusion front $\overline{\mbox{Prob}(x,t|0,0)}$
in the case of  a small average potential 
drop per unit length $2\delta >0$. The argument is as in
the symmetric case, except that one must distinguish
$x>0$ from $x<0$, which correspond respectively to the
starting point being on a descending, $(P^{+})$ or ascending
$(P^{-})$ renormalized bond at scale $\Gamma$. One thus uses
the formula:

\begin{eqnarray}
\overline{\mbox{Prob}(x,t|0,0)} =
\frac{1}{\overline{l}_\Gamma} \left[ 
\theta(x) \int_x^{+\infty} dl P_\Gamma^{+}(l) 
+ 
\theta(-x) \int_{-x}^{+\infty} dl P_\Gamma^{-}(l)  \right]
\end{eqnarray}

This yields, in the scaling limit where
$\Gamma$ is large while $\lambda = l/\Gamma ^2$ and $\gamma \equiv  \Gamma \delta$ are 
both fixed but arbitrary, the generalization of Eq(\ref{kesten}):
\begin{eqnarray} \label{biased}
&& \overline{\mbox{Prob}(x,t|0,0)} = \frac{\sigma}{T^2 (\ln t)^2} 
q\left( X= \frac{\sigma x}{T^2 (\ln t)^2}, \gamma = T \delta \ln t\right) \\
&& \mbox{with}  \nonumber \\
&& q(X,\gamma) = \left( \frac{\gamma}{\sinh \gamma} \right)^2 
\left[ \theta (X) LT^{-1}_{s \to X}\;
\frac{1}{s}\left( 1- \frac{\kappa 
e^{-\gamma }}{\kappa \cosh \kappa - \gamma\sinh \kappa}\right) \right. 
\nonumber \\
&& + \theta (-X) LT^{-1}_{s \to - X}\;\left. \frac{1}{s}
\left( 1- \frac{\kappa e^\gamma }{\kappa \cosh \kappa + \gamma\sinh \kappa}
\right) \right],
\end{eqnarray}
with $\kappa \equiv \sqrt{s+\gamma ^2}$. 
From this expression we can compute the  moments. One finds:

\begin{eqnarray}
\overline{\langle x(t) \rangle} & = & \frac{1}{8 \delta^2 \sinh[\gamma]^2} (
  \sinh[4 \gamma] - 6 \gamma \cosh[2 \gamma] +\sinh[2 \gamma]  ) \\
\overline{\langle x(t)^2 \rangle} & = &
\frac{1}{16 \delta^4 \sinh[\gamma]^2 }
( \cosh[6 \gamma] - 10 \gamma \sinh[4 \gamma] + 3 \cosh[4 \gamma] 
+ 18 \gamma^2  \cosh[2 \gamma]  \\
&&\;\;  - 12 \gamma \sinh[2 \gamma] + \cosh[2 \gamma] + 2 \gamma^2 -5 )
\nonumber
\end{eqnarray}
Note that $\overline{\langle
 x(t) \rangle} \approx \frac{3}{5} \delta \Gamma^3$ for small
$\gamma= \delta \Gamma$, a form implied by scaling and analyticity in $\delta$.

One can also perform the Laplace inversion. 
For $\gamma < 1$ let us introduce the roots 
$\alpha^{\pm}_{n}(\gamma)$ ($n=0,1,...$) of the equation:

\begin{eqnarray}
\alpha^{\pm}_{n}(\gamma) {\rm cotan} (\alpha^{\pm}_{n}(\gamma)) = \pm \gamma
\qquad \hbox{with} \qquad  n \pi < \alpha^{\pm}_{n}(\gamma) < (n+1) \pi
\label{anlaplace}
\end{eqnarray}
For $\gamma > 1$, the root $\alpha^{+}_{0}(\gamma)$ does not exist,
but is replaced by the positive root $\tilde{\alpha}^{+}_{0}(\gamma)$
of the equation $\tilde{\alpha}^{+}_{0}(\gamma) \coth(\tilde{\alpha}^{+}_{0}(\gamma)) =  \gamma$.
In terms of these roots, the Laplace inversion gives
\begin{eqnarray} \label{kestendrift}
q(X,\gamma) = \theta(X)
\sum_{n = 0}^{\infty} c^{+}_{n}(\gamma) e^{- X s_n^+(\gamma)}
+
\theta(- X)
\sum_{n = 0}^{\infty} c^{-}_{n}(\gamma) e^{- |X| s_n^-(\gamma)}
\end{eqnarray}
where 
\begin{eqnarray} \label{snlaplace}
&& s_n^{\pm}(\gamma)=\gamma^2+ (\alpha^{\pm}_{n}(\gamma))^2 \\
&& c_n^{\pm}(\gamma)=\left(\frac{\gamma}{\sinh \gamma} \right)^2
\frac{ 2 (-1)^{n+1} (\alpha^{\pm}_{n}(\gamma))^2 e^{\mp \gamma} }
{ \sqrt{\gamma^2+ (\alpha^{\pm}_{n}(\gamma))^2} (\gamma^2
+ (\alpha^{\pm}_{n}(\gamma))^2 \mp \gamma )}
\end{eqnarray}
except for the term $n=0$ in the domain $X>0$ and $\gamma>1$ for which
\begin{eqnarray}
&& s_0^{+}(\gamma>1)=\gamma^2- (\tilde{\alpha}^{+}_{0}(\gamma))^2  \\
&& c_0^{+}(\gamma>1)=\left(\frac{\gamma}{\sinh \gamma} \right)^2
\frac{ 2  (\tilde{\alpha}^{+}_{0}(\gamma))^2 e^{- \gamma} }
{ \sqrt{\gamma^2- (\tilde{\alpha}^{+}_{0}(\gamma))^2} (\gamma
+(\tilde{\alpha}^{+}_{0}(\gamma))^2-\gamma^2)}
\end{eqnarray}
Note that $s_0^{+}(\gamma)$ is an analytic function of
$\gamma$ despite its definition by two domains. 
In the limit of small $\gamma$, we recover of course
the symmetric case Eq. (\ref{kesten}) using 
$\alpha^{\pm}_{n}(\gamma \to 0) = (1+2n) \frac{\pi}{2}$.
For large $\gamma \gg 1$, i.e., $T \ln t \gg 1/\delta$,
we have $\tilde{\alpha}^{+}_{0}(\gamma) \simeq
\gamma (1-2 e^{-2 \gamma}+...)$ and thus
\begin{eqnarray} \label{largegamma}
&& s_0^{+}(\gamma \gg 1) \simeq  4 \gamma^2 e^{-2 \gamma} \\
&& c_0^{+}(\gamma \gg 1) \simeq  4 \gamma^2 e^{-2 \gamma}
\end{eqnarray}
whereas all other coefficients in exponentials are much bigger
since $s_n^{\pm}(\gamma) > \gamma^2$.
In the regime $\gamma \gg 1$, the distribution is thus heavily 
concentrated to the right 
of the origin and reduces to the simple exponential:
\begin{eqnarray} 
\overline{\mbox{Prob}(x,t|0,0)} \approx \theta (x) \exp [-x/\overline{x(t)}] / 
\overline{x(t)}  \label{asymptotic}
\end{eqnarray}
with the mean displacement 
\[
\overline{x(t)} \approx t^{2 \delta T} / (4 \delta ^2). 
\]

We can now compare with known results
\cite{sinai,ledou_1d}: for fixed $0< \mu< 1$ 
the variable $\tilde{x} = x/t^\mu$ is distributed with
a half-sided Levy probability density $L_{\mu}(\tilde{x}^{-1/\mu}) 
d \tilde{x}^{-1/\mu}$ where $L_{\mu}(z) = LT^{-1}_{s \to z} 
e^{- C_{\mu} s^\mu}$. Our asymptotic result (\ref{asymptotic}) 
reproduces correctly the the small $\mu$ limit 
of this Levy front \cite{sinai,ledou_1d} with the correct prefactor
$C_\mu$.

\section{Motion of thermal averages: returns to the origin and jumps}

\label{thermal averages}

We now study ``recurrence'' properties of the Sinai model.
One must carefully distinguish between the {\it effective 
dynamics} (i.e.,  the walker jumping between valley bottoms)
and the {\it real} dynamics. In this section we concentrate on the
{\it effective dynamics}. This amounts, as we will see, to 
studying the fine structure of the motion of the (thermal) packet.
Asking similar question for a single particule requires 
a study in the presence of absorbing walls 
and will be discussed 
in the next section. We will also study the zero crossings of 
the ``running average''
\begin{eqnarray}
\Xi(t) \equiv \frac{1}{t} \int_0^t x(\tau) d\tau
\end{eqnarray}
which is an approximation to the thermal average.

While in a single ``run'' in a given environment the walker 
typically crosses its starting point many times while trapped in a 
valley,  averaging over many runs in the same environment yields 
a $\langle x(t) \rangle$ which crosses $x_0$ exactly once each time 
the bond on which $x_0$ lies is decimated, since this causes 
its valley bottom to cross $x_0$.

We will first ask what is the fraction $M_k(t)$ of starting points, $x_0$,
for which the {\it  thermally averaged
position} $\langle x(t) | x(0)=x_0 \rangle$ has crossed $x_0$ exactly $k$ times
up to time $t$. Since the effective dynamics consists in putting
the particule at the lowest point of the decimated bond, the origin and 
the particule remain in the same bond at all times. The 
probability of crossing the origin---i.e., the starting point---between 
$0$ and $t$ exactly $k$ times is thus
the fraction of sites which belong to bonds {\it which have
changed orientation} exactly $k$ times between $0$ and $t$.
In particular, the probability of no return to the origin
is $M_0(t)$ and is equal to the probability that the
bond containing the origin has never been decimated. 

\subsection{Number of returns to the origin: symmetric case}

\label{sub:returns}

To compute $M_k(t)$ we use two equivalent methods, which we
both describe as they will be useful in the remainder of the paper.

{\it First method}:
Let $N_{\Gamma}(k,\eta)$ be the probability that the bond containing 
the origin has a rescaled barrier
$\eta = (F-\Gamma)/\Gamma$ at $\Gamma$ and has switched its orientation $k$ times
up to scale $\Gamma$. It is normalized as $\sum_{k=0}^{+\infty} \int_0^{+\infty}
d \eta N_{\Gamma}(k,\eta) = 1$ and it satisfies :

\begin{eqnarray}
&& \left(\Gamma \partial_\Gamma - (1+ \eta) \partial_\eta - 1 \right)
N_{\Gamma}(k,\eta) = 2 P_{\Gamma}(0) N_{\Gamma}(k,.) *_\eta P_{\Gamma}(.)  \\
&& - 2 P_{\Gamma}(0) N_{\Gamma}(k,\eta) + 
N_{\Gamma}(k-1,0) P_{\Gamma}(.)*_\eta P_{\Gamma}(.) \nonumber
\end{eqnarray}
Introducing the generating function 
$\hat N_{\Gamma}(z,\eta)= \sum_{k=0}^{+\infty} N_{\Gamma}(k,\eta) z^k$,
we obtain using the fixed point solution $P_{\Gamma}(\eta)=e^{-\eta}$
\begin{eqnarray}
\left(\Gamma \partial_\Gamma - (1+ \eta) \partial_\eta - 1 \right)
\hat N_{\Gamma}(z,\eta) = 2 \hat N_{\Gamma}(z,.) *_\eta e^{-\eta}
 - 2 \hat N_{\Gamma}(z,\eta) + z \hat N_{\Gamma}(z,0) \eta e^{-\eta}
\end{eqnarray}
We look for a solution of the form

\begin{eqnarray} \label{62a}
&&\hat N_{\Gamma}(z,\eta) = \Gamma^{-\Phi(z)} (a(z)+b(z) \eta  ) e^{-\eta}
\end{eqnarray}
and  find a quadratic equation $(\Phi(z)-1) (\Phi(z)-2)=1+z$ so that
\begin{eqnarray}
\Phi(z)= \frac{3 - \sqrt{5 + 4 z}}{2}
\label{phiz}
\end{eqnarray}

{\it Second method}: this consists of associating with each bond 
a set of 
auxiliary variables $m(k)$ counting, respectively, the number of sites
on the bond  which
have changed orientation exactly $k$ times since $t=0$.
The RG rules for these variables upon decimation of bond (2) read 
(see Fig \ref{fig2})
\begin{eqnarray}
&& m'(k) = m_1(k) + m_2(k-1) + m_3(k)  \\
&& m'(0)= m_1(0) + m_3(0).
\end{eqnarray}
Introducing the generating function 
$m(z)=\sum_{k=0}^{\infty} m(k) z^k$ one finds, for a fixed $z$
the RG rule
\begin{eqnarray}  \label{rulez}
m'(z) =  m_1(z) + z m_2(z) + m_3(z)
\end{eqnarray}
A method to analyze such rules is to write the RG equation
for the bond joint distribution $P(\eta,m)$. This is
hard to solve, however the RG equation for the first moment
$c(\eta) = \int_m m P(\eta,m)$ can be solved. Interestingly,
the type of combination rule (\ref{rulez}) under RSRG has been studied
in \cite{danfisher_rg2} in the context of quantum spin chain models.
We have recalled that analysis in
the Appendix \ref{rulesymmetric}, and generalized it to the recursion relation
$m'=a m_1+b m_2+c m_3$, which we will extensively need in the
present problem.
In particular, for $z=0$, which
corresponds to computing the probability of no return to the origin
$M_0(t)$, the rule is simply
$m' = m_1 + m_3$. Remarkably, this is exactly the 
the same as the one for the magnetization 
in the RTFIC \cite{danfisher_rg2}. Thus there is an
interesting relation between the magnetization of the RTFIC
and persistence properties in the Sinai model.

Since we are interested in the fraction of initial conditions with
$k$ crossings, we need the ratio $\frac{m(z)}{\overline{l}_{\Gamma}}$
where the characteristic length
$\overline{l}_{\Gamma}$ grows as $\Gamma^2$. 
The results from \cite{danfisher_rg2} and Appendix \ref{rulesymmetric}
is that the ratio $\frac{m(z)}{\overline{l}_{\Gamma}}$ decays as
$\Gamma^{-\Phi(z)}$ with $\Phi(z)=(3-\sqrt{5+4z})/2$
in agreement with the first method (\ref{phiz}).

{\it Results: returns to the origin and multifractality}:
From the above result we can extract several consequences.
First, setting $z=0$ we directly find that
the probability that a thermally averaged trajectory
does not return to its starting point decays, in terms of
$\overline{l}(t) \sim (T \ln t )^2$, as:

\begin{eqnarray}
M_0(t) \sim \overline{l}(t)^{-\overline{\theta}}  
\qquad \mbox{with}\qquad \overline{\theta} = \frac{3-\sqrt{5}}{4}.
\end{eqnarray}
Second, it is natural to introduce the rescaled number
of returns to the origin:

\begin{eqnarray}
g=\frac{k}{\ln \Gamma}=\frac{k}{\ln (T \ln t) }
\end{eqnarray}
and to define the generalized persistence
exponent $\overline{\theta}(g)$
characterizing the asymptotic decay of the probability 
distribution of $g$:
\begin{eqnarray}
\mbox{Prob}(g) \sim  \overline{l}(t)^{-\overline{\theta}(g)}
\end{eqnarray}

We now compute $\overline{\theta}(g)$ from the above 
generating functional (\ref{62a}). By
definition:
\begin{eqnarray}
\int_0^{\infty} d\eta \hat N_{\Gamma}
(z,\eta) \propto \ln \Gamma \int_0^{\infty} dg
~ z^{g \ln \Gamma} \Gamma^{-2\overline{\theta}(g)}=
\ln \Gamma \int_0^{\infty} dg ~~
e^{- \ln \Gamma \left( 2\overline{\theta}(g)-g\ln z \right)}
\end{eqnarray}
Since we know that $N_{\Gamma}(z) \sim \Gamma^{- \Phi(z)}$
we then obtain, using the saddle point method, that
$\Phi(z)=2\overline{\theta}(g^*(z))-g^*(z)\ln z$ where  $g^*(z)$
is the solution of $2 \overline{\theta}~'(g^*(z))=\ln z$.
Properties of Legendre transforms thus give that, reciprocally,
the exponent $\overline{\theta}(g)$ is given by
$2 \overline{\theta}(g)=\Phi(z^*(g))+g \ln z^*(g)$
where $z^*(g)$ is the solution of $\Phi'(z^*(g))=- {g \over {z^*(g)}}$.
We find  $z^*(g)=2g\left(g+\sqrt{g^2+5/4}\right)$ and thus:
\begin{eqnarray}
 \label{multi} 
\overline{\theta}(g) = \frac{g}{2} \ln\left[2 g\left(g+
\sqrt{g^2+\frac{5}{4}} \right)\right] + 
\frac{3}{4} - \frac{g}{2}
- \frac{1}{2} \sqrt{g^2+\frac{5}{4}}.
\end{eqnarray}
The exponent $\overline{\theta}(g)$ is a positive convex function :
it decays from $\overline{\theta}(g=0)=\frac{3-\sqrt{5}}{4}$
(for $g=0$ we of course recover the value found previously
when studying the probability of no return to the origin up to $\Gamma$)
to $\overline{\theta}(\frac{1}{3})=0$, and then grows again for $g>1/3$.
This implies that:

\begin{eqnarray}
g = \frac{k}{\ln (T \ln t)} \rightarrow \frac{1}{3}  
\qquad \hbox{with probability 1 at large time}
\label{g13}
\end{eqnarray}

All of the moments of g will be dominated by the typical behavior;
i.e. $\langle g^m\rangle\equiv 3^{-m}$ for all m. The full dependence on $g$
of the $\overline{\theta}(g)$ function describes the 
{\it tails} of the probability distribution of the number of returns
of $\langle x(t)\rangle$, i.e., the large deviations.

{\it Returns to the origin for the running average}:

For a given walker, $\Xi(t) \equiv \frac{1}{t} \int_0^t 
x(\tau) d\tau$ will typically behave like $\langle x(t) \rangle$.  We 
conjecture that
the probability of $k=g \ln(\ln t)$ sign changes of $\Xi(t)$ up to time $t$
decays with the same exponent $\overline{\theta}(g)$ for $g \leq \frac{1}{3}$.
For larger $g$, the behavior is dominated by rare valleys with closely
spaced and almost degenerate minima on opposite sides of the origin
which yield extra sign changes in $\Xi(t)$. We now estimate the
contribution of these rare events. 

We are interested in situations in which the number of
zero crossings $N(t)$ of $y(t)=\int_0^t x(\tau)\; d\tau$
is much greater than those of $\langle x(t) \rangle$ whose
statistics we know. The dominant contributions are from
configurations of the random potential for which the valley bottom
in which the origin lies is split in two halves on opposite side
of the origin for a very long time. In such configurations the
valley has two minima at $x_{+}>0$ and $x_{-}<0$ with a small
free energy difference $T \epsilon$,  separated by a
barrier of some height $\Gamma_0$. The key point is
that if the mean rate of change $\langle dy/dt\rangle
\approx \langle x\rangle_{\mbox{valley}} \approx
(x_{+} + x_{-} e^{-\epsilon})/(1+e^{-\epsilon})$
happens to be very close to zero, $y(t)$ will change sign an anomalously
large number of times. To estimate the corresponding
number of crossings, one can consider, crudely,
that $y(t)$ performs a biased random walk, with steps 
of order $x_{\pm} e^{\Gamma_0/T}$
and an average drift $e^{\Gamma_0/T} \langle x\rangle_{\mbox{valley}}$ 
(since the typical time between jumps is $e^{\Gamma_0/T}$).
Using well-known results for biased random walks,
we can estimate that if the particle is trapped in
the double valley for a time $\tau \gg e^{\Gamma_0/T}$,
the typical number of zero crossings of $y(t)$
in that time interval will be:
\begin{eqnarray}
N(\tau) \sim \min\left( \frac{x_{\pm}}{\langle x\rangle_{\mbox{valley}}}, 
\sqrt{\tau e^{-\Gamma_0/T}} \right)
\end{eqnarray}
i.e., with $\tau$ cutting off  a quantity inversely proportional to
$\langle x\rangle_{\mbox{valley}}$. 
But the distribution of $w=\langle x\rangle_{\mbox{valley}}/x_{\pm}$ is
constant near zero, with density of order
$1/\Gamma_0$, because of the distribution of $\epsilon$.
Thus we can focus on valleys with the smallest $\Gamma_0$
since these will produce the largest number of crossings.

We must now estimate also the probability that such 
an atypical valley survives for a long  time $\tau$.
For that, we need that neither
segment on either side of the origin be decimated in the
RG for time $\tau$, which happens with probability 
$1/\ln^2 \tau$. For the contribution of these to
the distribution of $N(t)$ we thus have, ignoring constants:
\begin{eqnarray}
\mbox{Prob}(N(t)  =  N) & \sim & 
\int^t \frac{d \tau}{\tau (\ln \tau)^3}
\left[ \int_{\tau^{-1/2}} dw \delta\left(N-\frac{1}{w}\right)
+ \int_{0}^{\tau^{-1/2}} dw \delta(N-\sqrt{\tau}) \right] \nonumber \\
& \sim & 
\frac{1}{N^2 (\ln N)^2} \theta(\sqrt{t} - N)
\end{eqnarray}
the dominant contribution coming from the first term
in the square bracket. The first moment of $N$ is (barely)
finite, but higher moments grow with time as:
\begin{eqnarray}
\overline{N(t)^{\alpha}} \sim 
\frac{t^{\frac{\alpha-1}{2}}}{(\ln t)^2}.
\end{eqnarray}
Thus these type of events completely dominate the
distribution of the more-than-typical $N(t)$ tail.
For $g \equiv N/\ln \ln t > 1/3$ the distribution is
therefore {\it not} multifractal.

On the anomalously small $N$ side of the distribution, the
type of events which might be troublesome appear to be rare
enough not to cause $\rm{Prob}(g)$ for $g<1/3$ to differ
from that using $\langle x(t)\rangle$ instead of $\Xi(t)$. The 
result (\ref{multi}) for $\langle x(t)\rangle$
should thus hold also for $\Xi(t)$ for $g<1/3$.

\subsection{Distribution of the sequence
of returns to the origin: symmetric case}

\label{distribreturn}

We now study a more refined quantity concerning the statistics
of returns to the origin of the thermal average $\langle x(t)\rangle$ 
in the Sinai model.
It turns out to be possible  to obtain the {\it full probability
distribution of the complete sequence } of the times
$\Gamma_1=T \ln t_1, ...\Gamma_k=T \ln t_k$
of successive returns to the origin. This is possible because
of the remarkable property that every time the thermal packet crosses
the origin, it ``loses its memory'' of the past.
 
We consider the probability $D_{\Gamma,\Gamma'}(\eta)$
that a bond has barrier $\eta$ at $\Gamma$ and has had
its last change of orientation at scale $\Gamma'$.
Its evolution equation reads
\begin{eqnarray}
\left(\Gamma \partial_\Gamma - (1+ \eta) \partial_\eta - 1 \right)
D_{\Gamma,\Gamma'}(\eta)
 = 2 P_{\Gamma}(0) D_{\Gamma,\Gamma'}(.) *_\eta P_{\Gamma}(.) 
- 2 P_{\Gamma}(0) D_{\Gamma,\Gamma'}(\eta)
\end{eqnarray}
with the initial condition at $\Gamma=\Gamma'$
given by $D_{\Gamma',\Gamma'}(\eta)=P_{\Gamma'}(.) *_\eta P_{\Gamma'}(.)$.
As stated above, this initial condition is independent of
previous history because at each decimation the bond is 
chosen afresh.
Since $P_{\Gamma}(\eta)=e^{-\eta}$, it is natural
to look for a solution of the form
$D_{\Gamma,\Gamma'}(\eta)=(A_{\Gamma,\Gamma'} +B_{\Gamma,\Gamma'} \eta) e^{-\eta}$.
This is found in terms of $\alpha=\Gamma/\Gamma'$ as:
\begin{eqnarray}
&& A_{\Gamma,\Gamma'}= { 1 \over {\lambda_+ - \lambda_-}} 
\left( \alpha^{-\lambda_-} - \alpha^{-\lambda_+} \right) 
\qquad {\rm with} ~~ \lambda_{\pm}={ {3 \pm \sqrt 5} \over 2} \\
&& B_{\Gamma,\Gamma'}= { 1 \over {\lambda_+ - \lambda_-}}
 \left[ (\lambda_+ -1)\alpha^{-\lambda_-} + (1-\lambda_-)\alpha^{-\lambda_+} \right]
\end{eqnarray}
The probability for a bond to be decimated at $\Gamma$  
given that its last decimation occurred at $\Gamma'$
is $\rho(\Gamma,\Gamma')=- \partial_\Gamma 
\int_0^{\infty} d \eta D_{\Gamma,\Gamma'}(\eta)
 = {1 \over \Gamma} \rho(\alpha=\Gamma/\Gamma')$
with:
\begin{eqnarray}
\rho(\alpha)=  {1 \over \alpha} { 1 \over {\lambda_+ - \lambda_-}} 
\left( \alpha^{-\lambda_-} - \alpha^{-\lambda_+} \right).
\end{eqnarray}

Thus we have obtained the probability distribution
$\Pi_{t'}(t)$ for the time $t$ of
next return to the origin (in the effective dynamics) 
given that the last return was at a time $t'$; this
exhibits ``aging behavior'' in $\alpha= {{\ln t} \over {\ln t'}}$ 
\begin{eqnarray}
\Pi_{t'}(t) dt = {dt \over { t \ln t}} 
{ 1 \over {\lambda_+ - \lambda_-}} \left( 
\left({{\ln t} \over {\ln t'}}\right)^{-\lambda_-} - 
\left({{\ln t} \over {\ln t'}}\right)^{-\lambda_+} \right).
\end{eqnarray}

For the sequence of successive returns,
the picture we  obtain is therefore very simple:
the sequence of scales $\{\Gamma_k\}$ at which 
the successive changes of orientation of a given bond occur
is {\it a multiplicative Markovian process} constructed
with the simple rule $\Gamma_{k+1}=\alpha_{k} \Gamma_k$
where $\{\alpha_k\}$ are independent identically 
distributed random variables
of probability distribution $\rho(\alpha)$. 
As a consequence, $\Gamma_k=\alpha_{k-1} \alpha_{k-2} 
\cdots \alpha_{2} \Gamma_1$ is simply the product of random variables,
so that we obtain using the central limit 
theorem that
\begin{eqnarray}
\lim_{k \to \infty} \left({ {\ln \Gamma_k} \over k} \right)
=\langle \ln \alpha\rangle =3
\end{eqnarray}
and we thus recover that the number $k$
of changes of orientation grows as $\ln \Gamma = \ln \ln t$ 
and that the rescaled variable $g={k \over {\ln (T \ln t)}}$
is equal to $1/3$ with probability $1$, as in
(\ref{g13}).

\subsection{Number of jumps up to time $t$
for the effective dynamics}

\label{secnumberjumps}

In this section we study the behavior of the total
number of jumps of the thermal averaged position 
$\langle x(t)\rangle$ at large times. We introduce the number $m(n)$
of starting points on a bond
such that the effective---i.e., $\langle x(t)\rangle$---walker
jumps exactly $n$ times between 
$0$ and $t$ for the effective dynamics. We will use $m$ to denote 
various auxiliary variables and trust that such 
local varying usage will not be confusing. The RG rules
for these auxiliary variables upon decimation of bond 2 read
(see Fig. \ref{fig2})
\begin{eqnarray}
&& m'(n) = m_1(n-1) + m_2(n-1) + m_3(n)  \\
&& m'(0)=m_3(0)
\end{eqnarray}
Introducing the generating function 
$m(z)=\sum_{n=0}^{\infty} m(n) z^n$ one finds the RG rule
$m'(z) =  z m_1(z) + z m_2(z) + m_3(z)$.
We thus find (see Appendix \ref{rulesymmetric}) 
that the ratio $m(z)/\overline{l}_\Gamma$ decays as $\Gamma^{ - \Phi(z)}$
where $\Phi(z)$ is now the solution of the  equation:
\begin{eqnarray}  \label{hyper}
0=\left({\Phi(z) \over 2} \right) U(-z, \Phi(z) ,1 )
- U(-z, 1+\Phi(z) ,1) 
\end{eqnarray}
in terms of the hypergeometric function $U(a,b,z)$.

Performing the same saddle point analysis as in the previous section
we find that the rescaled variable 
$G=n/{\ln \Gamma}=n/\ln(T \ln t)$ for the number $n$ of jumps up to time $t$,
has a multifractal distribution
\begin{eqnarray}
\mbox{Prob}(G) \sim  \overline{l}(t)^{-\omega(G)}
\end{eqnarray} 
where the exponent $\omega(G)$ is given by Legendre transform as
$2 \omega(G)=\Phi(z^*(G))+G \ln z^*(G)$
where $z^*(G)$ is the solution of 
$\Phi'(z^*(G)) = - G/z^*(G)$. Note that 
for $G=0$ (which corresponds to $z=0$) one has 
$\omega(G=0)=\Phi(z=0)=2$.
As in the previous section, the asymptotic value $G_a$ that $G$
takes with probability one at large times is determined by
the minimum of $\omega(G)$ where 
$\omega'(G_a)=0$ which 
corresponds to $z^*(G_a)=1$. Thus $G_a=-z^*(G_a) \Phi'(z^*(G_a))
=-\Phi'(z=1)$. Differentiating (\ref{hyper}) and using
$\Phi(z=1)=0$ we find:
\begin{eqnarray}   \label{G43}
G_a=\frac{U_1(-1,1,1)}{U(-1,0,1)/2-U_2(-1,1,1)} =  4/3
\end{eqnarray}
where $U_1(a,b,z) \equiv \partial_a U(a,b,z)$ and 
$U_2(a,b,z) \equiv \partial_b U(a,b,z)$.

A similar method can be used to compute the
the {\it joint distribution} $P(G,g)
\sim  \overline{l}(t)^{-\overline{\theta}(G,g)}$
of the two rescaled
variables $G=n/\ln(T \ln t)$ and $g=k/\ln(T \ln t)$
where $n$ and $k$ are respectively the total number
of jumps and the number of returns to the origin,
and hence the associated decay exponent $\overline{\theta}(G,g)$.
As an example of this application, we give the
large time limit (valid with probability 1) of the
the total rescaled number of jumps $G_g$,
{\it conditioned on} a fixed rescaled number $g$ of returns
to the origin:
\begin{eqnarray}
 G_g=\frac{1+z(g)}{3-2\Phi (z(g)) } \left[ U_1(-1,1+\Phi (z(g)),1)
+\frac{1-z(g)-\Phi (z(g))}{1+z(g)} U_1(-1,\Phi (z(g)),1)  \right]
\end{eqnarray}
where $z(g)=2 g \left(g+\sqrt{g^2+\frac{5}{4}}\right)$ 
and $\Phi(z)= \frac{3-\sqrt{5+4 z}}{2}$. Note that for
$g=1/3$ one recovers $G_{1/3}=4/3$ as expected.

\subsection{Correlations of the jumps}

\label{sub:jumpcorrelations}

In this section we will obtain some information
about the statistical properties of the sequence of
the directions and times of successive jumps.
We will define a {\it jump forward} as a
jump {\it in the same direction} as the previous one,
and a {\it jump backward} as a jump in the {\it opposite direction}
than the previous one. Note that a jump backward 
necessarily involves a return to the origin due to the properties
of the RG procedure. 
The directions of successive jumps exhibit
strong correlations since we have  
found in the previous sections that
the total number of jumps behaves as $n \sim {4 \over 3} \ln \ln t$,
whereas the number of backward jumps (returns to the origin)
behaves as $k \sim {1 \over 3} \ln \ln t$. Thus in the
effective dynamics the walker is substantially more likely 
to jump in the same direction as the previous jump.
This is simply because the barrier of a bond which 
has just been created by decimation and the resulting combination
of three bonds is higher than that of a typical bond at that 
scale and thus it is less likely to be decimated than the other
bond encompassing the valley in which the walker rests.

We first compute the
stationary distribution of the number of jumps forward
made since the last return to the origin, i.e., the  
probabilities $\{c_p\}$ that a walker (at a given time)
has made $p$ successive jumps forward
since its last jump backward. This can be obtained by
introducing $m(p)$ as the number of initial points on a bond
such that the walker has jumped exactly $p$ times
since the last passage over the origin for the effective dynamics,
normalized as $\sum_{p=0}^{\infty}  m(p) = l$
where $l$ is the length of the bond. 
The RG rules upon decimation of bond 2 read
(see Fig. \ref{fig2}) $m'(p) = m_1(p-1) + m_3(p)$, $m'(0)= l_2+m_3(0)$.
The generating function thus has the rules
$m'(z) =  z m_1(z) + l_2 + m_3(z)$ 
where $l_2=\sum_{p=0}^{\infty}  m_2(p)=m_2(z=1)$ is the length of bond (2).
Similar methods as above then yield the generating function
$c(z)=\sum_{p=0}^{\infty} c_p z^p$:
\begin{eqnarray}
c(z) ={{ U(-z,-1,1) + U(1-z,0,1)} \over {3(U(-z,0,1)+U(-z,1,1))}}
\label{hyper}
\end{eqnarray}
in terms of hypergeometric functions, normalized to $c(z=1)=1$.
From (\ref{hyper}) one gets the $c_p$, e.g.
$c_0=(1 + U(1,0,1))/6 =0.23394...$,
$c_1= 0.17492...$, $c_2=0.13356...$, $c_3=0.1029...$, $c_4=0.07958...$
etc. Since $c(z)$ has a pole for $z \approx 1.2884$,
we obtain that 
\[
c_p \sim \exp(-0.25343 p)
\]
for large $p$.
Also, at any given (large) time
the {\it average number of jumps forward} made since
the last jump backward is $N^{\text{av}}_{\text{forward}} =
c'(z=1) = 3.3975649..$.

Next, we study the jump time dependence of backward and
forward jumps. In Appendix \ref{conditional} we
compute the conditional probabilities
$\rho_{\Gamma,\Gamma'}^f$ (respectively $\rho_{\Gamma,\Gamma'}^b$)
to make a forward jump (respectively backward) at $\Gamma$
given that the last jump  occurred
at $\Gamma'$. These are scaling functions  of the ratio 
$\alpha=\frac{\Gamma}{\Gamma'} = \frac{\ln t}{\ln t'}$,
i.e., $\rho_{\Gamma,\Gamma'}^{f,b} d \Gamma = \rho^{f,b}(\alpha) d \alpha$
with:
\begin{eqnarray}
&& \rho^{f}(\alpha) = \frac{1}{2 \alpha^3} \left(5-(\alpha+2) e^{-(\alpha-1)} \right)  \\
&& \rho^{b}(\alpha) =  \frac{1}{2 \alpha^3} \left(5-(\alpha^2+2\alpha+2)  
e^{-(\alpha-1)} \right).   \label{rho}
\end{eqnarray}
Integrating over $\alpha$ we recover, as expected from (\ref{g13}) and
(\ref{G43}), the total probabilities of the next jump
begin  a forward or a backward jump as
$\rho^{f}=\int_1^{\infty} d \alpha \rho^{f}(\alpha) = \frac{3}{4}$
and $\rho^{b}= \frac{1}{4}$. Note that $\rho(\alpha)= \rho^{f}(\alpha) 
+\rho^{b}(\alpha)$ gives the total probability that the next jump
($f$ or $b$) occurs at $\Gamma = \alpha \Gamma'$.

We now study the statistical properties of the full sequence
of the times of successive jumps ($\Gamma_1=T \ln t_1$, \ldots
$\Gamma_k=T \ln t_k$). Contrary to the sequence
of the times of {\it backward} jumps 
studied in Section \ref{distribreturn} which was simply a multiplicative
Markovian process, there are persistent correlations in
the full sequence of jumps which makes it much harder to analyze.
Indexing each sequence by whether each jump is forward ($f$) or
backward ($b$) we need to introduce the following set of conditional
probabilities:
\begin{eqnarray}  \label{building}
\rho_k^{b f\ldots f b}(\Gamma_{k+1},\Gamma_k,\ldots  \Gamma_1|\Gamma_0)
\end{eqnarray}
that, given that a backward jump
occurred at $\Gamma_0$, there are exactly $k$ forward jumps
occurring at times $\Gamma_k,\ldots \Gamma_1$ before the next backward jump
occurs at $\Gamma_{k+1}$. These conditional probabilities are the
elementary building blocks of the full measure for the sequence of jumps,
since once a backward jump occurs, as was noted in 
Section \ref{distribreturn},  the process starts afresh.
Thus the full measure is  simply a product of the above
terms (\ref{building}).
We have computed the first terms of the set of conditional
probabilities in the Appendix \ref{correlations2}.
We obtain, for instance, that, given that the previous
jump was backward, the probability that the next jump is 
backward is ${- e \over 2} Ei(-1) = 0.298174$
and forward is $0.701826$.
This result is different from above where we did not assume
that the previous jump was backward.

\subsection{Number of returns to the origin: biased case}

\label{sub:numberreturnsbias}

We now study the returns to the origin of the thermal
averaged position $\langle x(t)\rangle$ in the case where a small bias is
applied. Then one expects that the number of returns is
finite, since eventually the packet will leave the
vicinity of the origin. However if the bias is small
the number of return is large and universal results
can be obtained in the limit $\delta$ small, $t$
large with $\gamma = \delta T \ln t$ fixed.

The method
consists again in introducing auxiliary variables $m^{\pm}(k)$
to count the number of initial points on a $(\pm)$ renormalized
bond at $\Gamma=T \ln t$
which have changed exactly $k$ times  orientation up to time $t$.
The RG rules for these variables upon decimation of bond (2) are
\begin{eqnarray}
m^{\pm}(k) = m_1^{\pm}(k-1) + m_2^{\mp}(k-1) + m_3^{\pm}(k)
\end{eqnarray}
for $k \ge 1$ and $m^{\pm}(0)=m_3^{\pm}(0)$ for $k=0$.
Introducing the generating functions 
$m^{\pm}(z)=\sum_{k=0}^{\infty} m^{\pm}(k) z^k$ one finds the RG rule
\begin{eqnarray}
m^{\pm}(z) =  m_1^{\pm}(z) + z m_2^{\mp}(z) + m_3^{\pm}(z).
\end{eqnarray}
The calculation of the mean value $\langle m^{\pm}(z)\rangle$ is performed
in  Appendix \ref{rulesbiased} and gives:
\begin{eqnarray}
\langle m^{\pm}(z)\rangle  = 
\delta^{-\psi(z)} ( A_z(\gamma) + e^{\pm \gamma} \sinh(\gamma) 
\partial_\gamma A_z(\gamma))
\label{mz}
\end{eqnarray}
with $\gamma=\delta \Gamma$, $\psi(z)= \frac{1}{2} (1+\sqrt{5+4 z})$ and
$A_z(\gamma)= K_z Q_{\psi(z)-1}(\coth \gamma)$
in term of the Legendre function $Q_{\nu}(z)$.
The constant $K_z$ depends {\it a priori} on $z$ in a nonuniversal way.
From this we obtain the generating function of
the probabilities $p_{\gamma}(k) = M_k(t)$ that the
averaged position $\langle x(t)\rangle$ has returned exactly $k$ times 
to the origin up to time $t$. It is simply given
(since initial conditions
are uniformly distributed) as the generating function of
the total number of initial conditions with $k$ 
returns divided by the total length
and thus reads
\begin{eqnarray}
\sum_{k=0}^{\infty} p_{\gamma}(k) z^k
={ {\langle m^+(z)\rangle  + \langle m^-(z)\rangle 
} \over {\overline{l}_\Gamma}}
 = \delta^{2 - \psi(z)} {\cal M}_z( \Gamma \delta)
\label{genretbias}
\end{eqnarray}
with the scaling function:
\begin{eqnarray}
{\cal M}_z(\gamma) = \frac{2 K_z}{\sinh(\gamma)^2}
\left[Q_{\psi(z) - 1}(\coth(\gamma)) - 
\coth(\gamma) Q'_{\psi(z) -1}(\coth(\gamma)) \right]
\end{eqnarray}
and normalization implies that $K_{z=1}=1/2$. 
This function has the following asymptotic behaviors. 
For small $\gamma$ it behaves as a power law
${\cal M}_z(\gamma) \simeq K_z \sqrt{\pi} 2^{1-\psi(z)} 
\frac{\Gamma[\psi(z)]}{\Gamma[\psi(z) + 1/2]} (1+\psi(z)) \gamma^{\psi(z)-2}$
which allows one to recover the results of Section
\ref{sub:returns} in the limit of a vanishing bias $\delta \to 0$.
For large $\gamma \to \infty$ it goes to a constant 
${\cal M}_z(\infty) = K_z$ consistent
with a finite total number of returns.

Setting $z=0$, we obtain the probability that $\langle x(t)\rangle$
has not returned to the origin up to time $t$:
\begin{eqnarray}
M_0(t) = p_{\gamma}(k=0) \sim \delta^{2 \overline{\theta}} {\cal M}_0(\gamma)
\end{eqnarray}
where the exponent $2 \overline{\theta} = \frac{3-\sqrt{5}}{2}$ 
coincides with the exponent $\beta$ of the magnetization 
of the RTFIC \cite{danfisher_rg2}. Note however, that although the
scaling function ${\cal M}_0(\gamma)$ in the particular case
of $z=0$, which corresponds to the probability
of no return, is closely related to the scaling function of
the magnetization of the RTFIC, it is not identical to it.
The probability of no returns of the running average of $x$, $\Xi(t)$, will
have the same asymptotic behavior as $\langle x(t)\rangle$. 

It is interesting to estimate the distribution of the
total number of returns. This is achieved by studying the
limit $\Gamma \to \infty$ in (\ref{genretbias}). We obtain that
the probabilities $M_k(t) = p_{\gamma=\infty}(k)$, that the 
thermally averaged position $\langle x(t)\rangle$ 
return exactly $k$ times to the origin between $t=0$ and $t=+ \infty$
have  generating function
\begin{eqnarray}
\sum_{k=0}^{\infty} p_{\gamma=\infty}(k) z^k =\delta^{2-\psi(z)} 
{\cal M}_{z}(\infty ). 
\end{eqnarray}
It is thus natural to introduce the rescaled variable $g=\frac{k}{(-\ln \delta)}$
and to look for the exponent characterizing the behavior for small $\delta$.
One finds, by an analysis similar to that of Section \ref{sub:returns}
\begin{eqnarray}
\mbox{Prob}(g) \sim \delta^{2 \overline{\theta}(g)}
\end{eqnarray}
with the same exponent $\overline{\theta}(g)$ and multifractal behavior as
in (\ref{multi}). The same reasoning as in Section \ref{sub:returns}
leads to the result that $g$ is equal to $\frac{1}{3}$ with probability
one for small $\delta$ 
and thus that the total number $k$ of returns to the origin in the
presence of a small bias is
\begin{eqnarray}
k \approx \frac{\vert \ln \delta \vert }{3}.
\end{eqnarray}

\subsection{ Distribution of the sequence of 
returns to the origin: biased case}

\label{sequencebias}

As in the symmetric case, it is possible to obtain the
full probability distribution of the sequence $\{\Gamma_1,\Gamma_2, \ldots\}$
of successive returns to the origin. 
However in the case with drift in direction $(+)$, 
this sequence is finite with probability 1 since
there is a finite probability that the particle never
returns to the origin if it is on the right of its starting point.
Therefore the probability $\Pi_{k}(\Gamma_1\ldots \Gamma_{k})$ 
that the particle returns exactly $(k)$ times to the origin
from $t=0$ to $t=\infty$ and that these returns take place at scale
$(\Gamma_1,\ldots, \Gamma_k)$ can be decomposed as the product

\begin{eqnarray}
\Pi_{k}(\Gamma_1\dots \Gamma_{2k}) = \rho^{+}(\infty,\Gamma_{k})
\rho^{-}(\Gamma_{k},\Gamma_{k-1}) \rho^{+}(\Gamma_{k-1},\Gamma_{k-2})\ldots 
\end{eqnarray}
where $\rho^{\pm}(\Gamma , \Gamma')$ are the conditional probabilities
that the particle returns to the origin
at $\Gamma$ given that the last return to the origin occurred at scale
$\Gamma'$ in the direction $(\pm)$, and where $\rho^{+}(\infty,\Gamma')$
represents the probability that the particle never
returns to the origin after the last passage to the origin
occurred at $\Gamma'$ in the $(+)$ direction

\begin{eqnarray}
\rho^{+}(\infty,\Gamma')=1-\int_{\Gamma'}^{\infty} 
d\Gamma \rho^{+}(\Gamma,\Gamma'). 
\end{eqnarray}

We have computed these probabilities in  Appendix
\ref{distribiased}. They are most naturally written in terms
of the reduced variables $y=\coth \gamma$, $y'=\coth \gamma'$
with  $1<y<y'$ (where $\gamma= \delta \Gamma = T \delta \ln t$
and $\gamma'= \delta \Gamma' = T \delta \ln t'$) as:
\begin{eqnarray}
\rho^{\pm}(\Gamma , \Gamma') d\Gamma = \tilde{\rho}^{\pm}(y,y') dy =
\frac{y \mp 1}{y' \pm 1} \left[ Q_{\phi-1}(y)
P_{\phi-1}(y')-Q_{\phi-1}(y')
P_{\phi-1}(y) \right] dy
\label{pq}
\end{eqnarray}
with $\phi = \frac{1}{2} (1 + \sqrt{5})$
and  $P_\nu$ and $Q_\nu$  Legendre
functions. One finds, as expected, that
$\int_1^{y'} dy \tilde{\rho}^{-}(y,y')= 1$.
On the other hand, the probability that
the thermally averaged position $\langle x(t)\rangle$ never crosses
the origin again after having crossed it at $\Gamma'$
is:
\begin{eqnarray}
 \rho^{+}(\infty,\Gamma')=
1-\int_{\Gamma'}^{\infty} d\Gamma \rho^{+}(\Gamma,\Gamma') 
= 2 \frac{P_{\phi-1}(y')}{y'+1}
\end{eqnarray}
For $\gamma' \to 0$ this probability vanishes as:
\begin{eqnarray}
\rho^{+}(\infty,\Gamma')= 2  \frac{ \Gamma(\phi-{1 \over 2})}
 {\sqrt{\pi} \Gamma(\phi)} \gamma'^{2-\phi}
\end{eqnarray}
while it goes to $1$ for large $\gamma'$ as:
\begin{eqnarray}
\rho^{+}(\infty,\Gamma')= 1- \frac{e^{-4 \gamma'}}{4} = 1 - \frac{1}{4 {t'}^{2 \mu}}
\end{eqnarray}
where $\mu=2 \delta T$. 
The factor $1/t'^{2 \mu}$ can be understood with a simple argument.
Since the particle
having crossed the origin at $t'$ belongs to a renormalized bond just 
being created, its barrier is distributed not with 
$P^{+}_{\Gamma'}(\zeta)$ (\ref{solu-biased}), but with 
$P^{\text {new}+}_{\Gamma'}(\zeta)=P^{+}_{\Gamma'}(.)*_{\zeta} P^{+}_{\Gamma'}(.)
=(u^+_{\Gamma'})^2 \zeta e^{- u^+_{\Gamma'} \zeta}
\sim
\zeta/t'^{2 \mu} e^{- \zeta \mu/(T t'^{\mu}) }$;
this is depleted near the origin which is the key point.
For a return to  occur after $t'$, the new renormalized bond has to be 
decimated in the future,
and the dominant contribution comes from the times near $t'$, i.e., we
have to compute the probability that the two independent barriers 
of the neighboring bonds each distributed with $P^-_{\Gamma'}$ 
are bigger than the new (+) renormalized 
bond at $\Gamma'$: this probability is simply $\frac{(u^+_{\Gamma'})^2}
{(u^+_{\Gamma'}+2u^-_{\Gamma'})^2}$ 
and thus behaves as $1/(4 t'^{2 \mu})$ at large $t'$. These events
are thus responsible for the dominant behavior of 
$\rho^{+}(\infty,\Gamma')$ found above.

\section{Return to the origin of a single walker, first passage times
and meeting time of two walkers}

\label{truedynamics}

\subsection{Probability of no return to the origin for a
single walker}

\label{return:single}

We now compute, in the presence of a small bias, the probability
$N^{+}(t)$ (respectively $N^{-}(t)$) 
that a {\it single} walker has remained all the time to the right 
(respectively to the left) of its starting point---the bias
being by convention to the right.
These probabilities are found by placing
an absorbing boundary at $x=0$ as discussed in
Section \ref{boundaryrg}.
We note that the probability distributions
$E^{\pm}_\Gamma(l)$ of the length $l$ of the absorbing zone
satisfying the RG equation (\ref{brsrgdrift}) has 
initial condition $\delta(l-1)$ (counting the first
infinitely deep bond in Fig. \ref{figboundary} as  length $1$ by convention)
and it is the weight of this $\delta$-function part that 
determines the desired no-return probability.
For finite $\Gamma$, $E^{\pm}_\Gamma(l)$ takes the form:
\begin{eqnarray}
E^{\pm}_\Gamma(l) = \delta(l-1) \int_{0}^{+\infty} d\zeta R^{\pm}_\Gamma(\zeta)
+ \text{regular part}
\end{eqnarray}
where $R^{\pm}_\Gamma(\zeta)$ is the probability that the first
descending bond (in Fig. \ref{figboundary} for the (+) case) has never been
decimated up to scale $\Gamma$ and has barrier $F=\Gamma+\zeta$
at $\Gamma$. The total weight of the delta function 
$r^{\pm}_\Gamma = \int_{0}^{+\infty} d\zeta R^{\pm}_\Gamma(\zeta)$ decays to
zero as the regular part of $E^{\pm}_\Gamma(l)$ converges towards
the fixed point determined in Section \ref{finitesolu}.
We  obtain the probabilities $N^{\pm}(t)$ 
from $N^{\pm}(t) \sim
r^{\pm}_{\Gamma= T \ln t}~~$.

The evolution equation for $R^{\pm}_\Gamma(\zeta)$ reads
\begin{eqnarray}
 (\partial_{\Gamma}-\partial_\zeta) R_{\Gamma}^{\pm}(\zeta)
 = -P_{\Gamma}^{\mp}(0) R_{\Gamma}^{\pm}(\zeta) + P_{\Gamma}^{\mp}(0)  
R_{\Gamma}^{\pm}(.) *_\zeta P_{\Gamma}^{\pm}(.) 
\end{eqnarray}
Setting $R_{\Gamma}^{\pm}(\zeta)=r_{\Gamma}^{\pm} P_{\Gamma}^{\pm}(\zeta) $
we obtain
$\partial_{\Gamma} \ln (r_{\Gamma}^{\pm} )
 = - u_{\Gamma}^{\pm} = \partial_{\Gamma} \ln (u_{\Gamma}^{\mp} )$
where we denote $u_{\Gamma}^{\pm} = u_{\Gamma}^{\pm}(0)$ 
in (\ref{solu-biased-u}). Thus $r_{\Gamma}^{\pm} \sim u_{\Gamma}^{\mp}$
which yields:
\begin{eqnarray}
N^{+}(t) \sim \frac{2 \delta}{1 - e^{-2 \delta \Gamma}} \\
N^{-}(t) \sim \frac{2 \delta}{e^{2 \delta \Gamma} - 1} 
\end{eqnarray}
In the biased case, one finds that there is a finite 
probability $N^{+}(t) \sim 2 \delta$ 
of never returning to the origin
if the particle starts in the direction of the drift, whereas
if it starts against the drift, the probability of not returning
to the origin up to time $t$ decays as $N^{-}(t) \sim 2 \delta e^{- 2 \delta \Gamma}
\propto t^{- \mu}$. This corresponds to the probability that
the origin happens to belong to a ``trap'' impeding the drift of waiting
time larger than $t$, as was discussed in \cite{ledou_1d}. 
We note that the calculation of these persistent probabilities
in the Sinai model  is
similar to the calculation of the endpoint magnetization 
in the RTFIC \cite{danfisher_rg2}.

In the symmetric case $\delta=0$ we thus find that the probability $N(t)$
that a {\it single walker} has {\it never} crossed its starting point 
$x(0)=x_0$ between $0$ and $t$ decays at large time as
\begin{eqnarray}
N(t) \sim \overline{l}(t)^{-\theta} \ \ \ \hbox{with} \ \ 
\theta=\frac{1}{2}
\end{eqnarray}
with $\overline{l}(t) = (T \ln t)^2$. Note that the
persistence exponent obtained here in the presence of disorder
is different from the result $\theta_{\rm{pure}}=1$
for the pure diffusion problem
where the probability of no return to the origin up to time $t$
decays simply as $\frac{1}{\overline{l}(t)} \sim \frac{1}{\sqrt t}$.
It is also significantly
larger than the persistence exponent $\overline{\theta}=\frac{3 - \sqrt{5}}{4}
=0.190983.. $ for thermally averaged trajectories 
obtained in Section \ref{sub:returns}.

\subsection{First passage times in an infinite sample}

\label{sub:firstpassage}

\begin{figure}[thb] 
\centerline{\fig{8cm}{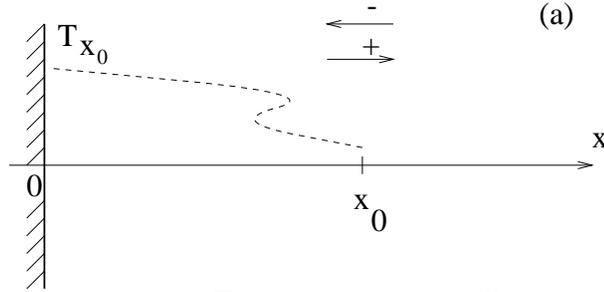}}
\caption{\label{figfirst} \narrowtext 
The first passage time $T_{x_0}$ at $x=0$ of a 
walker starting at $x_0$ is obtained from the survival
probability in the presence of an absorbing
boundary at $x=0$.}
\end{figure}

We now compute the distribution of the first passage
time $T_{x_0}$ at $x=0$ of a walker which start at
$x=x_0$ at $t=0$. The method consists in placing an
absorbing boundary at $x=0$ and studying the 
probability $S_{x_0}(\Gamma)$ that this walker
has survived up to scale $\Gamma = T \ln t$
as illustrated in (\ref{figfirst}).
We use the method of decimation in the presence
of a boundary discussed in Section \ref{boundaryrg}.
The probability $S_{x_0}^{\pm}(\Gamma)$ that the 
walker starting at $x_0$ is still 
alive at $\Gamma$ in the presence of a $(\pm)$ drift
is equal to the probability that the
length $l_1$ of the absorbing zone in 
near the boundary (see Figure \ref{figboundary}) is
smaller than $x_0$ at $\Gamma$, which is:
\begin{eqnarray}
S_{x_0}^{\pm}(\Gamma) = \int_0^{x_0} dl_1 E_\Gamma^{\mp}(l_1) 
\end{eqnarray}
in terms of the function $E_\Gamma^{\mp}(l_1)$ studied
in Section \ref{finitesolu}.
The probability that the first passage time $T_{x_0}$
is such that $\Gamma < T \ln T_{x_0} < \Gamma + d \Gamma$
is equal to the probability $\sigma_{x_0}^{\pm}(\Gamma)$ 
that the walker is absorbed between $\Gamma$ and $\Gamma + d \Gamma$
and is simply obtained:
\begin{eqnarray}
\sigma_{x_0}^{\pm}(\Gamma) = - \partial_{\Gamma} S_{x_0}^{\pm}(\Gamma)
= - \partial_{\Gamma} \int_0^{x_0} dl E_\Gamma^{\mp}(l)
\label{sigmapm}
\end{eqnarray}

\subsubsection{Symmetric case}

In the symmetric case $\delta=0$ we can rewrite (\ref{sigmapm})
in terms of the distribution $E(\lambda)$
of the rescaled variable $\lambda=\frac{l_1}{\Gamma^2}$
given by (\ref{elambda}) and obtain
\begin{eqnarray}
\sigma_{x_0}(\Gamma) = 
 - \partial_{\Gamma} \int_0^{\frac{x_0}{\Gamma^2}} d \lambda E(\lambda) 
= \frac{2 x_0}{\Gamma^3} 
E \left(\frac{x_0}{\Gamma^2}\right).
\end{eqnarray}
The distribution of first passage time $T_{x_0}$ is 
naturally obtained in terms of the rescaled variable
$w = T \ln T_{x_0}/ \sqrt{x_0} $ which is
a random variable distributed as
\begin{eqnarray}
s(w)= {2 \over w^3 } E\left({1 \over w^2}\right) 
= {2 \over w^3 } \sum_{n=-\infty}^{+\infty} e^{- {\pi^2 \over w^2} (n+\frac12)^2}
= \frac{2}{\sqrt{\pi} w^2}  \sum_{m=-\infty}^{+\infty} (-1)^m 
e^{- m^2 w^2}.
\end{eqnarray}
This distribution has the following asymptotic behavior. It behaves
as $s(w) \sim (4/w^3) \exp[-\pi^2/(4 w^2)]$ for small $w$ as the
smaller passage times are strongly suppressed. However it has a broad tail
for large $w$ and decays as $s(w) \sim \frac{2}{\sqrt{\pi} w^2}$. In particular
its first moment diverges: $\overline{w} =  T \overline{\ln T_{x_0}}
/\sqrt{x_0}  = + \infty$.
One can relate this tail to the result of the
previous Section concerning the probability $N(t)$ that the
walker never crosses $0$. In general one expects that
$N(t) \sim C(x_0)/\Gamma$ for a walker starting at $x_0$,
and that $C(x_0)$ is non-universal for fixed $x_0$. Here we
find coming from the other limit in the scaling regime 
$x_0/\Gamma^2$ fixed but small, that the behavior of $C(x_0)$ at large
$x_0$ should be universal as $C(x_0) \sim \sqrt{x_0}$ since we 
find here  $N(t) 
\sim \int_{\Gamma/\sqrt{x_0}}^{+\infty} dw/w^2 \sim \sqrt{x_0}/\Gamma$.

\subsubsection{Biased case}

The Laplace transform
 with respect to $x_0$ of the survival probability reads
\begin{eqnarray}
\int_0^{\infty} dx_0 e^{- p x_0} S_{x_0}^{\pm}(\Gamma) =
\frac{ E_\Gamma^{\mp}(p)}{p}  
 =  \frac{u_\Gamma^{\mp}}{p u_\Gamma^{\mp}(p)}. 
\end{eqnarray}
Introducing again the scaling variables $\gamma=\delta \Gamma$, 
$X_0=\frac{x_0}{\Gamma^2}$, the Laplace inversion gives
\begin{eqnarray}
S_{x_0}^{\pm}(\Gamma) =1-\sum_{n = 0}^{\infty} C^{\mp}_{n}(\gamma) 
e^{- X_0 s_n^{\mp}(\gamma)}
\end{eqnarray}
where $s_n^{\pm}(\gamma)$ have been introduced in (\ref{snlaplace}),
and where
\begin{eqnarray}
 C_n^{\pm}(\gamma)=\left(\frac{\gamma}{\sinh \gamma} \right)
\frac{ 2  (\alpha^{\pm}_{n}(\gamma))^2 e^{\mp \gamma} }
{ (\gamma^2+ (\alpha^{\pm}_{n}(\gamma))^2)
 (\gamma^2+ (\alpha^{\pm}_{n}(\gamma))^2 \mp \gamma )}
\end{eqnarray}
except for the term $n=0$ in the domain $(+)$ and $\gamma>1$ where
\begin{eqnarray}
C_0^{+}(\gamma>1)=  \left(\frac{\gamma}{\sinh \gamma} \right)
\frac{ 2  (\tilde{\alpha}^{+}_{0}(\gamma))^2 e^{- \gamma} }
{ (\gamma^2- (\tilde{\alpha}^{+}_{0}(\gamma))^2) (\gamma
+(\tilde{\alpha}^{+}_{0}(\gamma))^2-\gamma^2)}
\end{eqnarray}

For $\gamma \gg 1$, we find using (\ref{largegamma}) that the survival
probability $S_{x_0}^{-}(\Gamma)$ in the presence 
of a bias towards the absorbing boundary has a simple exponential
dependence on $x_0$
\begin{eqnarray}
S_{x_0}^{-}\left(\Gamma \gg \frac{1}{\delta} \right) \simeq 
1-e^{- x_0 4 \delta^2 e^{-2 \gamma} } =
1-e^{- x_0 \mu^2/(T^2 t^{\mu}) }
\end{eqnarray}
whereas in the case of a bias in the direction away from
the absorbing boundary, we find

\begin{eqnarray}
S_{x_0}^{+}\left(\Gamma \gg \frac{1}{\delta} \right) \simeq 
1- e^{-x_0 \delta^2} \sum_{n = 0}^{\infty} && \frac{ 4 \gamma n^2 \pi^2 }
{ (\gamma^2+n^2 \pi^2)  (\gamma^2+n^2 \pi^2 +\gamma)} 
e^{- \frac{x_0}{\Gamma^2} n^2 \pi^2 }.
\end{eqnarray}

In the limit $\Gamma \to \infty$, we obtain the probability
that a particle reaches the point at a distance $x_0$
from its starting point in the far region against the drift
\begin{eqnarray} 
\lim_{\Gamma \to \infty} (1 - S_{x_0}^{+}(\Gamma))
= \frac{1}{\sqrt{\pi} (x_0 \delta^2)^{3/2}} e^{-x_0 \delta^2}
\end{eqnarray}
which coincides, in the limit $\mu=2\delta T \to 0$,
with the exact result in the regime $0<\mu<2$---corresponding
to anomalous 
diffusion---\cite{tanaka,monthusyor} which reads (for $T=1$):
$ \pi^{3/2} \frac{\Gamma(\frac{\mu}{2})^2}{\Gamma(\mu)} 
\frac{1}{1-\cos(\pi \mu)} x_0^{-3/2} e^{- \mu^2 x_0/4}$.

\subsection{Distribution of the maximum position}

The above calculation also yields the distribution of the 
maximum position $x_{\max}(t) = \max_{0<t'<t} x(t')$ for
a particle starting from $x=0$ at $t=0$ in the presence
of a $+$ (i.e., along the positive direction) or $-$ bias:

\begin{eqnarray}
{\rm Prob}^{\pm}(x_{\max}(t) \le x_m) = S^{\mp}_{x_m}(\Gamma= T \ln t)
= \int_0^{x_m} dl E^{\pm}_{\Gamma= T \ln t}(l)
\end{eqnarray}
and thus the boundary probabilities defined in
(\ref{elambda},\ref{finitex}) $E^{\pm}_{\Gamma= T \ln t}(x_m)$
correspond exactly to the distribution of the maximum
of the Sinai walk.

In the symmetric case we thus recover via the RG a result 
derived by Golosov \cite{golosov2}. It is given
in explicit form in (\ref{elambda}). In addition we
obtain, in the presence of a small bias, the explicit
form:

\begin{eqnarray}
{\rm Prob}^{\pm}(x_{\max}(t) = x_m) = E^{\pm}_{\Gamma= T \ln t}(x_m)
= \sum_{n=0}^{+\infty}
C^{\pm}_{n}(\gamma) s_n^{\pm}(\gamma)
e^{- x_m s_n^{\pm}(\gamma)}
\label{maximum}
\end{eqnarray}
with the conventions of the previous section.

\subsection{Probability that two particules do not meet up to time $t$}

\label{twoparticles}

In this section we compute the distribution of the meeting time $T_L$
for two particules starting respectively, at $x=0$ and $x=L$.
The RSRG method is well suited
to compute this quantity which may be hard to get by
other means. We call $1$ and $2$ the two particles, $1$ starting from
$x=0$ at $t=0$ and $2$ from $x=L$. We compute the
probability $F_L(\Gamma)$ that the two particles 
have not yet met at time $t$ with $\Gamma = T \ln t$. 
The probability $1-F_L(\Gamma)$ that they have met
is equal to the probability that the segment $[0,L]$
is included in a single renormalized valley at $\Gamma$.
The distribution $V_\Gamma(l)$ of the length $l$ of the
valleys is given as $V_\Gamma(l) = P_\Gamma^{+} *_l P_\Gamma^{-}$.
The probability that both $0$ and $L$ belong to the same valley
at scale $\Gamma$ is simply:
\begin{eqnarray}
1-F_L(\Gamma) = \frac{1}{\overline{l}_{\Gamma}} 
\int_L^{+\infty} dl ~(l-L) ~ V_\Gamma(l). 
\end{eqnarray}
This leads to the following expression for the
Laplace transform with respect to $L$
\begin{eqnarray}
\int_0^{\infty} dL e^{-pL} F_L(\Gamma) = 
&& \frac {   1- P_{\Gamma}^+ (p) P_{\Gamma}^- (p)} 
 { p^2  \overline{l}_{\Gamma} } =
\left(\frac{ \delta^2 }{ p  \sinh^2 (\Gamma \delta) } \right)
\frac{ \sinh^2 \left(\Gamma \sqrt{p+\delta^2}\right) }
{ p  \cosh^2 \left(\Gamma \sqrt{p+\delta^2}\right)+\delta^2 }
\end{eqnarray}

In the symmetric case $\delta=0$, we find 
$F_L(\Gamma)=f\left(\lambda=\frac{L}{\Gamma^2}\right)$ with the
scaling function
\begin{eqnarray}
f(\lambda)=LT^{-1}_{s \to \lambda} \left( \frac{\tanh^2 (s)}{ s^2}\right)
& = & 
\int_0^{\lambda} dx \sum_{-\infty}^{+\infty} 
\left( 2 x +\frac{1}{\pi^2\left(n+\frac{1}{2} \right)^2} \right) 
e^{- x \pi^2\left(n+\frac{1}{2} \right)^2} \\
& = & \sum_{-\infty}^{+\infty} \left[ 
3 {{1-e^{- \lambda \pi^2\left(n+\frac{1}{2} \right)^2}} 
\over {\pi^4\left(n+\frac{1}{2} \right)^4}}
 -2 {{\lambda e^{- \lambda \pi^2\left(n+\frac{1}{2} \right)^2}}
 \over {\pi^2\left(n+\frac{1}{2} \right)^2}} \right].
\end{eqnarray}
The probability density $H_L(\Gamma)$ 
that the two particles meet between $\Gamma$ and $\Gamma+d\Gamma$ 
is $H_{L}(\Gamma) = -\partial_\Gamma F_L(\Gamma)$. 
We thus obtain that the meeting time is 
$T_L=\exp[w \sqrt{L}/T]$ where $w=\Gamma_L/\sqrt{L}$
is a random variable distributed as 
\begin{eqnarray}
h(w)={2 \over w^3 } f'\left({1 \over w^2} \right) 
= {2 \over w^3 }  \sum_{-\infty}^{+\infty} 
\left( {2 \over w^2} +\frac{1}{\pi^2\left(n+\frac{1}{2} \right)^2} \right) 
e^{-  {\pi^2 \over w^2} \left(n+\frac{1}{2} \right)^2} 
\end{eqnarray}

Note that in the effective dynamics,
once the two particle meet they remain together
at all later times. In the real dynamics, rare events as
explored in Section \ref{sub:rareevents}, can cause them to  
again split for a limited amount of time in distinct
wells separated by a distance of order $\ln^2 t$, with a
probability of order $1/\ln t$.

\section{Two time diffusion front in the Sinai model and aging properties}

\label{sectwotime}

In this section we will study the two-time quantity
\[
P(x,t,x',t')\equiv \overline{\mbox{Prob}
(xt,x't'|00)},
\]
i.e., the probability,
over the ensemble of random landscapes and thermal fluctuations,
that a particle starting at 
$x=0$ at $t=0$, be successively at $x'$ at $t'$ and at $x$ at $t$.
Note that it is normalized as $\int dx' dx P(x,t,x',t')=1$
and thus it is different---due to the averaging over the  
landscape---than the conditional probability
$\overline{\mbox{Prob}(xt|x't',00)}$ 
that the particle be at $x$ at $t$, knowing
that it was at $x'$ at $t'$ and at $x=0$ at $t=0$.

The average of the two-time probability contains a lot of information about 
the dynamics after letting the system evolve from $t=0$
to $t=t'\equiv t_w$, i.e., the {\it aging dynamics}. We study 
$P(x,t,x',t')$ in the limit where both $t$ and $t'$ are 
large. There are several time regimes, according to the
precise way that the double limit $t',t \to \infty$
is taken, and we obtain  analytic expressions
for the scaling form of $P(x,t,x',t')$ in each of
these regimes. We also study 
\[
Q(y,t,t') \equiv \int dx' P(x'+y,t,x',t'),
\]
i.e., the distribution of the displacements $y=x-x'$ between
$t'$ and $t$. Finally, as explained below
in \ref{subsecgeneral}, we simultaneously obtain results
for a ``two-temperature'' evolution.

Some properties of the quantity $P(x,t,x',t')$ were
investigated previously in \cite{laloux_pld_sinai}, by a
numerical simulation and qualitative arguments.
Here we obtain instead detailed {\it exact results} for
this quantity. Whenever they can be compared these results
are found in agreement with the conclusions
of \cite{laloux_pld_sinai}.

Before presenting the analytical results, let us
first give a discussion of the various regimes 
studied in the following sections.

\subsection{Discussion of the various regimes}
\label{subsecgeneral}

\label{subsubsecsymmetric}

One can distinguish 
two main regimes for large $t$ and $t'$, which we
discuss in the symmetric case.

\medskip

(i) ``{\it scaling regime}'': $t-t' \sim t \sim t'^\alpha$, 
with $\alpha>1$:

\medskip

This  first regime is $t \sim t'^\alpha$ with a fixed
$\alpha = \frac{\ln t}{\ln t'} >1$. This regime was
called the ``diffusion regime'' in \cite{laloux_pld_sinai}.
In this regime, typically the bond containing the origin can been
decimated between $\Gamma'= T \ln t'$ and $\Gamma=T \ln t$
and thus motion can occur. 
$P(x,t,x',t')$, obtained below by iterating the RG
from $\Gamma'$ to $\Gamma$, takes a scaling form
in the rescaled position variables. We thus define
\[
X=\frac{x}{\Gamma^2},
\] 
and since there are two 
possible choices for the rescaled $x'$, we
define 
\[
X'=\frac{x'}{\Gamma'^2}
\] 
and 
\[
\tilde{X}'=\frac{x'}{\Gamma^2}=X' \alpha^2.
\]
Choosing to scale $x$ and $x'$ by
the same factor $\Gamma^2$, the
scaling form $P_\alpha(X,\tilde{X}')$ for the two-time probability 
distribution reads:
\begin{eqnarray}   \label{scalingtwotime}
P(x,t,x',t') \sim \frac{1}{(T \ln t)^4} ~~
P_{\alpha = \frac{\ln t}{\ln t'}}\left(
X = \frac{x}{(T \ln t)^2} ,
\tilde{X}' = \frac{x'}{(T \ln t)^2} \right).
\end{eqnarray}
This diffusion front simplifies in the 
two limits $\ln t \approx \ln t'$ ($\alpha=1$) and 
$\ln t \gg \ln t'$ ($\alpha\rightarrow +\infty$). First 
for $\Gamma=\Gamma'$ one must have:
\begin{eqnarray}
P_{\alpha=1}(X,\tilde{X}')= q(\tilde{X}')~\delta(X- \tilde{X}') 
\end{eqnarray}
where $q(\tilde{X}')$ is the Kesten distribution (\ref{kesten})
obtained previously (note that $X' = \tilde{X}'$ for
$\alpha=1$). An interesting feature is that
the delta function component of the two-time diffusion front
at $x=x'$ persists even for $\ln t > \ln t'$:
\begin{eqnarray}  \label{defaging}
P_\alpha(X,\tilde{X}')
= D_\alpha(\tilde{X}') ~ \delta(X-\tilde{X}')
 + \tilde{P}_\alpha(X,\tilde{X}')
\end{eqnarray}
where $\tilde{P}$ is a smooth function of its arguments.
This property was suggested in \cite{laloux_pld_sinai}.
Here we find that it arises naturally in the RSRG,
since there is a finite probability that 
the bond which contains the origin (starting point)
will have its lowest point unchanged by the renormalization
between $\Gamma'$ and $\Gamma$ (note that the bond can grow
but only on one side).

\begin{figure}[thb]  
\centerline{\fig{8cm}{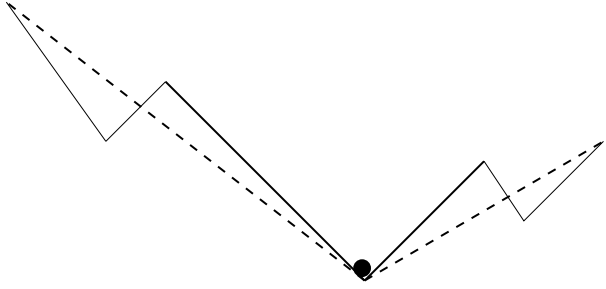}}
\caption{\label{undecimated} 
\narrowtext Fraction of walkers which do not move 
between $t'$ and $t$.}
\end{figure}

\noindent
This implies that a finite fraction of
particles, $D(t,t') = D_\alpha \equiv 
\int_{\tilde{X}'} D_\alpha(\tilde{X}')$
remain at the bottom of a valley (their renormalized valley at
$\Gamma'$) and
do not move appreciably
(i.e., by less than $O(\ln^2 t)$) between $t'$ and $t$.

Finally, for very separated times, i.e., large $\alpha$, the
time evolutions of $X$ at $t$ and $X'$ at $t'$ decouple and
one recovers again the Kesten distribution (\ref{kesten}). One has:

\begin{eqnarray}
D_{\alpha \to \infty}(\tilde{X}') \to 0 \qquad
\tilde{P}_{\alpha \to \infty}(X,\tilde{X}') \to 
 \alpha^2 q( \alpha^2 \tilde{X}') ~
q(X)
\end{eqnarray}

In the next section \ref{sub:singular} we will explicitly
compute $D_{\alpha}(\tilde{X}')$. In section
\ref{sub:full} we will compute the full smooth part, which is
more complicated.

\bigskip

(ii) ``{\it quasi-equilibrium regime}'': $t-t' \sim t'^\alpha$, $\alpha<1$:

\medskip

The second regime is for 
$t-t' \sim t'^\alpha$ with fixed $\alpha = \frac{\ln (t-t')}{\ln t'} < 1$.
[This definition of $\alpha$ is consistent with the
previous one in Ref. 
\cite{laloux_pld_sinai}.]

In the second regime, the typical situation is that
the thermal packet at $t'$ is  well equilibrated
in a valley with the
packet  of width of $O(1)$.
In this regime, there is typically no motion on scales
larger than $O(1)$ between $t'$ and $t$ as the particle
is near the bottom of a valley.
Motion on larger scales will thus be dominated by rare events,
which we now analyze.

First, there is some probability that the valley to which the
origin belongs undergoes a decimation resulting in a jump between $t'$ and $t$.
Although this jump is large (the walker will 
jump to the bottom of a deeper valley a distance of order 
$\overline{l}(\Gamma ') \sim \Gamma '^2$ away) the
probability that it occurs is of order 
the probability that one of the barriers of the valley at time 
$t'$ is less than $\Gamma$, which is itself of order
$(\Gamma -\Gamma ')/\Gamma ' \sim t'^{1-\alpha} \sim
\exp(- (1-\alpha) \Gamma')$ in this regime and thus negligible.
\begin{figure}[thb]  
\centerline{\fig{8cm}{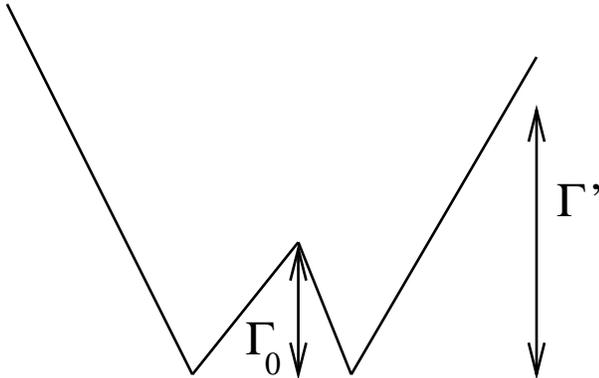}}
\caption{\label{figwell}
\narrowtext Well with two almost degenerate minima
(potential difference of order $O(1)$ and a barrier $\Gamma_0$
which contribute to the inwell equilibrium dynamics.}
\end{figure}
The behavior will instead be dominated by rare
configurations (but less rare than the previous ones)
in which the valley at time $t'$ has 
two almost degenerate minima, separated by a barrier $\Gamma_0$,
as represented in Fig. \ref{figwell}.
Jumping between such minima  
persists even for $t' \to \infty$. The motion between these
minima is equilibrium motion, since typically $\Gamma_0 < \Gamma'$
and the packet are already well equilibrated. In this limit the
statistics of the infinitely deep valley 
potential becomes that of a random walk restricted to have
$U_i-U_{\mbox{valley}-\min}>0$ 
\cite{sinai,laloux_pld_sinai}

Thus in this second regime $t-t' \sim t'^\alpha$ with fixed $\alpha
=\frac{\ln(t-t')}{\ln t'} < 1$
we will find that the diffusion front for relative displacements
$Q(y,t,t')$ has, in addition to
a $\delta$-function part for the rescaled variable 
$\tilde{y} =y/\ln^2(t-t')$ (of weight almost equal to $1$
which corresponds to the typical valleys),
an additional---subdominant by $\frac{1}{\ln(t,t')}$~~---smooth part:

\begin{eqnarray}  \label{inwellsmooth}
\tilde{Q}(y,t,t') \sim \frac{T}{(T \ln (t-t'))^3}
f_\alpha\left[ \frac{y}{(T \ln (t-t'))^2} \right]
\end{eqnarray}
where the function
$f_\alpha$ is universal. This result is obtained in section
(\ref{sub:inwell}).

\bigskip

(ii) ``{\it crossover regime}'': $t \sim t'$, $\alpha=1$:

\medskip
Finally, the matching between the regimes (i) and (ii)
as $t-t' \sim t'$ is studied in Section \ref{crossover}.

Before closing this section, it is interesting to note that
by computing the two-time diffusion front,
we obtain simultaneously the answer to the problem of evolution
of two independent particles
in the same environment, each seeing a thermal noise with
a {\it different temperature} $T$, for the first particle 
with trajectory $x(t)$,
and $T'<T$ for the second with trajectory $x'(t)$. If the
two particles start from the same point 0 at $t=0$, it is clear from
the considerations of the effective dynamics that
the distribution of their respective rescaled
positions
$X= x/(\ln t)^2$ and $\tilde{X}'= x'/(\ln t)^2$ 
will be given by $P(x,x',t) \sim P_{\alpha=T/T'}(X,\tilde{X}')$
with the same scaling function as in (\ref{scalingtwotime}).
[Although in the aging problem the thermal noises are
identical between $t$ and $t'$, this does not make a 
difference at large times for rescaled quantities.]
Note that it should be easier to measure the dependence on
$T$ rather than $\ln t$ as the latter in practice cannot be varied 
over a wide range.

\subsection{Singular part of the two time diffusion front: symmetric case }
\label{sub:singular}

\subsubsection{ Probability $D_{\alpha}$ of staying within a well from
$t'$ to $t$}

We start by computing the probability for a particle to stay
within the bottom of a valley between $t'$ and $t$:

\begin{eqnarray}
D_{\alpha} = \int_{-\infty}^{\infty} d \tilde{X}' D_\alpha(\tilde{X}').
\end{eqnarray}

We compute the fraction of walkers $D_{\Gamma, \Gamma'}(\zeta)$
which are on a bond $F=\Gamma+ \zeta$ at $\Gamma$ and 
have not moved from $\Gamma'$ to $\Gamma$. This means that
(i) this bond has not been decimated, (ii) one of its neighbors
has not been decimated either (the neighbor in the same
valley, i.e., the right neighbor for a descending bond and
left neighbor for an ascending one). Thus the bond has been able
to grow only on one side. As a result $D$ satisfies the RG equation:

\begin{eqnarray}
(\partial_\Gamma - \partial_\zeta) D_{\Gamma,\Gamma'}(\zeta) 
= - P_\Gamma(0) D_{\Gamma,\Gamma'}(\zeta)
+ P_\Gamma(0) ( P_\Gamma(.) *_\zeta D_{\Gamma,\Gamma'}(.) 
- D_{\Gamma,\Gamma'}(\zeta))
\end{eqnarray}
The second and third terms on the r.h.s come from the allowed decimations
of the neighbor. Integration over $\zeta$ shows the total loss of weight from
the forbidden decimations: (ii) from the first term in the r.h.s. 
and (i) from the boundary term at $\zeta=0$.
We are interested in large $\Gamma'$, and thus we can assume that
$P_\Gamma$ has reached its fixed point value (\ref{solu}). Thus one obtains,
in the rescaled variables $\eta=\frac{\zeta}{\Gamma}$
and $\alpha=\frac{\Gamma}{\Gamma'}$
\begin{eqnarray}  \label{decadix}
(\alpha \partial_\alpha - (1+\eta)
\partial_\eta  + 1)  D_\alpha(\eta)  =\int_0^{\eta_1} e^{-\eta_1} 
D_\alpha(\eta-\eta_1).
\end{eqnarray}
The initial condition $D_{\alpha=1}(\eta)$
corresponds to the probability that a walker be on a bond 
with $F=\Gamma'(1 + \eta)$ at $\Gamma'$ and, since this probability 
is proportional to the length of the bond, it is obtained from (\ref{solu}) as
\begin{eqnarray}
D_{\alpha=1}(\eta)= \frac{1}{\overline{\lambda}} 
\int_0^{\infty} d\lambda  \lambda P(\eta,\lambda)
=\frac{1}{3} \left(1 + 2 \eta \right) e^{-\eta}.
\end{eqnarray} 

The solution of (\ref{decadix}) reads:
\begin{eqnarray}  \label{soludecadix}
D_\alpha(\eta) = e^{-\eta} \frac{1}{3 \alpha^2} 
\left(5 + \frac{2}{\alpha-1} e^{1-\alpha} \right)
- e^{- \alpha \eta} \frac{2}{3(\alpha-1)} e^{1-\alpha}.
\end{eqnarray}
We note that this barrier distribution is a combination of two exponential
factors, the expected one $\exp(-F/\Gamma)$, and the other
one $\exp(-F/\Gamma')$ which represents the ``memory'' from
the condition at $t'$. This will be the typical
form for barrier distributions which we will encounter
in all aging calculations.

We then obtain from (\ref{soludecadix})
$D(t,t') = D_{\alpha} = \int_{0}^{\infty} d\eta D_\alpha(\eta)$, i.e.,
the probability that a walker has not moved from $t'$ to $t$, which
takes the remarkably simple universal form:
\begin{eqnarray}  \label{dtt}
D(t,t')
=\left(\frac{\ln t'}{\ln t}\right)^2
\left(\frac{5}{3} - \frac{2}{3} e^{-\left(\frac{\ln t}{\ln t'} 
-1\right)}\right).
\end{eqnarray}

The behavior of (\ref{dtt}) for close times $\alpha = \frac{\ln t}{\ln t'}$
near $1$ is dominated by valleys about to be decimated at $\Gamma'$.
Expanding (\ref{dtt}) yields $D(t,t') \sim 1 - \frac{4}{3} (\alpha-1)$.
The factor $4/3$ is consistent with the most probable
number of jumps growing as $4/3 \ln t$ found earlier.
Let us note that $- \partial_{t'} D(t,t')$ with 
$D(t,t')$ given in (\ref{dtt}), also represents the
scaled distribution of the first passage time $t'$ at the bottom of the
renormalized  valley where the particle is at $t$. This is consistent
with the result of Golosov \cite{golosov2}.

\subsubsection{Weight $D_\alpha(\tilde{X}')$ of the delta function 
component of
$P_\alpha(X,\tilde{X}')$ }

\label{secdeltacomponent}

To compute the full singular part $D_\alpha(\tilde{X}')$
in (\ref{defaging}) we simply have to extend the
previous calculation keeping track of the
length of the bond. Since we are not interested in the
length at $\Gamma$ the length
only appears at $\Gamma'$ as a parameter in the initial 
condition. The final results reads:

\begin{eqnarray}  \label{definitiond}
&& D_{\alpha} (\tilde{X}') = 
\alpha^2 d_{\alpha}(\alpha^2 \vert \tilde{X}' \vert ) \nonumber \\
&& d_{\alpha} (X') = LT^{-1}_{p' \to |X'|} {1 \over { \alpha^2 p'}} \left( 
1-{ {a(p')} \over {b^2(p')}} +{ {a(p') (1-b(p'))} \over {b^2(p')}}
e^{-(\alpha-1)b(p')} \right) \label{resultsingular}
\end{eqnarray}

One can check that for $\alpha=1$, one recovers the Laplace transform
of the Kesten distribution (\ref{kesten}), 
${\hat d}_{1} (p')= {\hat q}(p') 
= {1 \over {  p'}} \left(1-{ {a(p')} \over {b(p')}} \right)$. Also
one recovers ${\hat d}_{\alpha} (p'=0) = { 1 \over 2} D_{t,t'}$
with $D(t,t')$ given in (\ref{dtt}) (with a factor ${1 \over 2}$
corresponding to the total probability restricted to
the half axis $\tilde{x}'>0$).

This result (\ref{resultsingular})
can be explicitly Laplace inverted  in the limit of a 
large $\alpha=\frac{\ln t}{\ln t'}$, where one can neglect the exponential
term, yielding

\begin{eqnarray}
d_{\alpha}(X') \sim \frac{2}{\alpha^2}
\int_{X'}^{+\infty} d \lambda'
\lambda' P(\lambda') = D_{\alpha} \Psi(X')
\end{eqnarray}
where $D_{\alpha} \sim 5/(3 \alpha^2)$ from (\ref{dtt}) 
is the total weight of particles remaining in their wells
and  $\Psi(X')$ is the normalized distribution
of their positions:
\begin{eqnarray}
\Psi(X') = \frac{6}{5} \sum_{n=-\infty}^{+\infty}
\frac{(-)^n}{\pi^3 (n+ \frac{1}{2})^3 } \left(1+ \pi^2 
\left(n+ \frac{1}{2}\right)^2 |X'|\right)
e^{- \pi^2 (n+ \frac{1}{2})^2 |X'| }
\end{eqnarray}
We note that
compared to the Kesten distribution $q(X'=x'/\Gamma'^2)$,
$\Psi(X'=x'/\Gamma'^2)$ has more weight towards the larger
values of $x'/\Gamma'^2$. This is a consequence of the
fact that the farther the particle goes the more
likely it is to be in a deep well where it is 
likely to remain longer without further motion.

Finally, it is instructive to estimate also the
singular part of the averaged {\it conditional probability}
$\overline{\mbox{Prob}(x t|x' t' , 0 0)}$. Using a similar method it is
found to be:
\begin{eqnarray}
{\hat D}^{\mbox{cond}}_{\alpha}(X') 
= \frac{{ d}_{\alpha}(X')}{q(X')}
\end{eqnarray}
where $d_{\alpha}(X')$ is the function defined 
in (\ref{resultsingular}).

\subsection{Probability of staying within a well: biased case}

\label{sub:singularbias}

Next we obtain the probability $D(t,t')$
that a walker does not move substantially between $t$ and $t'$
(i.e., does not jump to a new valley bottom) 
in the presence of a small bias. This can be computed
by extending to the biased case the direct method 
of section \ref{sub:singular}, or from the more
general approach presented in the next section.
Here we only quote the end result:
\begin{eqnarray}
 D_{\Gamma,\Gamma'}  =   \frac{1}{\sinh ^2 \gamma}
\left[ ( 2 \sinh^2 \gamma'+1-\gamma' \coth \gamma')  -
e^{-(\gamma-\gamma')\coth \gamma'} \cosh \gamma
 \left(\cosh \gamma'-\frac{\gamma'}{\sinh \gamma'} \right) \right]
\label{singularagingbias}
\end{eqnarray}
where 
\[
\gamma\equiv \delta T \ln t\quad \mbox{and}\quad
\gamma'\equiv \delta T \ln t'.
\]
This formula is exact in the small bias scaling limit
$t, t' \to \infty$, $\delta \to 0$, with fixed $\gamma$ and
$\gamma'$.

This formula (\ref{singularagingbias}) is interesting
as it exhibits a crossover between two different aging 
scaling functions corresponding to the symmetric model
and the directed model, respectively.
In the limit $\delta \to 0$, i.e., $\gamma \to 0$,
$\gamma' \to 0$ with a fixed ratio $\gamma/\gamma' = \alpha =
\frac{\ln t}{\ln t'}$ Eq. (\ref{dtt}) is recovered.
In the opposite limit in which
both $\gamma \gg 1$ and $\gamma' \gg 1$,
a nontrivial scaling limit exists when $\gamma - \gamma'$
is kept fixed, i.e., $t/t'$ fixed, and the 
above expression simplifies to:
\begin{eqnarray}  \label{directed}
D(t,t') \sim e^{-2 (\gamma-\gamma')} = \frac{L(t')}{L(t)}   
\quad \mbox{with}\quad L(t) = \overline{l}_\Gamma \sim t^{\mu}
\end{eqnarray}
using $\mu=2 \delta T$. 
This coincides with the small $\mu$ limit
of the aging form of the directed model 
(formula (51) 
of \cite{laloux_pld_sinai}) which can be written 
as $D(t,t') = H[(t'/t)^\mu]$ with $H[z] = \left(\frac{\sin(\pi 
\mu)}{\mu \pi}\right) 
\int_0^{z} dy (1-y^{1/\mu})^{-\mu}$. When $\mu \to 0$
the function $H$ becomes exactly $H[z] = z$,
and one recovers (\ref{directed}).

\subsection{Two-time diffusion front: full analysis}
\label{sub:full}

\subsubsection{Sketch of the method}

\label{sketch}

To compute the two time diffusion front 
$\overline{\mbox{Prob}(xt,x't'|00)}$ in the general biased case,
we need to introduce quantities associated with bonds which
keep track of their {\it endpoints} and for which a RG equation
can be written. We thus define
$\Omega^{++}_{\Gamma,\Gamma'} (\zeta,x_L,x_R ; x_L',x_R')$
(resp. $\Omega^{-+}_{\Gamma,\Gamma'} (\zeta,x_L,x_R ; x_L',x_R')$)
as the probability that the origin belongs to a descending
bond with ends $[-x_L',x_R']$ at $\Gamma'$
and to a descending (resp. ascending) bond of barrier $\zeta=F-\Gamma$
and with ends $[-x_L,x_R]$ at $\Gamma$. Similar definitions
hold with $\Omega^{+-}$ and $\Omega^{--}$ for an ascending bond
at $\Gamma'$. From these quantities one can recover the
two-time diffusion front for $x'>0$:
\begin{eqnarray}
&& \overline{\mbox{Prob}(x,t,x'>0,t'|0,0)} = 
\theta(x) \int_0^{\infty} d\zeta \int_0^{\infty} dx_L \int_0^{\infty} dx_L'
\Omega^{++}_{\Gamma,\Gamma'} (\zeta,x_L,x_R=x ; x_L',x_R'=x') \\
&& + \theta(-x) 
\int_0^{\infty} d\zeta \int_0^{\infty} dx_R \int_0^{\infty} dx_L'
\Omega^{-+}_{\Gamma,\Gamma'} (\zeta,x_L=-x,x_R ; x_L',x_R'=x') 
\nonumber 
\end{eqnarray}
and similarly for $x'<0$:
\begin{eqnarray}
&& \overline{\mbox{Prob}(x,t,x'<0,t'|0,0)} =
\theta(x) \int_0^{\infty} d\zeta \int_0^{\infty} dx_L \int_0^{\infty} dx_R'
\Omega^{+-}_{\Gamma,\Gamma'} (\zeta,x_L,x_R=x ; x_L'=-x',x_R') \\
&& + \theta(-x) 
\int_0^{\infty} d\zeta \int_0^{\infty} dx_R \int_0^{\infty} dx_R'
\Omega^{--}_{\Gamma,\Gamma'} (\zeta,x_L=-x,x_R ; x_L'=-x',x_R') 
\nonumber 
\end{eqnarray}

The four RG equations for the four quantities
$\Omega^{+ \epsilon'}_{\Gamma,\Gamma'} (\zeta,x_L,x_R ; x_L',x_R')$
and $\Omega^{- \epsilon'}_{\Gamma,\Gamma'}(\zeta,x_L,x_R ; x_L',x_R')$, with
$\epsilon'= \pm 1$ can be written in a compact form:

\begin{eqnarray}
&& \left(\partial_{\Gamma}-\partial_\zeta \right)
 \Omega^{\pm \epsilon'}_{\Gamma,\Gamma'} (\zeta,x_L,x_R ; x_L',x_R')
=-2 P^{\mp}_{\Gamma}(0) \Omega^{\pm \epsilon'}_{\Gamma,\Gamma'} 
(\zeta,x_L,x_R ; x_L',x_R') \nonumber \\
&& 
+ \int_{l_1>0,l_2>0,y>0} P^{\mp}_{\Gamma}(0,l_2) P^{\pm}_{\Gamma}(.,l_1) *_{\zeta}
\Omega^{\pm \epsilon'}_{\Gamma,\Gamma'} (.,y,x_R ; x_L',x_R')
\delta \left(x_L-(y+l_1+l_2) \right) \nonumber \\
&& + \int_{l_2>0,l_3>0,y>0} 
P^{\mp}_{\Gamma}(0,l_2) P^{\pm}_{\Gamma}(.,l_3) *_{\zeta}
\Omega^{\pm \epsilon'}_{\Gamma,\Gamma'} (.,x_L,y ; x_L',x_R')
\delta \left(x_R-(y+l_2+l_3) \right) \nonumber  \\
&& + \int_{l_1>0,l_3>0,y_1>0,y_2>0}
P^{\pm}_{\Gamma}(.,l_1) *_{\zeta} P^{\pm}_{\Gamma}(.,l_3) 
\Omega^{\mp \epsilon'}_{\Gamma,\Gamma'} (0,y_1,y_2 ; x_L',x_R') \nonumber \\
&& \qquad \delta \left(x_L-(y_1+l_1) \right) \delta \left(x_R-(y_2+l_3) \right)
\label{rgomega}
\end{eqnarray}

These equations must be solved with the
following initial conditions at $\Gamma=\Gamma'$:
\begin{eqnarray}
&&  \Omega^{\epsilon \epsilon'}_{\Gamma',\Gamma'} (\zeta,x_L,x_R ; x_L',x_R')
=\delta_{\epsilon, \epsilon'} \delta(x_L-x_L') \delta(x_R-x_R') 
\omega^{\epsilon'}_{\Gamma'} (\zeta,x_L',x_R')
\label{omegainit}
\end{eqnarray}
where $\omega^{\epsilon'}_{\Gamma'} (\zeta,x_L',x_R')$ is the probability that the origin
belongs at $\Gamma'$ to a bond (ascending if $\epsilon'=-1$ and
descending if $\epsilon'=+1$) 
with barrier  $\zeta=F-\Gamma'$
and of ends $[-x_L',x_R']$ at $\Gamma'$ :
\begin{eqnarray}
\omega^{\epsilon'}_{\Gamma'} (\zeta,x_L',x_R') = \int_0^{\infty} dl' { 
{P^{\epsilon'}_{\Gamma'}(\zeta,l') }
\over {\overline{l}_{\Gamma'}} 
} \delta(l'-(x_L'+x_R')) =
{ 1 \over {\overline{l}_{\Gamma'}}}
P^{\epsilon'}_{\Gamma'}(\zeta,x_L'+x_R')
\end{eqnarray}

Note that all the primed quantities---those at the earlier time---enter
only via the initial conditions on the $\Omega$'s.
These equations (\ref{rgomega}), together with the 
initial condition (\ref{omegainit})
are solved explicitly using Laplace transforms in Appendix
\ref{app:twotime}. For the symmetric case
the explicit expression for the Laplace transform of
the full distribution $\overline{\mbox{Prob}(xt,x't'|00)}$ 
with respect to $x$ and $x'$ is given in 
(\ref{resultsymm1},\ref{resultsymm2}). In the next two subsections
we give explicit expressions for some simpler quantities.

\subsubsection{Some results for the symmetric case}

\label{sub:resultsymmetric}

We first give the explicit expression for the
distribution $Q(y,t,t')$ of relative displacements 
$y=x(t)-x(t')$, with $Q(y,t,t') = \int dx' \overline{P(x'+y,t,x',t'|00)}$.
This distribution takes the scaling form:

\begin{eqnarray}
Q(y,t,t') \sim   \frac{1}{(T \ln t')^2}
Q_{\alpha= \frac{\ln t}{\ln t'} }\left(Y = \frac{|y|}{(T 
\ln t')^2}\right).
\end{eqnarray}
Note that we have chosen $Y=\frac{y}{\Gamma'^2}$ as the scaling
variable here for convenience.

From Appendix \ref{app:twotime}
the Laplace transform
$\hat{Q}_{\alpha}(p) = \int_0^{\infty} dY e^{- p Y} Q_{\alpha}(Y)$
is found to be:
\begin{eqnarray}
\hat{Q}_{\alpha}(p) & = &
\frac{\tanh ( \sqrt{p} \alpha)}{ p \alpha^2} 
\left( -\frac{1}{2 \sqrt{p}}- \sqrt{p} +
\left(\frac{5 p}{6}+\frac{1}{2}\right) \coth \sqrt{p} 
+\sinh \sqrt{p} 
\left( \cosh \sqrt{p} -\frac{\sinh \sqrt{p}}{ \sqrt{p}}\right) \right)  
\nonumber \\
& + &\frac{\cosh \sqrt{p}}{p \alpha^2} 
\left( \frac{\sqrt{p}}{\sinh \sqrt{p}} + \frac{\sinh \sqrt{p}}{ 
\sqrt{p}}-\cosh \sqrt{p} \right)
\nonumber \\
& - &  \frac{1}{ 6 p \alpha^2 \cosh ( \sqrt{p} \alpha)} 
\left( \frac{3}{\cosh \sqrt{p}}+
\frac{ \sqrt{p}}{\sinh \sqrt{p}} \frac{p-3}{p-1} \right) \nonumber \\
& + & e^{-(\alpha-1)} \frac{(1- \sqrt{p} \coth \sqrt{p}) (1 +
 \sqrt{p} \tanh ( \sqrt{p} \alpha))}{ 3 \alpha^2 (p-1)} \nonumber \\
& + & \frac{e^{-(\alpha-1) \sqrt{p} \coth \sqrt{p}}}{ 
2 p \alpha ^2 \cosh ( \sqrt{p} \alpha)}
 \left( \cosh \sqrt{p} +\frac{1}{\cosh \sqrt{p}}- 
\frac{ \sqrt{p}}{\sinh \sqrt{p}} - \frac{\sinh \sqrt{p}}{ \sqrt{p}} \right)
\label{fullq}
\end{eqnarray}
Several properties of this expression can be checked
explicitly. First, from normalization on the half space 
$Y>0$ one has $\hat{Q}_{\alpha}(p=0) =\frac{1}{2}$. 
Then, the initial condition at $t=t'$
is $\hat{Q}_{\alpha=1}(p) =\frac{1}{2}$ since
$Q(y,t,t')=\delta(y)$. One can also recover
the singular part $D(t,t') \delta(y)$
of the distribution corresponding to walkers which have not 
moved appreciably between $t'$ and $t$. Indeed one finds
$\hat{Q}_{\alpha}(p \to +\infty)
\sim D_{\alpha}/2$ where $D_{\alpha}$ is given by (\ref{dtt}).
Finally, for very separated times $\alpha \to 
+ \infty$ one recovers the 
Kesten distribution (\ref{kesten}) since
the initial condition at $\Gamma'$ has been forgotten:
\begin{eqnarray}
\hat{Q}_{\alpha}( p = \frac{k}{\alpha^2}) \to 
\frac{1}{k} (1- \frac{1}{\cosh \sqrt{k}}).
\end{eqnarray}
From the above result (\ref{fullq}) one obtains the
moments of the relative displacements which take the
general scaling form:
\begin{eqnarray}
&& \overline{\langle |x(t)-x(t')|^n \rangle}   
\sim   (T \ln t')^{2 n} F_n\left[ \frac{\ln t}{\ln t'} 
\right]  \nonumber \\
&& F_n[ \alpha]= a_n( \alpha) + e^{1- \alpha} b_n( \alpha).
\label{momentdiff}
\end{eqnarray}
We give explicitly the form of the second moment:
\begin{eqnarray}
&& a_2( \alpha)=\frac{61  \alpha^4}{180} - \frac{4  \alpha}{5} 
+ \frac{47}{60} + \frac{2}{7  \alpha}
- \frac{409}{378  \alpha^2} \\
&& b_2( \alpha)= - \frac{2}{9} + \frac{8}{27  \alpha} + \frac{2}{5  \alpha^2}
\end{eqnarray}
while the first moment has a simpler expression:
\begin{eqnarray}
\overline{\langle |x(t)-x(t')| \rangle} \sim (T \ln t')^2 \left(
\frac{5}{12}  \alpha^2 - \frac{6}{5  \alpha}
+ \frac{221}{180  \alpha^2} - \frac{4}{9  \alpha^2} e^{-( \alpha-1)} 
\right).
\end{eqnarray}

At large $ \alpha$ one recovers Sinai's result (from (\ref{kesten})).
When $t$ and $t'$ are not too separated, i.e., $ \alpha \approx 1$
one finds:
\begin{eqnarray}
F_2[ \alpha] \sim \frac{272}{315} (\alpha-1)
\end{eqnarray}
and the motion is
much slower consistent with the aging of the system.

Again it is simple to recover the
behavior for relatively close times $\alpha \approx 1$.
Expanding the above (\ref{fullq}) one finds:
\begin{eqnarray}  
&& 2 \hat{Q}_{\alpha}(p) \sim 
1 - \frac{4}{3} (\alpha-1) + 
(\alpha-1) {\cal H}(p) + O((\alpha-1)^2) \nonumber \\
&& {\cal H}(p) = 
\frac{1}{\cosh \sqrt{p} \sinh \sqrt{p}}
\left(\sqrt{p} - \frac{1}{\sqrt{p}}\right)
+ \frac{1}{(\sinh \sqrt{p})^2} 
\label{qexpand}
\end{eqnarray}
The $1- 4 (\alpha-1) /3$, $p$-independent term represents
the probability that the particule does not jump
and has been discussed in Section \ref{sub:singular}.
The other term ${\cal H}(p)$ is the Laplace transform of
the probability that the jump is over a scaled distance $\lambda=Y$,
${\cal H}(\lambda) = \int d\lambda_1 ( 1 + 
\frac{\lambda_1}{\overline{\lambda}}) 
P(0,\lambda_1)  P(\lambda-\lambda_1)$ where the
first term corresponds to the bond containing the origin
being decimated and the second term corresponds to the 
neighboring bond in the same valley being decimated.
In Laplace variables  this gives 
${\cal H}(p)=( (-2 \partial_p P(\eta=0,p) + P(\eta=0,p) ) P(p) )$
which using (\ref{solu}) gives back the above result (\ref{qexpand}).

We also give the explicit expression for the normalized 
dimensionless correlation:

\begin{eqnarray}
&& \frac{\overline{\langle x(t) x(t') \rangle}}
{ \sqrt{\overline{\langle x^2(t) \rangle}}
\sqrt{\overline{\langle x^2(t') \rangle}} }
= \frac{72}{61 \alpha} -
\frac{40}{61 \alpha^2} -
\frac{180}{427 \alpha^3} +
\frac{2045}{1281 \alpha^4} \nonumber \\
&& +
e^{-( \alpha-1)}
\left( \frac{20}{61 \alpha^2} -
\frac{80}{183 \alpha^3} -
\frac{36}{61 \alpha^4} \right)
\end{eqnarray}
which decreases from $1$ to $0$ as $\alpha=\frac{\ln t}{\ln t'}$ goes from
$1$ to $+\infty$. Note that the decay as $1/\alpha$ for large
$\alpha$ is characteristic of the generic decay of 
corrections to asymptotics as $1/\Gamma$.

\subsubsection{Some results for the biased case}

\label{biased}

From Appendix \ref{app:twotime} one has in principle
an exact expression for the Laplace transformed
two-time diffusion front. It is, however, very complicated
and thus we give here only a few simpler quantities.

Given that the bias is towards $x>0$ and that
the starting point is $x=0$, the probability that 
the particle is on the side $x>0$ both at $t'$ and $t$
has the aging behavior:

\begin{eqnarray}
P^{++}(\Gamma,\Gamma') & = & \overline{\langle\theta(x(t))\theta(x(t'))
\rangle } =
\frac{1}{ 16 \sinh ^2 \gamma \sinh ^2 \gamma'}
(e^{2 \gamma} ( e^{2 \gamma'} - 2 \gamma' -1 )
\nonumber \\
& - & 2 \gamma ( e^{- 2 \gamma'} + 2 \gamma' -1 )
-2 e^{2 \gamma'} -e^{- 2 \gamma'}
+ 4 \gamma'^2+2 \gamma' +3 )
\end{eqnarray}
with $\gamma=\Gamma \delta$ and $\gamma'=\Gamma' \delta$,
while $P^{--}(\Gamma,\Gamma')$ is obtained by $\delta \to -\delta$.
The other possibilities,
$P^{+-,-+}(\Gamma,\Gamma')$ are obtained from
the single time identities
$P^{++}(\Gamma,\Gamma')+P^{+-}(\Gamma,\Gamma') = P^{+}(\Gamma)=
\overline{l}^+_{\Gamma}/\overline{l}_{\Gamma}$
and $P^{++}(\Gamma,\Gamma')+P^{-+}(\Gamma,\Gamma') = P^{+}(\Gamma')=
\overline{l}^+_{\Gamma'}/\overline{l}_{\Gamma'}$ where
$\overline{l}^{\pm}_{\Gamma'}$ are given in 
(\ref{lengthpm})

The explicit expression for the correlation $\langle x(t) x(t')\rangle$ 
is given by (\ref{corrb}).

\subsection{Full two-time aging function in a semi-infinite system}

There is one instructive situation
where it is simpler to obtain explicitly the full two-time
probability distribution $\overline{\mbox{Prob}(xt,x't'|00)}$,
even in the presence of a bias. This is the case 
of  Sinai diffusion on a semi-infinite axis $(0,+\infty)$
with a {\it reflecting boundary} at $x=0$ and possibly some drift
in the $(+)$ direction.

In this case the single time diffusion front is simply 
$\overline{\mbox{Prob}^{+}_R(x't'|00)} = E_{\Gamma'}^{+}(x')$,
i.e., the probability distribution of the length
$x'$ of the first renormalized bond near the boundary,
whose Laplace transform is given in equation (\ref{finitex}).
Similarly, the two-time diffusion front is equal to
$\overline{\mbox{Prob}^{+}_R(xt,x't'|00)} = E^{+}_{\Gamma,\Gamma'}(x,x')$,
i.e., the probability that the 
first bond near the boundary has  length $x'$ at $\Gamma'$
and length $x$ at $\Gamma$. Its RG equation is given by 
(\ref{brsrgdrift})
\begin{eqnarray} 
\partial_\Gamma  E_{\Gamma,\Gamma'}^{+}(x,x') = 
P_{\Gamma}^{-}(0,.) *_x E_{\Gamma,\Gamma'}^{+} (.,x') *_x \int_0^{\infty} 
d\zeta' P_{\Gamma}^{+}(\zeta' ,.) -
E_{\Gamma,\Gamma'}^{+}(x,x') \int_0^{\infty} dl' P_{\Gamma}^{-}(0,l')
\end{eqnarray}
but with the initial condition 
\[
E_{\Gamma',\Gamma'}^{+}(x,x') 
= \delta(x-x') E_{\Gamma'}^{+}(x')
\] 
at $\Gamma=\Gamma'$.
Note that within the effective dynamics this system
has the flavor of a directed model, since 
$x(t)-x(t')$ is always positive. It is thus convenient to
define the double Laplace transform
$\hat{E}_{\Gamma,\Gamma'}^{+}(p,p') 
=\int_0^{+\infty} dx' e^{-p'x'} \int_{x'}^{\infty} dx e^{-p(x-x')}  
E_{\Gamma,\Gamma'}^{+}(x,x')$
and using the fixed point solution (\ref{solu-biased})
for $P_{\Gamma}^{\pm}$ together with the properties (\ref{equpm})
the above RG equation simplifies into 
$\partial_\Gamma  \ln \hat{E}_{\Gamma,\Gamma'}^{+}(p,p') = 
 \partial_\Gamma  \ln \frac{u^+_{\Gamma}(0)}{u^+_{\Gamma}(p)}$.
Using the above initial condition, we obtain the factorized
form $\hat{E}_{\Gamma,\Gamma'}^{+}(p,p') = E_{\Gamma'}^{+}(p')
Q^{+}_{\Gamma,\Gamma'}(p)$ i.e., in real space:
\begin{eqnarray}
\overline{\mbox{Prob}^{+}_R(xt,x't'|00)} = 
E^{+}_{\Gamma'}(x') Q^{+}_{\Gamma,\Gamma'}(x-x')
\end{eqnarray}
where $E^{+}_{\Gamma'}(x')$ is by
(\ref{maximum}) and  the distribution of the
relative displacements $y=x(t)-x(t')$ is
\begin{eqnarray} 
Q^+_{\Gamma,\Gamma'}(y) 
=LT^{-1}_{p \to y} \  e^{-\delta (\Gamma-\Gamma')}
 \frac{ \sinh (\delta \Gamma') }{\sinh (\delta \Gamma)}
\ \frac{\sqrt{p+\delta^2} \coth(\Gamma' \sqrt{p+\delta^2})-\delta)}
{\sqrt{p+\delta^2} \coth(\Gamma \sqrt{p+\delta^2})-\delta} 
\end{eqnarray}

In the symmetric case $\delta=0$, Laplace inversion gives
the two-time front:
\begin{eqnarray} 
Q(y,t,t') & = & \frac{2}{\Gamma^2}
 \sum_{n=0}^{\infty} \pi 
\left(n+\frac{1}{2}\right) \frac{\Gamma'}{\Gamma}
{\rm cotan} 
\left( \pi \left( n+\frac{1}{2}\right) 
\frac{\Gamma'}{\Gamma}\right)
e^{-\frac{y^2}{\Gamma^2} \pi^2 (n+\frac{1}{2})^2}   
\nonumber \\
& + & \frac{\Gamma'}{\Gamma} \left( \delta(y) - \frac{2}{\Gamma'^2}
 \sum_{m=1}^{\infty} \pi m \tan \left(\pi m \frac{\Gamma}{\Gamma'} \right)
e^{-\frac{y^2}{\Gamma'^2} \pi^2 m^2}  \right)
\end{eqnarray}
with $\Gamma=T \ln t$ and $\Gamma'=T \ln t'$.

In the biased case one finds:
\begin{eqnarray} 
Q^{+}_{\Gamma,\Gamma'}(y) 
& = & \frac{e^{\gamma'} \sinh(\gamma')}{e^{\gamma} \sinh(\gamma)}
\left[ \frac{2}{\Gamma^2} 
\sum_{n=0}^{+\infty} d_n(\gamma,\gamma')
e^{- \frac{y}{\Gamma^2} s_n^+(\gamma)} \right. 
\nonumber \\
& + & \left.  
\delta(y) -
\frac{2}{\Gamma'^2} \sum_{m=1}^{+\infty} 
\frac{\pi^2 m^2}{
\pi m\; {\rm cotan}(\pi m \frac{\gamma}{\gamma'}) - \gamma'}
e^{- \frac{y}{\Gamma'^2} (\gamma'^2 + m^2 \pi^2 ) } \right]
\end{eqnarray}
where the $s_n^{+}(\gamma)$ are given in (\ref{snlaplace})
and the $d_n(\gamma,\gamma')$ are given in terms of
the $\alpha^{+}_n(\gamma)$ defined in (\ref{anlaplace}) via
\begin{eqnarray}
d_n(\gamma,\gamma') = \alpha^{+}_n(\gamma)^2
\frac{\alpha^{+}_n(\gamma) {\rm cotan}\left(\frac{\gamma'}{\gamma}
\alpha^{+}_n(\gamma) \right) - \gamma }{
\alpha^{+}_n(\gamma)^2 + \gamma^2 - \gamma}
\end{eqnarray}
except for the term $n=0$ in the domain $\gamma>1$ where
\begin{eqnarray}
d_0(\gamma,\gamma') = \tilde{\alpha}^{+}_0(\gamma)^2
\frac{\tilde{\alpha}^{+}_0(\gamma) {\rm coth}
\left(\frac{\gamma'}{\gamma}
\tilde{\alpha}^{+}_0(\gamma) \right) - \gamma }{
\tilde{\alpha}^{+}_0(\gamma)^2 - \gamma^2 + \gamma}
\end{eqnarray}
with $\tilde{\alpha}^{+}_0$ given in (\ref{anlaplace}).
In the limit $\gamma= \Gamma \delta = T \delta \ln t \gg 1$
$\gamma'= \Gamma' \delta = T \delta \ln t' \gg 1$, one finds
the simple aging form:
\begin{eqnarray} 
Q^{+}_{\Gamma,\Gamma'}(y) \sim
\frac{t'^\mu}{t^\mu} \delta(y) +
\left(1 - \frac{t'^\mu}{t^\mu}\right) \frac{\mu^2}{T^2 t^{\mu}}
e^{- y \mu^2/(T^2 t^\mu)}
\label{diffbias}
\end{eqnarray}
with $\mu=2 \delta T$. Equation (\ref{diffbias}) coincides exactly with
the small $\mu$ limit of the
two-time diffusion front in the directed model
\cite{footnote2}. As noted above
the model with a reflecting wall is in effect asymptotically
to a directed model.
In the presence of a bias,  in the long-time limit $\gamma, \gamma' \gg1$
it gives the same results as the full model
discussed in the previous section, as the influence of the
wall vanishes in that limit.

\subsection{Dynamics within a well}
\label{sub:inwell}

We now study the dynamics in the time regime $t-t' \sim t'^\alpha$
with fixed $\alpha<1$. As was discussed in \ref{subsecgeneral}
this is dominated by renormalized valleys at $\Gamma'$ 
with two
degenerate minima $U_1$ and $U_2$ with $U_1-U_2$ of order 
$O(T)$ as in Fig. \ref{figwell}. 
In a typical valley many such degeneracies may exist
on small scales $y=x-x'$ of order $1$, with non-universal
statistics, but we are interested in rare valleys where such degeneracies
exist on scales $y \sim \Gamma'^2$  with barriers 
$\Gamma_0 \sim \Gamma'$; the distribution of these is universal.

We  introduce the probability density 
$R_\Gamma(\zeta,l,x,\Gamma_0)$ that a renormalized bond
at scale $\Gamma$ has length $l$, barrier $\zeta=F-\Gamma$
and has a secondary minimum,
degenerate in energy with the absolute minimum
(i.e., the lower edge of the bond) and separated from it
by a distance $x$ and a barrier $\Gamma_0$.
The calculation of this quantity is performed in
Appendix \ref{app:inwellbiased}. We find the
simple decoupled form:
\begin{eqnarray}
R_\Gamma(\zeta,l,x,\Gamma_0) =  \theta(\Gamma - \Gamma_0) 
\theta(l-x) P_\Gamma(\zeta,l-x) \frac{1}{\Gamma_0^4} \hat{r}(x/\Gamma_0^2) 
\end{eqnarray}
with
\begin{eqnarray}
\hat{r}(X) = 4 \sum_{n=1}^{\infty} n^2 \pi^2 
\left( 2 X n^2 \pi^2  - 3 \right)
 e^{- X n^2\pi^2 }.
\label{soludecoupled}
\end{eqnarray}
We have written for simplicity $R$ in unrescaled variables,
but the expression is of course valid only in the scaling regime 
$\zeta \sim \Gamma_0 \sim \Gamma$, $x \sim l \sim \Gamma^2$
(see Appendix for details).  Note that its total normalization
is $\int_0^{+\infty}  d\zeta \int_0^{\Gamma} d\Gamma_0 \int_0^{+\infty}  dl  \int_0^{l} dx 
R_\Gamma(x,\Gamma_0) \sim 1/\Gamma$
as expected since it corresponds to a rare event; with 
$T\ll\Gamma$, the density with double minima within 
$T$ of each other is $TR$.

We can now obtain the probability that a
Sinai walker will move by $y$ between $t'$ and $t$
for $t-t' \sim t'^{\alpha}$ with $\alpha<1$.
We need first the probability $K_{\Gamma'}(y,\Gamma_0)$ that the
the starting point happens to belong to a
renormalized {\it valley} at $\Gamma' = T \ln t'$ 
which possesses two degenerate minima separated by a distance $y$
and a barrier $\Gamma_0<\Gamma'$. Taking into account that each of the
two bonds forming the valley may be the one with the degenerate minima
(the probability that both have degenerate minima is negligible in the
scaling regime of interest) one gets, using (\ref{soludecoupled})
\begin{eqnarray}
K_{\Gamma'}(y,\Gamma_0) & = &
\frac{1}{\Gamma'^2}
\int_{l_1,l_2} (l_1+l_2) [ P_{\Gamma'}(l_1) R_{\Gamma'}(l_2,y,\Gamma_0)
+ P_{\Gamma'}(l_2) R_{\Gamma'}(l_1,y,\Gamma_0) ] 
\nonumber \\
& = & 2 (1 + \frac{y}{\Gamma'^2}) 
\frac{1}{\Gamma_0^4} \hat{r}(\frac{y}{\Gamma_0^2})
\label{K}
\end{eqnarray}

When the starting point belongs to such a 
valley (characterized by $y$ and $\Gamma_0$)
the walker at $\Gamma'$ is well equilibrated
(since $\Gamma_0 < \Gamma'$) and its 
thermal distribution is,
in a scaling sense, a sum of two delta function peaks
separated by a distance $y>0$ with weights
$p=1/(1 + e^{-w/T})$ (at the abolute minimum)
and $1-p$ (at the secondary minimum), where
$w$ represents the (free) energy difference
(of order $O(T)$) between the two minima.
When estimating the
distribution of $y=|x(t)-x(t')|$ one probes this
equilibrium distribution at $t$ and $t'$.
Denoting $\hat{\Gamma}=T \ln (t-t')$
it is clear that $y$ can be larger than
$O(1)$ only if $\hat{\Gamma} > \Gamma_0$
in which case it is equal to the separation, $y$,
of the two minima with  probability $2 p(1-p)$ and 
small otherwise.
Thus to obtain the distribution $\tilde{Q}(y,t,t')$ for
$y$ in the scaling regime, one must sum over
all barriers smaller than $\hat{\Gamma}=T \ln (t-t')$
(the larger ones contribute only to the already dominant
delta function part of $Q(y,t,t')$). Thus, using
(\ref{soludecoupled}), we find in the scaling regime
of fixed $\alpha=\frac{\ln (t-t')}{\ln t'} < 1$ and $y/(T \ln t')^2$:
\begin{eqnarray}
\tilde{Q}(y,t,t')  & = & C(T)  \int_0^{\hat{\Gamma}} 
d \Gamma_0 ~ K_{\Gamma'}(y,\Gamma_0) \\
& = & 2 \left[1 + \frac{y}{(T \ln t')^2}\right] 
\frac{T}{(T \ln (t-t'))^3} G\left[\frac{y}{(T \ln (t-t'))^2}\right]
\label{qwell}
\end{eqnarray}
where we have defined:
\begin{eqnarray}
G(X) = 4 \pi^2 \sum_{n=1}^{\infty} n^2  e^{-X n^2\pi^2 }
={1 \over {\sqrt \pi X^{3/2}}} 
\sum_{m=-\infty}^{+\infty} \left( 1 +{2 m^2 \over X} \right) 
 e^{-{m^2 \over X}}
\label {functionG}
\end{eqnarray}
Note that the factor $C(T)=2 \int_0^{+\infty} dw e^{-w/T}/(1 + e^{-w/T})^2=T$ arises
from the fact that the distribution of $w$ is constant around $w=0$.

The above result is consistent with previous observations
in the case of finite, but large, $t-t'$ where 
moments $\int dy y^k \tilde{Q}(y,t,t')$ were argued \cite{laloux_pld_sinai}
to grow as $(T \ln(t-t'))^{2 k-1}$ for $k>1/2$ and be bounded
for $k<1/2$. Here we obtain, in addition, the behavior for
more separated times with positive $(\ln(t-t')/\ln t') <1$.

In the biased case we
find similarly (see Appendix \ref{degeneratebiased})
\begin{eqnarray} \label{functionR}
R^{\pm}_\Gamma(\zeta,l,x,\Gamma_0) =  \theta(\Gamma - \Gamma_0) 
\theta(l-x) P^{\pm}_\Gamma(\zeta,l-x) \frac{1}{\Gamma_0^4} 
\hat{r}(x/\Gamma_0^2) e^{- x \delta^2}
\end{eqnarray}
where $\hat{r}(X)$ is the {\it same} function (\ref{soludecoupled})
as in the symmetric case. From this we obtain, as above,
the probability $K^{\pm}_{\Gamma'}(y,\Gamma_0)$ that the
the starting point happens to belong to a
renormalized valley at $\Gamma' = T \ln t'$ 
which possesses two degenerate minima separated by a distance $y$
and a barrier $\Gamma_0$. We find 
\[
K^{\pm}_{\Gamma'}(y,\Gamma_0) = 2 
\left(1 + \frac{y}{\overline{l}_{\Gamma'}}\right) 
\frac{1}{\Gamma_0^4} \hat{r}\left(\frac{y}{\Gamma_0^2}\right)  
e^{- y \delta^2}
\]
and thus, integrating over the barriers $\Gamma_0 < \hat{\Gamma}$
yields the distribution of displacements $y=|x(t)-x(t')|$ in 
the presence of
a small bias. We thereby obtain that in the scaling regime where
the three scaling variables
$y/(T \ln(t-t'))^2$, $\delta T \ln t'$ and 
$\alpha=\ln(t-t')/\ln t'$ are held fixed,
the distribution of displacements is dominated
by the rare events with valleys with two degenerate minima
and with
\begin{eqnarray}
\tilde{Q}^{\pm}(y,t,t') \sim 2
\left(1 + \frac{ \delta^2 y}{\sinh^2( \delta T \ln t')}\right) 
\frac{T}{(T \ln (t-t'))^3} 
G\left(\frac{y}{(T \ln (t-t'))^2}\right) e^{- y \delta^2}
\end{eqnarray}
where, interestingly, $G$ is the same function (\ref{functionG})
as in the symmetric case. This is 
because the rare events that dominate 
are  those in which the relevant part
of the landscape is almost symmetric.

\subsection{crossover at $t \sim t'$}
\label{crossover}

So far we have studied separately the regime $t-t' \sim t'^\alpha$ ($\alpha >1$)
and the regime $t-t' \sim t'^\alpha$ ($\alpha < 1$). For completeness
let us mention what happens (for the symmetric case)
in the crossover regime $t-t' \sim t'$.
In this regime, the distribution of displacements $y=|x(t) - x(t')|$
takes the form of a sum of two contributions.

First, it was found in (\ref{qexpand}) that in the
limit $\alpha \to 1^+$ this distribution is controlled
by barriers of order $\Gamma' \approx \Gamma$. For closer times,
$\Gamma - \Gamma' \sim O(T)$ one must consider
more precisely the jumping process over the barrier.
Asssociating a single relaxation time $\tau = e^{(\Gamma'+\epsilon)/T}$ 
with the barrier of height $\Gamma'+\epsilon$
one finds that the contribution of these events
to the the distribution of $y$ takes the form,
(in addition to a piece proportional to $\delta(y)$):
\begin{eqnarray}
Q_1(y,t,t') = f(t,t') \frac{1}{(T \ln t')^3} {\cal H}(Y = \frac{y}{(T \ln t')^2}) 
\label{previous}
\end{eqnarray}
where the Laplace transform of ${\cal H}(Y)$ was defined in (\ref{qexpand}) .
The coefficient $f(t,t')$ is obtained 
by noting that the probability that the particle jumps between $t'$ and $t$
is $(e^{- t'/\tau} - e^{- t/\tau})$ and that the distribution of $\epsilon$ 
is uniform around $\epsilon=0$ with density $1/\Gamma$:
\begin{eqnarray}
f(t,t') = \int_{-\infty}^{+\infty} d \epsilon ~ \left[ \exp( - e^{ - \epsilon/T}) - 
\exp( - \frac{t}{t'} e^{ - \epsilon/T} ) \right] = T \ln \frac{t}{t'}
\end{eqnarray}

The second contribution comes from the events
discussed in the previous subsection, corresponding to
degenerate wells (Fig. \ref{figwell}). These are dominant for $\alpha <1$
and they also give a contribution for $\alpha=1$, which must be added to
(\ref{previous}). The corresponding contribution $Q_2(y,t,t')$ to the distribution
of displacements is simply given by the limit $\alpha \to 1$
of equation (\ref{qwell}) which corresponds to setting 
$\ln(t-t') \approx \ln t'$).

Putting this all together we give the explicit
expressions for the second moment of $y$ in the various regimes
which can be obtained from $Q = Q_1 + Q_2$:

\begin{eqnarray}
&& \overline{\langle (x(t)-x(t'))^2 \rangle}  \approx \\
&& T (T \ln (t-t'))^3 \left[ \frac{8}{45} + \frac{48}{945} 
\left( \frac{\ln (t-t')}{\ln t'} \right)^2 \right]  \qquad t-t' \sim t'^{\alpha} ~,~
 \alpha<1 \\
&& T (T \ln t')^{3} ( \frac{8}{35} + \frac{272}{315} \ln(\frac{t}{t'}) )
\qquad t-t' \sim t' \\
&& (T \ln t')^{4} F_2\left[ \frac{\ln t}{\ln t'} \right]  \qquad t-t' \sim t'^{\alpha} ~,~
 \alpha > 1
\end{eqnarray}
where the function $F_2[\alpha]$ is given in (\ref{momentdiff}).

\subsection{Rare events in the single time-diffusion front}

\label{sub:rareevents}

In this Section we examine further the rare events which
produce subdominant corrections to the results from the
RSRG. As discussed in Section \ref{validity}, although
subdominant, these corrections give the principal
contribution to some observables when the leading 
contribution vanishes. This is the case for the 
{\it thermal width} of the diffusion front 
analyzed in \ref{secsinglediffusionsymm}, since this is zero in 
the effective dynamics. 
We now examine this in the symmetric model.

\begin{figure}[thb] 
\centerline{\fig{12cm}{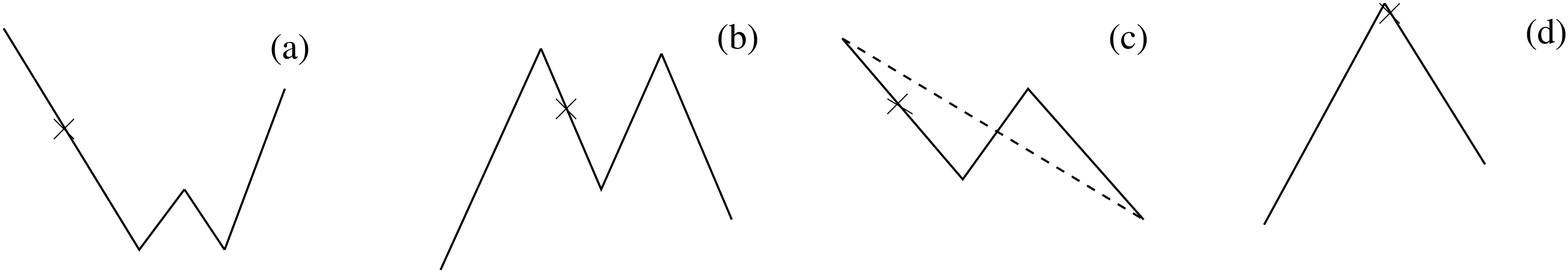}}
\caption{\label{figrare} \narrowtext Rare events which contribute
to to the thermal width of the diffusion front. The starting
point is indicated by a cross. (a) Valley with two degenerate minima.
(b) Almost degenerate barriers. (c) Valley just being decimated with a
barrier $\Gamma + \epsilon$. (d) Rare event of
higher order with the  starting point near the upper edge of a bond.}
\end{figure}

The possible rare events which contribute to the splitting of the
thermal packet are indicated in Fig. \ref{figrare}.
The most important ones, all occurring with probabilities
of order $1/\Gamma$, are the following. In case (a) the starting point belongs
to a valley with two degenerate minima: this is the equilibrium
situation already considered in the previous Section. 
In case (b) the splitting is due to the starting point
being in a valley with two almost degenerate barriers: at scales
when the packet overcomes the barriers, the packet will split
between the two wells located on either sides, an  intrinsically
nonequilibrium phenomenon. Note that if the packet is split at
$t$ (as in (b)) the probability that it remains split until
a later time $\tilde{t}$ decays as $(\ln t/\ln \tilde{t})^2$.
In case (c) the walker at $\Gamma$ belongs to a valley
with a barrier $\Gamma + \epsilon$, with $\epsilon \sim O(T)$ positive or 
negative.
In this case the thermal packet is already split at $\Gamma$ between 
two valleys. 
There are
of course other rare events: for instance the one illustrated as
case (d), when the starting point is near the 
upper edge of a bond; this also corresponds to an out of
equilibrium situation, but it occurs with a smaller probability
$O(1/\Gamma^2)$.

Let us estimate in more detail the probabilities
$Q_{\Gamma}^{(a)}(y)$, $Q_{\Gamma}^{(b)}(y)$ and
$Q_{\Gamma}^{(c)}(y)$
associated respectively with events (a), (b) and (c)
in Fig. \ref{figrare}, that the packet be split 
at $\Gamma$ with a fixed separation $y$ between 
the two parts of the packet. Let us start with events (a)
and (b) which can be treated similarly.

Events of type (a):
We have already computed in (\ref{K}) the probability
$K_{\Gamma}(y,\Gamma_0)$ that the origin belongs at $\Gamma$ to 
a renormalized valley having two degenerate minima separated by a distance $y$ 
and a barrier $\Gamma_0$. Integrating over the barrier $\Gamma_0$
one gets:
\begin{eqnarray}
&& Q_{\Gamma}^{(a)}(y) = \int_0^{\Gamma} 
d \Gamma_0 K_{\Gamma}(y,\Gamma_0)
=  2 \left(1 + \frac{y}{\Gamma^2}\right) 
\frac{1}{\Gamma^3} G\left( \frac{y}{\Gamma^2}\right)  
\label{qa}
\end{eqnarray}
where the scaling function $G$ has been introduced in (\ref{functionG}).

Events of type (b):
Here we need to compute the probability $Q_{\Gamma}^{(b)}(y)$ 
that the origin belongs to a configuration of type $(b)$ with
a distance $y$ between the two minima. We first compute
the probability $R_{\Gamma}(l_1,y_1)$ that a bond
 at scale $\Gamma$ has a length $l_1$ and two degenerate minima
separated by a distance $y_1$ (\ref{soludecoupled})
\begin{eqnarray}
R_{\Gamma}(l_1,y_1) = \int_0^{\infty} d\zeta \int_0^{\Gamma} d \Gamma_0 
R_{\Gamma}(\zeta,l_1,y_1,\Gamma_0) = P_{\Gamma}(l_1-y_1) 
\frac{1}{\Gamma^3} G\left( \frac{y_1}{\Gamma^2}\right) 
\end{eqnarray}
and thus:
\begin{eqnarray}
 Q_{\Gamma}^{(b)}(y)
&&  = \frac{2}{\Gamma^2}
\int_0^{\infty} dl_1 \int_0^{\infty} dl_2 \int_0^{l_1} dy_1
y_1 R_{\Gamma}(l_1,y_1) P_{\Gamma}(l_2) \delta(y-(l_1+l_2)) \nonumber \\
&& =\frac{2}{\Gamma^3 }
\int_0^{\frac{y}{\Gamma^2}} dY_1 Y_1  G( Y_1)
(P(.)*P(.))_{\lambda=\frac{y}{\Gamma^2}-Y_1}  
\label{qb}
\end{eqnarray}
in terms of the scaled $P$ from Eq. (\ref{pdel}).

In either case (a) or (b) the packet is split
between two wells and the thermal 
distribution can be written, in a scaling sense, as a sum of
two delta function peaks, of the form $p \delta(x-x_1) + (1-p) \delta(x-x_2)$
centered at each minima $x_1$ and $x_2$ with $x_1 <x_2$,
$|x_2-x_1|=y$. In case (a), as before $p_a=1/(1+e^{-w/T})$
where $w$ is the free energy difference between the two minima,
while in case (b) a simple estimate of the relative escape rates
in Fig. \ref{figrare} also leads to $p_b=1/(1+e^{-v/T})$ where $v$ is now the 
(effective free-)
energy difference between the maxima.

Thus we can estimate the dominant large time
behavior of the moments of the thermal width
coming from the contributions of (a) and (b) which simply add to give:
\begin{eqnarray}
\overline{\langle |x(t) - \langle x(t)\rangle |^k \rangle}_{(a + b)}
 \approx c_k(T)~ (T \ln t)^{2 k -1} 
\int_0^{+\infty} dY Y^k {\cal Q}^{(a + b)}(Y) 
\label{momab}
\end{eqnarray}
where the scaled distribution is
\begin{eqnarray}
{\cal Q}^{(a + b)}(Y) = 2 (1 + Y ) G(Y)  +2  \int_0^Y dY_1 Y_1  G( Y_1)  
(P(.)*P(.))_{\lambda=Y-Y_1}  
\end{eqnarray}
with $Y\equiv y/\Gamma^2$
The coefficients $c_k(T)$ can be computed using the fact the 
distributions of $w$ and of $v$ have constant density near $0$. This
gives 
\[
c_k(T) = \int_{-\infty}^{+\infty} 
dw (e^{- k w/T} + e^{-w/T})/(1 + e^{-w/T})^{k+1}= \frac{2}{k} T .
\]
Using the Laplace transformed expression 
\begin{eqnarray}
&& \int_0^{+\infty} dY Y {\cal Q}^{(a + b)}(Y) e^{- s Y} \nonumber \\
&& = 2 \partial_s 
\left( \sqrt{s} \coth(\sqrt{s}) 
+ \frac{1 + \cosh^2(\sqrt{s})}{\sinh(2 \sqrt{s})}
\left(\frac{2}{\sinh(2 \sqrt{s})} - \frac{1}{\sqrt{s}}\right) \right)
\label{lapab}
\end{eqnarray}

We must now consider the events (c) as shown in Fig. \ref{figrare}. 
The barrier that the particle must overcome to
leave the valley is $\Gamma + \epsilon$. This corresponds to a single
relaxation time $\tau = e^{(\Gamma + \epsilon)/T}$. Thus the probability 
that the particle is still in the valley at time $t= e^{\Gamma/T}$
is simply $p_c = \exp(-t/\tau) = \exp(- e^{-\epsilon/T})$.
The thermal distribution can then again be written, 
in a scaling sense, as $p_c \delta(x-x_1) + (1-p_c) \delta(x-x_2)$
where $x_1$ is the bottom of the valley being decimated and
$x_2$ the bottom of the new valley. The distribution of 
$y=|x_2-x_1|$ is simply: 

\begin{eqnarray}
 Q_{\Gamma}^{(c)}(y)
&&  = \frac{2}{\Gamma^2}
\int_0^{\infty} dl_1 \int_0^{\infty} dl_2 \int_0^{\infty} dl_3
(l_1 + l_2) P_{\Gamma}(l_1) P_{\Gamma}(\zeta=0,l_2) P_{\Gamma}(l_3) 
\delta(y-(l_2+l_3)) \nonumber \\
&& = \frac{1}{\Gamma^3 }
\int_0^{\frac{y}{\Gamma^2}}
d\lambda_2 (1 + 2 \lambda_2) P(\eta=0,\lambda_2) P(\frac{y}{\Gamma^2}  - \lambda_2)
\label{qc}
\end{eqnarray}
The contributions of (c) to the moments thus read:
\begin{eqnarray}
\overline{\langle |x(t) - \langle x(t)\rangle |^k \rangle}_{(c)}
 \approx d_k(T)~ (T \ln t)^{2 k -1} 
\int_0^{+\infty} dY Y^k {\cal Q}^{(c)}(Y) 
\label{momc}
\end{eqnarray}
where the scaled distribution reads in Laplace transform:
\begin{eqnarray}
\int_0^{+\infty} dY e^{-s Y} {\cal Q}^{(c)}(Y) = 
\frac{2}{\sinh(2 \sqrt{s})} ( 
{\rm coth} \sqrt{s} + \sqrt{s} - \frac{1}{\sqrt{s}} ) 
\label{lapc}
\end{eqnarray}
Using the fact that the distribution of $\epsilon$ is constant
near zero, one obtains the coefficients $d_k(T)$ for $k \ge 1$ as:
\begin{eqnarray}
&& d_k(T)= \int_{-\infty}^{+\infty} d\epsilon \left[
\exp(- k e^{-\epsilon/T}) (1 - \exp(- e^{-\epsilon/T}))
+ 
\exp(- e^{-\epsilon/T}) (1 - \exp(- e^{-\epsilon/T}))^k \right]  
\nonumber \\
&& = T \left[ \ln(1 + \frac{1}{k}) + \sum_{p=1}^{k} (- 1)^{1+p} C^p_k \ln(1+p) 
\right]
\end{eqnarray}
Note that the above argument can be made indentically
in the region $\epsilon <0$ and thus we have integrated 
$\epsilon$ from $-\infty$ to $+\infty$.

Our final result for the moments are obtained as the sum of (\ref{momab}) and 
(\ref{momc}) and can be computed using the Laplace transforms in 
(\ref{lapab}) and (\ref{lapc}).
Let us give the explicit resulting expression for the
the lowest moments:
\begin{eqnarray}
&& \overline{\langle |x(t) - \langle x(t)\rangle|\rangle} 
\approx \frac{2}{45} ( 68 + 41 \ln 2 ) T (T \ln t) \\
&& \overline{\langle x(t)^2\rangle - 
\langle x(t)\rangle^2} \approx \frac{4}{315} ( 95 + 68 \ln 2) T (T \ln t)^3
\end{eqnarray}

It would be interesting to measure these quantities in numerical
simulations and test these predictions.

Note that the above formula (\ref{momab},\ref{momc}) give
the leading behavior
for moments with $k>1/2$ which grow with time, while the
moments for $k<1/2$ are expected to be finite and non-universal
as in \cite{laloux_pld_sinai}. This can  be compared with
the work of Golosov \cite{golosov}, who 
showed the existence of an infinite time  
limit distribution for $y(t)=x(t) - \langle x(t)\rangle$
and gave an explicit formula in the case of
a continuum Brownian potential $U(x)$ (which
corresponds here to the limit $\sigma \to 0$
where additional universality holds). It is easy
to see from \cite{golosov}, as well as from more general
arguments, that this distribution has a tail $1/y^{3/2}$
at large $y$. 
Indeed, from Eqs. (\ref{functionG}, \ref{qa}, \ref{qb}) 
we see that for 
$t\rightarrow\infty$ so that $1 \ll y\ll \Gamma^2$,
\begin{eqnarray}
Q^{(a)+(b)+(c)}_\Gamma(y) \approx Q_\Gamma^{(a)}
(y)\approx\frac{2}{\sqrt\pi y^{3/2}}.
\end{eqnarray}
Thus moments $\langle y(t)^k\rangle$ for $k<1/2$ should
be finite and determined from short scales, while
moments for $k>1/2$ should diverge as $t\rightarrow\infty$.
Our results include the large but finite time 
behavior and thus go beyond those results of
\cite{golosov}.

To conclude, note that the rare events in (\ref{figrare}) 
which contribute to the width of the thermal packet
are also the one which play a dominant role in the aging dynamics
in the regimes $\alpha \leq 1$. We have seen that $(a)$ and $(c)$ 
are the one which contribute to $Q(y,t,t')$ in these regimes.
The event $(b)$ does not contribute to $Q(y,t,t')$
(since the particle cannot jump back to
the degenerate valley) but would
have to be considered to evaluate $P(x,t,x',t')$ in these regimes
as well. 

\section{Finite size properties of Sinai's model}

In this Section we apply the RSRG 
procedure to a finite size system with
various boundary conditions and obtain exact results
for the approach to equilibrium 
of several quantities.

\subsection{RG for a finite size system}

\label{finitesizerg}

\begin{figure}[thb] 
\centerline{\fig{14cm}{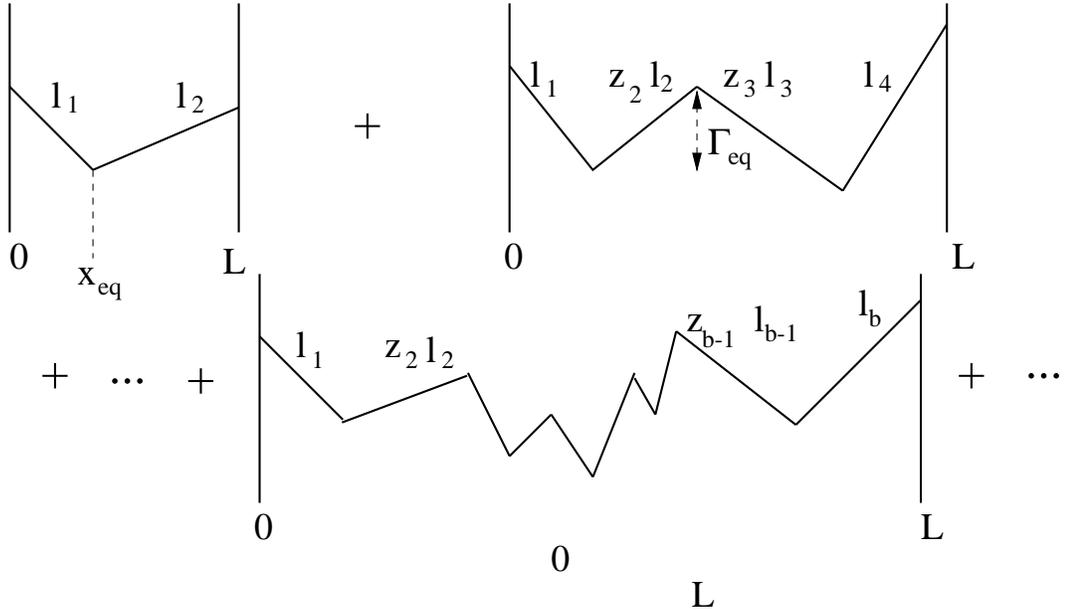}}
\caption{\label{figfinitegen} \narrowtext
Schematic of a finite size system of fixed length $L$
with reflecting boundaries (case RR).}
\end{figure}

In order to follow the general measure for a finite size
landscape one needs to introduce the set of functions
representing probabilities 
$N_{\Gamma,L}^b(l_1;\zeta_2,l_2;\zeta_3,l_3;\dots;\zeta_{b-1},l_{b-1};l_b)$
in an ensemble of systems of length $L$ with $b$  the number
of bonds and with barriers $\zeta_i=F_i - \Gamma$ and lengths $l_i$.
Note that we will not keep track of $\zeta_1$ and $\zeta_b$ as these
will effectively be $\pm\infty$ depending on the 
boundary conditions. The normalization condition 
reads:
\begin{eqnarray}  \label{measure}
Z_L = \sum_b \int_{\zeta_i,l_i} 
N_{\Gamma,L}^b(l_1;\zeta_2,l_2;..\zeta_{b-1},l_{b-1};l_b) = 1
\end{eqnarray}
Note that under decimation one follows separately 
$b=2,4,\ldots$ or $b=1,3,5\ldots$,
depending on the type of boundary conditions studied
here either reflecting (R) or absorbing (A). Let us write
the RG equation for a finite size system, choosing for
definiteness the case denoted RR below of two reflecting boundaries---as
explained in section \ref{boundaryrg}, in this case the first 
and last bonds have infinite barrier and cannot be decimated:
\begin{eqnarray} 
&& \left(\partial_\Gamma - \sum_{k=2}^{b-1} \partial_{\zeta_k} \right) 
N_{\Gamma,L}^b(l_1;\zeta_2,l_2;\ldots\zeta_{b-1},l_{b-1};l_b) = 
\nonumber \\
&&\int_{z,l+l'+l''=l_1} 
N_{\Gamma,L}^{b+2}(l;0,l';z,l'';\zeta_2,l_2;\ldots;
\zeta_{b-1},l_{b-1};l_b) + \nonumber \\
&& \sum_{k=2}^{b-1} 
\int_{z,l+l'+l''=l_k} N_{\Gamma,L}^{b+2}(l_1;\zeta_2,l_2;\ldots;\zeta_{k-1},
l_{k-1};z,l;0,l';\zeta_k-z,l'';
\zeta_{k+1},l_{k+1};\ldots;\zeta_{b-1},l_{b-1};l_b)
\nonumber \\
&& + \int_{z,l+l'+l''=l_b} N_{\Gamma,L}^{b+2}(l_1;\zeta_2,l_2;\ldots;
\zeta_{b-1},l_{b-1};z,l;0,l';l''). 
\label{rggeneral}
\end{eqnarray}
There exists a quasi-decoupled solution---for 
Laplace transformed distributions---of this equation (\ref{rggeneral})
(as was also found in the case of the RTFIC \cite{danfisher_rg3}),
which reads:
\begin{eqnarray} \label{decoupledrr}
&& N_{\Gamma,L}^b(l_1;\zeta_2,l_2;\ldots\zeta_{b-1},l_{b-1};l_b) 
= 
E_{\Gamma}^{+}(l_1) P_{\Gamma}^{-}(\zeta_2,l_2) P_{\Gamma}^{+}(\zeta_3,l_3)
... \nonumber \\
&& \ldots P_{\Gamma}^{-}(\zeta_{b-2}, l_{b-2}) 
P_{\Gamma}^{+}(\zeta_{b-1},l_{b-1}) E_{\Gamma}^{-}(l_{b}) 
\overline{l}_\Gamma \delta\left(L-\sum_{i=1}^{b} l_i\right) 
\quad {\rm b~~even} \ge 2
\end{eqnarray}
where we have allowed for a bias towards the right,
and $b$ is restricted to be even since we are
dealing with the RR case (see Figure \ref{figfinitegen}).
In this formula (\ref{decoupledrr})
$P_{\Gamma}^{\pm}(\zeta,l)$ are the bulk distributions 
satisfying (\ref{rg1}), $\overline{l}_\Gamma$ the average length
satisfying (\ref{rglength}),
and the $E_{\Gamma}^{\pm}(l)$ satisfy the semi-infinite 
boundary RG (\ref{brsrgdrift}). The integral of the measure
$Z_L$ over all variables in (\ref{measure}) satisfies:
\begin{eqnarray}
&& \int_0^{+\infty} dL e^{-p L} Z_L = 
\overline{l}_\Gamma \frac{E^{-}(p) E^{+}(p)}{1 - P^{-}(p) P^{+}(p)} 
= \overline{l}_\Gamma 
\frac{u^{+}(0) u^{-}(0)}{u^{+}(p) u^{-}(p) - U^{+}(p) U^{-}(p)} = \frac{1}{p} 
\nonumber \\
&& \label{normcheck}
\end{eqnarray}
where we have used (\ref{solu-biased}, 
\ref{finitex}), and thus
$Z_L=1$ so the finite 
size measure  is correctly normalized. 

In the case of two absorbing boundaries AA, the 
solution of the corresponding RG equations is obtained
by simply exchanging $+$ and $-$ in (\ref{decoupledrr})
($b$ remains even). In the case RA the solution reads:
\begin{eqnarray} \label{decoupledra}
&& N_{\Gamma,L}^b(l_1;\zeta_2,l_2;\ldots\zeta_{b-1},l_{b-1};l_b) = 
E_{\Gamma}^{+}(l_1) P_{\Gamma}^{-}(\zeta_2,l_2) 
P_{\Gamma}^{+}(\zeta_3,l_3) \\
&& \ldots P_{\Gamma}^{+}(\zeta_{b-2}, l_{b-2}) P_{\Gamma}^{-}
(\zeta_{b-1},l_{b-1}) E_{\Gamma}^{+}(l_{b}) 
\overline{l}_\Gamma \delta\left(L-\sum_{i=1}^{b} l_i\right)  
\quad b ~ \text{odd} \ge 3 
\end{eqnarray}
together with the term $b=1$, which corresponds to the
final state with a single $(+)$ bond over the
whole system (all particles having been absorbed
by the right boundary), and has for probability:
\begin{eqnarray} \label{distrieqra}
N_{\Gamma,L}^1 = \int_0^\Gamma d\Gamma' 
\int_{l_1,l_2,l_3} N_{\Gamma',L}^3(l_1;\zeta_2=0,l_2;l_3)
\end{eqnarray}
Finally, the solution in the AR case can be 
obtained by exchanging $+$ and $-$ 
in (\ref{decoupledra}).

\subsection{Evolution towards equilibrium in a system with
reflecting boundaries}

\label{secfinitesizediff}

Let us start by studying
the equilibrium and the approach to
equilibrium in the system of size $L$ with two reflecting
boundaries. The equilibrium state corresponds to the
large $\Gamma$ limit of the measure (\ref{measure}).
In that limit only the term $b=2$ (\ref{decoupledrr}) 
survives  and it corresponds to equilibrium
in a single renormalized valley. It
will be reached, as illustrated in Fig. \ref{figfinitegen},
when the last bulk bond is decimated at some 
$\Gamma = \Gamma_{eq} = T \ln t_{eq}$, with 
a certain sample to sample distribution 
for the equilibrium time $t_{eq}$, which we now compute.

\subsubsection{Distribution of equilibration time}

\label{sub:equilibriumtime}

The probability for a sample to reach equilibrium between $\Gamma_{eq}$
and $\Gamma_{eq}+d\Gamma_{eq}$, i.e., the probability that
the slowest relaxation time $t_{eq}$ be such that
$\Gamma_{eq}< T \ln t_{eq} < \Gamma_{eq} + d\Gamma_{eq}$, is
\begin{eqnarray} \label{distrieq}
\rho_{L}(\Gamma_{eq})=\partial_{\Gamma_{eq}}
\int_{l_1,l_2} N_{\Gamma_{eq},L}^2(l_1,l_2) =
\partial_{\Gamma_{eq}}
\left( \overline{l}_{\Gamma_{eq}} 
  E_{\Gamma_{eq}}^{+}(.) *_L E_{\Gamma_{eq}}^{-}(.) \right)
\end{eqnarray}
and using (\ref{finitex}), 
the Laplace transform with respect to the system size $L$ is
\begin{eqnarray} 
&& \int_0^{\infty} dL e^{-pL} \rho_{L}(\Gamma_{eq})
= \partial_{\Gamma_{eq}} \left( \frac{1}{
(p+\delta^2) \coth^2\left[\Gamma_{eq} \sqrt{p+\delta^2}\right] 
-\delta^2}\right).
\end{eqnarray}

For zero bias, one introduces the scaling variable 
\[
w=\frac{T \ln t_{eq}}{\sqrt{L}}
\]
and finds that
it is distributed as:

\begin{eqnarray}
P(w) & = & \frac{2}{w^3} \sum_{n=-\infty}^{+\infty} 
\left( 2 \pi^2 \left(n+\frac{1}{2}\right)^2 \frac{1}{w^2} -1 \right) 
e^{- \frac{\pi^2}{w^2} \left(n+\frac{1}{2}\right)^2 } \\
& = & \frac{2}{\sqrt{\pi} w^2} \sum_{m=-\infty}^{+\infty}
(-1)^m (1 - 2 m^2 w^2) e^{- m^2 w^2}
\label{rhoeqlast}
\end{eqnarray}

In the presence of a bias one can compute, e.g.,  the
average:

\begin{eqnarray}
&& \overline{T \ln t_{eq}}=\overline{\Gamma_{eq}}=
2 \sqrt{\frac{L}{\pi}} \int_0^1 du e^{- \delta^2 L u^2} 
\ln\left(\frac{1}{u}\right)
\end{eqnarray}

\subsubsection{Distribution of equilibrium position $x_{eq}$}

\label{sub:equilibriumposition}

The probability that the bottom of the single remaining 
equilibrated valley is at $x=x_{eq}$ can be obtained
as 
\begin{eqnarray}
Q_{L}(x_{eq})= N_{\Gamma \to \infty,L}^2(x_{eq},l_2) =
\left(\overline{l}_{\Gamma} 
E_{\Gamma}^{+}(x_{eq})  E_{\Gamma}^{-}(L-x_{eq}) \right)_{\Gamma \to \infty},
\end{eqnarray}
which leads simply to:
\begin{eqnarray} 
Q_{L}(x_{eq})= 
e^{+}(x_{eq})  e^{-}(L-x_{eq}) 
\end{eqnarray}
where 
\begin{eqnarray}
e^{\pm}(x)=LT^{-1}_{p \to x} (\frac{1}{u_{\Gamma=\infty}^{\pm}(p)})
=LT^{-1}_{p \to x} \frac{1}{\sqrt{p+\delta^2} \mp \delta}.
\end{eqnarray}
Thus we 
get $e^{+}(x)=e^{-}(x)+2\delta$ with
\begin{eqnarray}
e^{-}(x)=\frac{1}{\sqrt{\pi x}} e^{-x \delta^2} 
-\frac{2 \delta}{\pi }  \int_0^{\infty} dv 
\frac{e^{-x \delta^2 (1+v^2)}}{1+v^2}. \label{emp}
\end{eqnarray}
At small $x$ one has $e^{\pm}(x) \sim \frac{1}{\sqrt{x}}$
while for large $x$, $e^{+}(x) \approx 2\delta$
and $e^{-}(x) \sim e^{-x \delta^2}$. Thus
in the biased case with
$L \gg 1/\delta^2$ in 
equilibrium the particle is confined within
a distance $y=L-x_{eq} \sim 1/\delta^2$ near the left boundary
distributed as $2 \delta e^{-}(y)$.
In the symmetric case $\delta=0$
the equilibrium position is distributed over the
whole system as:
\begin{eqnarray}
Q_L(x_{eq}) = \frac{1}{\pi \sqrt{x_{eq} (L-x_{eq})} }
\label{xeq}
\end{eqnarray}
which has a simple probabilistic interpretation in terms of the
landscape random walk confined to $U(x)>U_{\min}=U(x_{eq})$ on
both sides of $x_{eq}$.

Finally, we  obtain the joint distribution of
equilibrium position $x_{eq}$ and equilibrium time
$\Gamma_{eq}=T \ln t_{eq}$:
\begin{eqnarray}
P_L(\Gamma_{eq},x_{eq}) = \partial_{\Gamma_{eq}} 
\left( \overline{l}_{\Gamma_{eq}} 
E^{+}(x_{eq}) E^{-}(L-x_{eq}) \right) 
\end{eqnarray}
where $E^{\pm}$ was computed in (\ref{finitex},\ref{elambda}).

\subsection{First passage times}

\subsubsection{With a reflecting boundary}

\label{sub:firstreflecting}

Let us compute the probability $S_{x_0,L}(\Gamma)$ that a walker 
starting at $x_0$ is still alive at $\Gamma$ in the presence of 
an absorbing boundary at $x=0$ and a reflecting boundary at $x=L$.
It can be expressed as an average over 
the measure AR (\ref{measure},\ref{decoupledra}) $S_{x_0,L}(\Gamma)=
\langle\theta(x_0-l_1)\rangle$. 
Thus its Laplace transform with respect to $L$ reads
\begin{eqnarray} 
\int_0^{\infty} dL e^{-pL} S_{x_0,L}(\Gamma) 
=  \frac{  P_{\Gamma}(p) }{ p E_{\Gamma}(p)}
 \int_0^{x_0} dl_1 e^{-p l_1} E_\Gamma (l_1).
\end{eqnarray}

In the particular case where the starting point
coincides with the reflecting boundary ($x_0=L$)
it is simpler to obtain the first passage time $T_{LL}$
at $x=0$. In that case, the probability to be absorbed coincides with
the probability $\rho_L(\Gamma)$ that the last decimation
occurs in the AR system at $\Gamma$; from (\ref{distrieqra}):
\begin{eqnarray} \label{distrieqra2}
\rho_{L}(\Gamma)= \partial_\Gamma N_{\Gamma,L}^1
= \overline{l}_{\Gamma} 
E_{\Gamma}^{+}(.) *_L P_{\Gamma}^{-}(0,.) *_L E_{\Gamma}^{+}(.).
\end{eqnarray}
In the symmetric case we thus obtain that 
the scaled first passage time variable 
$w = T \frac{\ln T_{LL}}{\sqrt{L}}$
is distributed as:
\begin{eqnarray} 
s(w) = \frac{2 \pi }{w^3} \sum_{n = -\infty}^{\infty} (-1)^n 
\left(n+\frac{1}{2}\right)
e^{- \frac{\pi^2}{w^2}  \left(n+\frac{1}{2}\right)^2} = \frac{2}{\sqrt \pi }
\sum_{m = -\infty}^{\infty} (-1)^m \left(m+\frac{1}{2}\right)
 e^{- w^2 \left(m+\frac{1}{2}\right)^2}.
\end{eqnarray}

We note that the distribution of $T_{LL}$ was obtained previously
by a completely different method in \cite{noskowicz}. Here
we also recover $\ln(\overline{(T_{LL})^q}) \sim q^2 L$, i.e.,
that the first passage time is a strongly
fluctuating quantity \cite{ledou_prl}.

\subsubsection{With an absorbing boundary}

\label{sub:firstabsorbing}

Let us now consider an absorbing boundary at $x=L$.
We first compute the probability
$p_0(x_0,L)$ that the walker starting at $x_0$ reaches
$x=0$ before $x=L$. Since the final state of the
AA system consists of two absorbing zones associated
with  each boundaries, the 
first one $[0,x_{eq}]$ and the second $[x_{eq},L]$
where $x_{eq}$ is distributed as in (\ref{xeq}). 
Thus, in presence of a bias applied in the direction
$\pm$ the result reads:
\begin{eqnarray} 
p^{\pm}_0(x_0,L) = \int_{x_0}^{L} dx e^{\mp}(x) e^{\pm}(L-x)
\end{eqnarray}
where $e{\mp}(x)$ was computed in (\ref{emp}).
In the symmetric case this gives 
$p_0(x_0,L) = \frac{1}{\pi} {\rm Arccos} \left(2 \frac{x_0}{L} - 1\right)$.

One can also compute the survival probability $S_{x_0,L}(\Gamma)$
of a walker starting at $x_0$ in the presence of two absorbing boundaries
at $x=0$ and $x=L$. It is obtained as an average over
the measure (\ref{decoupled}) of the finite size system AA as 
\begin{eqnarray}
S_{x_0,L}(\Gamma) = \overline{\langle\theta(x_0-l_1) \theta(L - l_1 -x_0)
\rangle}
\end{eqnarray}
and thus reads:
\begin{eqnarray} 
S_{x_0,L}(\Gamma)= 
LT^{-1}_{p \to L/\Gamma^2,q \to x_0/\Gamma^2} 
\frac{ \tanh (\sqrt{p+q}) \tanh (\sqrt p)}
{ q \sqrt p \sqrt{p+q} } \left(\frac{1}{\sinh^2(\sqrt p)}
-\frac{1}{\sinh^2(\sqrt {p+q})}\right).
\end{eqnarray}
Note that the distribution of times $t_{{\rm last}}$ at which
all particles have left a given AA sample is identical
to the one computed in (\ref{rhoeqlast})
as $\ln t_{{\rm last}} = \ln t_{eq}$.

\subsection{Averaged diffusion front }

\label{sub:diffusionfinite}

\label{secsemiinfinite}

We first discuss a system bounded by
two reflecting walls at $x=0$ and $x=L$. The full 
averaged diffusion front $\overline{{\rm Prob}_{0 L}(x,t \vert x_0,0)}$
for a walker starting at $x_0$ at $t=0$
is computed in  Appendix \ref{app:fullfinitesize}
for both biased and symmetric cases.

In the symmetric case it takes a
scaling form $\overline{{\rm Prob}_{0 L}(x,t \vert x_0,0)}
= \frac{1}{\Gamma^2} q_{\lambda}(X|X_0)$ as a function
of the rescaled variables $X=x/\Gamma^2$, 
$X_0=x_0/\Gamma^2$ and the rescaled length of the system
$\lambda=L/\Gamma^2$ where $\Gamma = T \ln t$.
The Laplace transform 
$\tilde{q}(p,p_0,q)=
\int_0^{+\infty} d\lambda \int_0^{\lambda} dX \int_0^{\lambda} dX_0 
e^{- (p X + p_0 X_0 + q \lambda)} q_{\lambda}(X|X_0)$
of the rescaled front is:
\begin{eqnarray}
\tilde{q}(p,p_0,q)= 
\frac{\tilde{P}_{p+p_0+q} \tilde{P}_{p+q} -
\tilde{P}_{p_0+q} \tilde{P}_{q}}{p_0 q (p+p_0+q) 
\tilde{E}_{p+p_0+q} \tilde{E}_q}
+
\frac{1}{p_0 q} \frac{\tilde{E}_{q+p}}{\tilde{E}_q} - 
\frac{1}{p_0 (p + p_0+q)}
\frac{\tilde{E}_{p_0+q}}{\tilde{E}_{p+p_0+q}} 
\end{eqnarray}
where $\tilde{P}_p=1/\cosh(\sqrt{p})$ 
and $\tilde{E}_p=\tanh(\sqrt{p})/\sqrt{p}$.
In the limit $L \to +\infty$ we can obtain 
the averaged front in a semi-infinite space
with a reflecting wall at $x=0$ as:

\begin{eqnarray}
\lim_{q \to 0} \left( q ~ \tilde{q}(p,p_0,q) \right) 
= \frac{\tilde{P}_{p+p_0} \tilde{P}_{p} -
\tilde{P}_{p_0}}{p_0 (p+p_0) \tilde{E}_{p+p_0}}
+ \frac{1}{p_0} \tilde{E}_{p}.
\end{eqnarray}

The corresponding formula for AA and RA are given 
in Appendix \ref{app:fullfinitesize}.

There is a case where the diffusion front
in a finite sample takes a particularly simple
form. This is when the starting point 
coincides with the reflecting boundary $x_0=0$.
The calculation of the Appendix simplifies
as one then has that $x=l_1$ where $l_1$ is the length of
the first bond. In the RR (or RA) case: 

\begin{eqnarray}
\overline{{\rm Prob}_{0 L}(x,t \vert x_0=0,0)} =
E^{+}_\Gamma(x) \phi^{R,A}_\Gamma(L-x) 
\end{eqnarray}
with
\begin{eqnarray}
\phi^R_\Gamma(x) & = & LT^{-1}_{p \to x} \frac{1}{p E^{+}_\Gamma(p)} \\
\phi^A_\Gamma(x) & = & LT^{-1}_{p \to x} 
\frac{P^{-}_\Gamma(p)}{p E^{-}_\Gamma(p)}
\end{eqnarray}

In the symmetric case one finds simply:

\begin{eqnarray}
&& \phi^R_\Gamma(x) = LT^{-1}_{p \to x} \frac{\Gamma \coth  
(\Gamma \sqrt p)}{\sqrt p}
=2 \sum_{n=0}^{\infty} e^{- n^2 \pi^2 \frac{x}{\Gamma^2} } \\
&& \phi^A_\Gamma(x) = LT^{-1}_{p \to x} \frac{\Gamma }{\sqrt p \sinh (\Gamma \sqrt p)}
 =2 \sum_{n=0}^{\infty} (-1)^n  e^{- n^2 \pi^2 \frac{x}{\Gamma^2} }.
\end{eqnarray}

\section{Results for Fokker Planck and associated Schr\"{o}dinger
operator}

\label{shrofokk}

It is also interesting to obtain results, via the RSRG, for the 
random Schr\"{o}dinger operator associated with the Sinai diffusion
problem. We first recall the connection between these
two problems. In this Section we set $T=1$.

\subsection{From Fokker Planck to Schr\"{o}dinger operator}

In a given environment $U(x)$ the probability distribution 
for the position of a particle $P(x,t|x_0,0)$
satisfies the Fokker-Planck equation (in the continuum):

\begin{eqnarray} 
&& \partial_t P(x,t|x_0,0)  = 
\partial_x (  \partial_x  + U'(x) ) P(x,t|x_0,0) ) =
- H_{FP} P(x,t|x_0,0)  \nonumber
\end{eqnarray}
with the initial condition $P(x,t \to 0|x_0,0) \to \delta(x-x_0)$.
As is well known, setting
\begin{equation}   \label{greens}
{\cal G}(x,t \vert x_0,0)= e^{(U(x)-U(x_0))/2 } P(x,t \vert x_0,0)
\end{equation}
one obtains the following imaginary time Schr\"{o}dinger equation 
for the Green function ${\cal G}(x,t \vert x_0,0)$
\begin{eqnarray} 
&& \partial_t {\cal G}(x,t \vert x_0,0) = 
\left(  \partial_x^2 + \frac{1}{2} U''(x) -\frac{1}{4} U'(x)^2 \right)
{\cal G}(x,t \vert x_0,0) = - H_{S} {\cal G}(x,t \vert x_0,0)
\end{eqnarray}
with the initial condition ${\cal G}(x,t \vert x_0,0)
\to \delta(x-x_0)$. This is the standard form for
the Schr\"{o}dinger operator $H_{S}$ associated with a diffusion
process. It can  be factorized as
$H_S=Q^{\dagger} Q$ with 
\[
Q=\partial_x + U'(x)/2
\]
and 
\[
Q^{\dagger}=- \partial_x + U'(x)/2,
\] and 
thus has a real positive spectrum $E_n$. 
The Fokker Planck operator $H_{FP}$ 
is non-hermitian but has the same
real positive spectrum, with 
right and left eigenfunctions
$\Phi_n^R(x)$ and $\Phi_n^L(x)$ 
associated with $E_n$. They are
related to the eigenfunctions 
$\psi_n(x)$ of the Schr\"{o}dinger operator
by $\Phi_n^R(x) = e^{- U(x)/2} \psi_n(x)$
and $\Phi_n^L(x) = e^{U(x)/2} \psi_n(x)$.

In the next two sections we use some of the
results obtained previously for the Sinai diffusion
process to obtain results for the Schr\"{o}dinger
and Fokker Planck operators.

\subsection{Averaged Green's function for the Schr\"{o}dinger operator}

\label{greenfunction}

Interestingly, one can obtain the averaged Green's function
of the Schr\"{o}dinger operator (\ref{greens}) 
from a slight variation of the previous 
calculation for the dynamics inside a well
of Section \ref{sub:inwell}. 
The physical reason is that
in Sinai's model the particle tends to jump to and
occupy lower accessible wells, with weight $e^{-U(x)}$
near the bottom. As a result one can show that,
due to the the exponential factor in (\ref{greens}),
the dominant contribution in the average over
disorder of (\ref{greens}) comes from rare 
configurations in which the point $x$
and the point $0$ are at about the same potential.
The calculation is sketched in  Appendix 
\ref{app:inwellbiased}. The result is
\begin{equation}
\overline {{\cal G}(x,0,t)}
={ 2 \over {\Gamma^5} } 
G \left( { x \over \Gamma^2}\right)
\label{greenf}
\end{equation}
with $\Gamma= \ln t$ with the
scaling function $G(X)$ given by (\ref{functionG}).

In the case of a small bias $\delta>0$ the result 
becomes:
\begin{equation}
\overline {{\cal G}(x>0,0,t)} = \overline {{\cal G}(x<0,0,t)} =
{2 \over \Gamma^5}  \left({ \Gamma \delta \over {\sinh (\Gamma \delta) }}\right)^2
e^{-\delta^2 |x|} G(|x|/\Gamma^2)
\label{biasedgreenf}
\end{equation}
which is valid in the usual scaling regime
$\Gamma \delta$ and $x/\Gamma^2$ fixed with $\Gamma= \ln t$
large. Note that this averaged Green's function Eq. (\ref{biasedgreenf}),
is closely related to the average Green's functions of a 
one-dimensional lattice fermion problem with random nearest
neighbor hopping, $t_n=\overline{t}+\delta t_n$ as 
computed recently by L. Balents and M.P.A. Fisher \cite{balents_fisher}.
In particular, the inverse Laplace transform of Eq. (\ref{biasedgreenf}),
$\overline{{\cal G}(x,x_0,E)}$ which is a function of the wave 
functions of ${ \cal H }_s$ at energy $E$, is equivalent to the 
Green's function of the Fermi problem at energy
$\epsilon=\sqrt E$ with $x-x_0\equiv n-n_0$ {\it even}, 
(corresponding to $\psi_R+\psi_L$ in terms of the right and 
left moving fermions of ref. \cite{balents_fisher}). This is 
related, in the Sinai problem, to the dominance of the averaged
Green's function by $x$ and $x_0$ both at bottoms of valleys
which correspond to even sites. The random hopping, $(-1)^n\delta t_n$ 
corresponds to $U'(x)/2$ in the Sinai problem.

By methods similar to those used in the present paper, one can 
obtain much more information about the statistical properties 
of eigenfunctions and Green's functions of both problems. These 
will be analyzed in ref. \cite{DSFWF}.

\section{conclusions}

\label{secconclusion}

In this paper we have developed a powerful real space
renormalization group (RSRG) method procedure for models of diffusion
in one-dimensional random potentials which belong
to the universality class of the Sinai model. This
method is simple to implement, physically 
transparent  and allows one to obtain
exact results for universal quantities.

The RSRG was first applied to recover, as a check of
its validity, the single time diffusion
front for the rescaled position $x(t)/(T \ln t)^2$
obtained previously  by Kesten and
Golosov \cite{kesten,golosov2}.
In addition we obtained the diffusion front in presence
of a small bias in the crossover region. 

The study of persistence properties, i.e., probabilities of
return to the origin  and their associated 
decay exponents, showed that in disordered systems
distinctions must be made between recurrence
properties of thermally averaged trajectories $\langle x(t)\rangle$ 
(exponent $\overline{\theta}$) and single particle trajectories
(exponent $\theta$). Nontrivial exponents (e.g.
$\overline{\theta}=(3-\sqrt{5})/4$) were obtained
for thermally averaged trajectories, a
novel and unexpected feature of the Sinai model.
The distribution of number of returns $k$ was found
to be strongly peaked in the rescaled variable, $g$,
at $g=k/\ln(T \ln t)=1/3$ but with  multifractal
tails characterized by an exponent
$\theta(g)$. It was shown that single-run
averages $\frac{1}{t} \int_0^t x(t') dt'$ obey the same scaling
for $g<1/3$, but
with  deviations on the larger than typical ($g>1/3$)
side of the distribution due to rare events which were
analyzed. We found  that at each return
to the origin, the thermally averaged trajectory loses
some memory of the past. This allowed us to compute exactly the
probability distribution of the complete sequence
of return times. By contrast the successive jumps
of $\langle x(t)\rangle$ exhibit persistent correlations which
we have studied in detail. Much of the analysis
was extended to the case of a small applied bias.

Single particle properties, such as return probabilities,
distributions of first passage times and of maximum displacement,
were obtained by studying the RSRG in the presence
of boundaries. The first meeting time distribution of two
independent particles was also obtained. Extensions
to large but finite size systems was studied, for
reflecting or absorbing boundary conditions. The distributions of
equilibration time and position, of the first passage
times in presence of boundaries and the
finite size diffusion front were all obtained.

A second set of results concerned aging dynamics.
The scaling form of the joint distribution of positions $x(t')$ 
and $x(t)$ at two times $t'<t$ was obtained explicitly in Laplace
transforms. This two-time diffusion front was
found to possess an overall $\ln t/\ln t'$ scaling,
with an interesting singularity at $x(t)=x(t')$.
Explicit expressions of several moments and correlation
functions were obtained. In the presence of a bias,
our single time diffusion results (the distribution 
of $x/t^\mu$ being related to a Levy distribution)
and our two-time aging results (with a $t/t'$ scaling) are consistent
with known exact results and with the phenomenological description
in terms of an effective directed model with an algebraic distribution
of waiting times. But in addition we have obtained the full crossover
between the symmetric and biased aging scaling forms.
Our aging results are also consistent with the numerical
simulations and qualitative arguments of \cite{laloux_pld_sinai}.

We have also obtained several quantities which
are controlled by rare events such as
renormalized valleys with degenerate minima 
or degenerate barriers. These can be studied systematically
as subdominant contributions in the RSRG. From them,
we computed the fluctuations in the thermal width of the single time
diffusion front (i.e., moments such as 
$\overline{\langle x^2(t)\rangle - \langle x(t)\rangle^2} 
\sim T (T \ln t)^3$), 
the two-time diffusion front in the quasi-equilibrium
regime (for $t-t' \sim t'^\alpha$, $\alpha<1$).

This work exhibits the relationships which exist
between the Sinai model and problems
such as quantum spin chains with disorder: both can be treated 
via very similar RSRG methods.  
Although observables of physical interest are often different
in each of these models, some interesting connections, 
have appeared---e.g., between 
persistence properties of Sinai model and magnetization in the 
random transverse field Ising model. The
RSRG methods enable one to consider this class of models in a unified
way. Since the method allows one to check its own range of 
validity, it may shed light on different universality classes.
The averaged imaginary time Green function of a related random 
Schr\"odinger problem was found as a side benefit. 

In conclusion,  the model studied here 
provides an all too rare  explicit example of a zero temperature
glassy fixed point where detailed non-equilibrium
quantities can be obtained. Qualitatively similar behavior should be expected
in systems where, as in Sinai's model, the dynamics consists
of jumps over large barriers between  partially equilibrated configurations.
The detailed understanding of physics  
in the simple one-dimensional
case studied here  perhaps encourages hope that new methods---exact
or approximate---based on similar ideas 
can be developed for more complex glassy systems.
As a start, we have already applied the methods introduced here to
more complex one-dimensional models, in particular the 
non-equilibrium dynamics of the classical random field
Ising model \cite{us_rfim}, as well as reaction diffusion models
with disorder \cite{us_rd}. Furthermore,
recent work on random quantum Ising models 
in two and three dimension \cite{RQI} suggest that in at least
some systems the type of behavior found here are not
particular to one dimension. 

\acknowledgments

One of us (DSF) acknowledges support from the National Science Foundation
via DMR9630064, DMS9304580, and Harvard University's MRSEC.

\appendix

\section{ Auxiliary variable RG rule, 
symmetric case}
\label{rulesymmetric}

In this Appendix we study the general RG rule 
$m' = a  m_1 + b m_2 + c m_3$ upon decimation of link 2
(see Fig. \ref{fig1}). We introduce the rescaled variable $\mu=m/{\Gamma^{\psi}}$
where $\psi$ is an unkown exponent 
and look for the fixed point joint probability distribution 
$P(\eta,\mu)$ that is a solution of
\begin{eqnarray}
&& 0=\left( (1+ \eta) \partial_\eta + 1 +
\psi \left(\mu \partial_\mu + 1 \right) \right) P(\eta,\mu) \nonumber \\
&& +\int_0^{\infty} d\mu_1 \int_0^{\infty} d\mu_2 
\int_0^{\infty} d\mu_3 P(0,\mu_2) P(.,\mu_1) *_{\eta} P(.,\mu_3)
\delta \left( \mu-(a\mu_1+b\mu_2+c\mu_3) \right)
\end{eqnarray}
We have of course $P(\eta)=\int_0^{\infty} d\mu  P(\eta,\mu)=e^{-\eta}$.
The equation for the first moment $C(\eta)={1 \over P(\eta) } 
\int_0^{\infty} d\mu \mu   P(\eta,\mu)$ reads
\begin{eqnarray}
0=(1+\eta)\partial_\eta C(\eta) - (\eta + \psi) C(\eta)
+ (a + c) \int_0^\eta d\eta_1 C(\eta_1) + b C(0) \eta
\end{eqnarray}
It is useful to differentiate this equation with respect to $\eta$
to obtain
\begin{eqnarray}
0=(1+\eta)\partial^2_\eta C(\eta) + (1- \eta - \psi) \partial_\eta C(\eta)
+ (a + c - 1 ) C(\eta) + b C(0) 
\end{eqnarray}
and to keep the boundary condition 
 $C'(0)=C(0) \psi$ at $\eta=0$.

For $a+c-1 \neq 0$ (the case for all the physical quantities discussed
in this paper), it is convenient to set
$y=1+\eta$ and $T(y)=C(\eta)+\frac{b}{a+c -1} C(0)$
so that that $T(y)$ satisfies the confluent hypergeometric
equation 
\begin{eqnarray}
0=y \partial_y^2 T +(B-y) \partial_y T - A T(y)
\end{eqnarray}
where $B=2-\psi$ and $A=1-a-c$, together with the boundary condition
$T'(y=1)=\psi {A \over {A-b}} T(y=1)$.
Since we are looking for a well-behaved (i.e., not exponentially growing)
solution at $\eta=\infty$, we find that $T(y)$ has to be proportional
to the confluent hypergeometric function
$U(A,B,y)$. To satisfy the boundary condition at $y=1$,
we obtain, using the functional relation $U'(A,B,1)=U(A,B,1)-U(A,B+1,1)$,
that the exponent $\psi$ has to satisfy the equation
\begin{eqnarray}
0=\left(1-\psi { {a+c-1} \over {a+b+c-1}} \right) U(1-a-c,2-\psi,1)-U(1-a-c,3-\psi,1)
\label{hypergeo}
\end{eqnarray}
Note that in the particular case where $A=1-a-c=-1$,
the function $U(-1,B,y)$ reduces to the linear function $-B+y$,
and the equation for $\psi$ is simply quadratic
$\psi(\psi-1)=1+b$ yielding $\psi=(1+\sqrt{5+4b})/2$
as in \cite{danfisher_rg1}.

In the text we use the ratio $m/\overline{l}_{\Gamma}$
which decays as $m/\overline{l}_{\Gamma} \sim \Gamma^{-\Phi}$
with $\Phi = 2 - \psi$. Both exponents $\psi$ and $\Phi$
depend explicitly on the coefficients $a,b,c$,  of the RG rule.

\hbox{}

\section{Auxiliary variable RG rule, 
biased case}
\label{rulesbiased}

We consider the auxiliary variables $m^{\pm}$ that evolve with
the RG rules 
$m^{+} = a^{+}  m_1^{+} + b^{+} m_2^{-} + c^{+} m_3^{+}$ 
upon decimation of an ascending link 2 and
$m^{-} = a^{-}  m_1^{-} + b^{-} m_2^{+} + c^{-} m_3^{-}$ 
upon decimation of a descending link 2.
We introduce the joint probability distributions $P^{\pm}_{\Gamma}(\zeta,m)$ 
that evolve as
\begin{eqnarray}
(\partial_{\Gamma} - \partial_\zeta ) P^{\pm}_{\Gamma}(\zeta,m) & = &
P^{\pm}_{\Gamma}(\zeta,m) \int_0^{\infty} dm_2 
( P^{\pm}_{\Gamma}(0,m_2) - P^{\mp}_{\Gamma}(0,m_2)  ) \nonumber \\
& + & \int_0^{\infty} dm_1 \int_0^{\infty} dm_2 
\int_0^{\infty} dm_3 P^{\mp}_{\Gamma}(0,m_2) 
\nonumber \\ 
& &  P^{\pm}_{\Gamma}(.,m_1)*_{\zeta} P^{\pm}_{\Gamma}(.,m_3)
\delta \left( m-(a^{\pm}  m_1 + b^{\pm} m_2 + c^{\pm} m_3) \right).
\end{eqnarray}
We have  $P^{\pm}_{\Gamma}(\zeta)=\int_0^{\infty} dm 
P^{\pm}_{\Gamma}(\zeta,m) = u^{\pm}_{\Gamma} e^{-z u^{\pm}_{\Gamma} }$.
The equation for the first moments $C^{\pm}_{\Gamma}(\zeta)={1 \over P^{\pm}_{\Gamma}(\zeta) } \int_0^{\infty} dm \ m   P^{\pm}_{\Gamma}(\zeta,m)$
is
\begin{eqnarray}
 (\partial_{\Gamma} - \partial_\zeta ) C^{\pm}_{\Gamma}(\zeta)=
u^{+}_{\Gamma} u^{-}_{\Gamma} \left[ 
z ( b^{\pm} C^{\mp}_{\Gamma}(0) - C^{\pm}_{\Gamma}(\zeta) )  + 
(a^{\pm}+c^{\pm}) \int_0^z d\zeta' C^{\pm}_{\Gamma}(\zeta') \right] 
\end{eqnarray}
with $u^{+}_{\Gamma} u^{-}_{\Gamma}={ {\delta^2} \over {\sinh^2(\Gamma \delta)}}=1/\overline{l}_\Gamma$

We study the  simpler particular
case when $a^{+}+c^{+}=2$ , $a^{-}+c^{-}=2$ , $b^{+}=b^-=b$.
Then the solutions $C^{+}_{\Gamma}(\zeta)=C^{-}_{\Gamma}(\zeta)=C_{\Gamma}(\zeta)$
are simply linear in $\zeta$ :  $C_{\Gamma}(\zeta)=A_{\Gamma}
+ \zeta B_{\Gamma}$ and the coefficients satisfy
\begin{eqnarray}
&& B_{\Gamma}=\partial_{\Gamma} A_{\Gamma}  \\
&& \partial^2_{\Gamma} A_{\Gamma}= (1+b) u^{+}_{\Gamma} u^{-}_{\Gamma} A_{\Gamma} 
= (1+b) { {\delta^2} \over {\sinh^2(\Gamma \delta)}} A_{\Gamma}
\label{agamma}
\end{eqnarray}

For $\delta=0$ we have already seen in Appendix \ref{rulesymmetric} that 
the auxiliary variable $m$ grows as $\Gamma^{\psi(b)}$ with
$\psi(b)=(1+\sqrt{5+4b})/2$. Indeed for $\delta=0$,
$A_{\Gamma} \propto \Gamma^{\psi(b)}$ is a solution of (\ref{agamma}).
For $\delta >0$ following \cite{danfisher_rg2}
we thus look for a solution of the scaling form:
\begin{eqnarray}
A_{\Gamma}=\delta^{-\psi(b)} A^{(b)}(\gamma=\delta \Gamma)
\end{eqnarray}
where $A^{(b)}(\gamma)$ satisfies the equation:
\begin{eqnarray}
\partial_{\gamma}^2 A^{(b)}(\gamma)=
{ {1+b} \over {\sinh^2(\gamma)}} A^{(b)}(\gamma)
\end{eqnarray}
with the boundary condition $A^{(b)} (\gamma) \propto \gamma^\psi(b)$ 
as $\gamma \to 0$. Introducing the new variable $y=\coth \gamma$,
we obtain the differential equation for the Legendre functions:
\begin{eqnarray}
(y^2-1) { {d^2 A^{(b)}} \over {dy^2}}+2y { {d A^{(b)} } \over {dy}}
-(1+b) A^{(b)}=0.
\label{legendreq}
\end{eqnarray}
The solution for $A^{(b)}(\gamma)$ with the above boundary condition is:
\begin{eqnarray}
A^{(b)}(\gamma) & = & K Q_{\psi(b)-1}(\coth \gamma)  \nonumber \\
& = & K' \tanh(\gamma)^{\psi(b)} 
F\left( 
\frac{\psi(b)+1}{2} , \frac{\psi(b)}{2}, \psi(b)+\frac{1}{2} , 
\tanh(\gamma)^2 \right) 
\end{eqnarray}
with $K' = K \sqrt{\pi} 2^{-\psi(b)} \frac{\Gamma(\psi(b))}{\Gamma(\psi(b)+1/2)}$
where $K$ is a non-universal constant.
The asymptotic behaviors are
  $A^{(b)}(\gamma) \simeq K' \gamma^{\psi(b)}$ at small $\gamma$,
and $A^{(b)}(\gamma) = 2 K' \frac{\Gamma[\psi(b) + 1/2] 
}{\Gamma[(\psi(b) + 1)/2] \Gamma[\psi(b)/2]} \gamma$ at large
$\gamma$. We can now compute the mean values of the variables $m^{\pm}$
\begin{eqnarray}
&& \langle m^{\pm}\rangle=\int_0^{\infty} d\zeta \int_0^{\infty} dm\; m P^{\pm}_{\Gamma}(\zeta,m)
\int_0^{\infty} d\zeta  P^{\pm}_{\Gamma}(\zeta)  C^{\pm}_{\Gamma}(\zeta) 
= A_{\Gamma}+ { { \partial_\Gamma A_{\Gamma} } \over {u^{\pm}_{\Gamma}}}
\end{eqnarray}
yielding (\ref{mz}) in the text.

\section{Correlation of times and directions
of successive jumps}

\label{correlations}

\subsection{Conditional probabilities of times of jumps forward and
backward}

\label{conditional}

In this Appendix we compute the
conditional probabilities $\rho^{(f)}_{\Gamma \Gamma'}$ to make a jump forward
at $\Gamma$ (respectively a jump backward) given that the
last jump occurred at $\Gamma'$.
We  define $D_{\Gamma,\Gamma'}(F)$ as the probability to be 
on a descending bond of barrier $F$ given that the last jump
of the effective dynamics occurred at $\Gamma'$ (this jump was
necessarily in the (+) direction since the walker is on a descending bond).
The initial condition is thus given by:
\begin{eqnarray}
&& D_{\Gamma',\Gamma'}(F) = K \int_{\Gamma}^{\infty} dF_1 \int_{0}^{\infty} dl_1
\int_{0}^{\infty} dl_2 \int_{\Gamma}^{\infty} 
dF_3 \int_{0}^{\infty} dl_3 (l_1+l_2) \nonumber \\
&& P_{\Gamma}(F_1,l_1) 
 P_{\Gamma}(\Gamma,l_2) 
 P_{\Gamma}(F_3,l_3) \delta 
\left(F-(F_1+F_3-\Gamma) \right)  \label{condinit}
\end{eqnarray}
Indeed, the bond must be  a new bond made, at 
$\Gamma'$,  out of three bonds,
and the origin of the random walk must have been on either the
first or the second bond in order to satisfy the condition
that the last jump  occurred at scale $\Gamma'$.
The normalization constant $K$ has to be choosen to ensure that
$\int_{\Gamma}^{\infty} D_{\Gamma',\Gamma'}(F) =1$.
Introducing the rescaled variables $\eta=\frac{F-\Gamma}{\Gamma}$
and $\alpha=\frac{\Gamma}{\Gamma'}$, we obtain that
$D_{\alpha}(\eta)$ evolves according to
\begin{eqnarray}
\left(\alpha \partial_\alpha - (1+ \eta) \partial_\eta - 1 \right)
D_{\alpha}(\eta) = - 2  D_{\alpha}(\eta) + 
\int_0^{\eta} d\eta' D_{\alpha}(\eta') e^{-(\eta-\eta')}
\end{eqnarray}
with the initial condition at $\alpha=1$ given from (\ref{condinit})
by $D_{\alpha=1}(\eta)=\left( \frac{\eta}{2}+\frac{\eta^2}{4} \right) e^{-\eta}$.
The solution reads
\begin{eqnarray}
D_{\alpha}(\eta) & = & A_{\alpha} e^{-\eta} +
(B_{\alpha}+C_{\alpha} \eta) e^{-\alpha \eta} \\
A_{\alpha} & = & \frac{1}{2 \alpha^2} \left[ 5+ \left( \frac{\alpha^2+2 \alpha-2)}{(\alpha-1)^2}\right)
 e^{-(\alpha-1)} \right]  \\
B_{\alpha} & = & -\frac{1}{2} \left( \frac{\alpha}{\alpha-1}\right)^2 
e^{-(\alpha-1)} \\
C_{\alpha} & = & -\frac{1}{2} \left( \frac{\alpha}{\alpha-1}\right)
 e^{-(\alpha-1)}
\label{solucond}
\end{eqnarray}
The probability to make a forward jump at $\Gamma$ (i.e., in the (+) direction)
given that the last jump occured at $\Gamma'$ (and by convention
was in the (+) direction) is
\begin{eqnarray}
\rho_{\Gamma,\Gamma'}^{(f)}=P_{\Gamma}(\Gamma) \int_{\Gamma}^{\infty} dF 
D_{\Gamma,\Gamma'}(F) =\frac{1}{\Gamma} 
\int_0^{\infty} d\eta D_{\alpha}(\eta) 
\label{bienexplique}
\end{eqnarray}
since the probability that the neighboring bond
is decimated at $\Gamma$ is $P_{\Gamma}(\Gamma)=1/\Gamma$.
On the other hand the probability to make a backward jump at $\Gamma$ (i.e.,
in the (-) direction)
given that the last jump occurred at $\Gamma'$ is
\begin{eqnarray}
\rho_{\Gamma,\Gamma'}^{(b)}= 
D_{\Gamma,\Gamma'}(\Gamma) =\frac{1}{\Gamma} D_{\alpha}(0)
\end{eqnarray}
which is the probability to decimate the bond
we are interested in. 
The total probability to jump at $\Gamma$ in any direction 
given that the last jump occurred at $\Gamma'$ must satisfy
$\rho_{\Gamma,\Gamma'}= \rho_{\Gamma,\Gamma'}^{(f)} +\rho_{\Gamma,\Gamma'}^{(b)}
=- \partial_{\Gamma} \int_{\Gamma}^{\infty} dF D_{\Gamma,\Gamma'}(F)$.
These expressions, after substituting the above solution (\ref{solucond})
yield the formulae (\ref{rho}) given in the text.

\hbox{}

\subsection{\bf Correlations in the sequence of times
of successive forward and backward jumps}

\label{correlations2}

A full calculation of all terms is
quite involved and goes beyond the present paper.
Here we indicate only the result for the
two first elementary building blocks for the 
many jump correlations. The first one is

\begin{eqnarray}
\rho_0^{bb}(\Gamma_1|\Gamma_0) d\Gamma_1 =
\rho_0^{bb}(\alpha_1) d \alpha_1 = 
\frac{d \alpha_1}{ \alpha_1^3} \left(2 - 
(1 + \alpha_1) e^{-(\alpha_1-1)} \right),
\label{rhobb}  
\end{eqnarray}
which is a scaling function of $\alpha_1 = \Gamma_1/\Gamma_0$.
Intermediate calculations also yield the probability that the
second jump occurs at $\Gamma_1$ and is a forward jump
given that the first one occurs at $\Gamma_0$ and is backward.
\begin{eqnarray}
\rho_0^{fb}(\Gamma_1|\Gamma_0) d\Gamma_1 =
\rho_0^{fb}(\alpha_1) d \alpha_1 = 
\frac{d \alpha_1}{ \alpha_1^3} \left(2- e^{-(\alpha_1-1)} \right).  
\label{rhofb}  
\end{eqnarray}

The second elementary building block is given by:

\begin{eqnarray}
&& \rho_1^{bfb}(\Gamma_2,\Gamma_1|\Gamma_0) d\Gamma_2 d\Gamma_1 =
\rho_1^{bfb}(\alpha_1,\alpha_2) d \alpha_1 d \alpha_2 = 
\frac{d \alpha_1 d \alpha_2 \left(2- e^{-(\alpha_1-1)} \right)}
{ \alpha_1^3 \alpha_2^3 (2- e^{-(\alpha_1-1)})} \nonumber \\
&& \left(4- e^{-(\alpha_1-1)} - 2 (\alpha_2+1) e^{-(\alpha_2-1)} 
- (\alpha_2+1) { {e^{-(\alpha_2+\alpha_1-2)} } \over {\alpha_1-1}} 
+(\alpha_2\alpha_1+1) { {e^{-(\alpha_2\alpha_1-1)} } 
\over {\alpha_1-1}} \right) 
\label{rhobfb}  
\end{eqnarray}
which is a scaling function of $\alpha_1=\Gamma_1/\Gamma_0$ and
$\alpha_2=\Gamma_2/\Gamma_1$.

\section{distribution of sequences of returns to the origin: biased case}

\label{distribiased}

To compute the conditional probabilities $\rho^{\pm}(\Gamma , \Gamma')$
of returns to the origin defined in the text (Section \ref{sequencebias})
we consider the probability $D^{\pm}_{\Gamma,\Gamma'}(\zeta)$
that a bond has barrier $z$ at $\Gamma$ 
and has not changed  orientation since the scale $\Gamma'$
where it was created. Its RG equation is
\begin{eqnarray}
\left( \partial_\Gamma -  \partial_\zeta   \right)
D^{\pm}_{\Gamma,\Gamma'}(\zeta)
 = 2 P^{\mp}_{\Gamma}(0) D^{\pm}_{\Gamma,
\Gamma'}(.) *_\zeta P^{\pm}_{\Gamma}(.) 
- 2 P^{\mp}_{\Gamma}(0) D^{\pm}_{\Gamma,\Gamma'}(\zeta)
\end{eqnarray}
with the initial condition $D^{\pm}_{\Gamma',\Gamma'}(\zeta)= 
\left(u^{\pm}_{\Gamma'} \right)^2 \zeta e^{- \zeta u^{\pm}_{\Gamma'}  }$.
Since $P^{\pm}_{\Gamma}(\zeta)= u^{\pm}_{\Gamma}  e^{- \zeta u^{\pm}_{\Gamma}  }$,
the solution has the following form
\begin{eqnarray}
D^{\pm}_{\Gamma',\Gamma'}(\zeta)= \left( A_{\Gamma,\Gamma'} 
+ \zeta B_{\Gamma,\Gamma'}  \right) {
 {u^{\mp}_{\Gamma} }
 \over {u^{\mp}_{\Gamma'} }}
\left(u^{\pm}_{\Gamma} \right)^2 e^{- \zeta u^{\pm}_{\Gamma}  }
\end{eqnarray}
where the coefficients are
\begin{eqnarray}
&& A_{\Gamma,\Gamma'}= \delta^{-1} \left( Q_{\phi-1}(\coth \gamma)
P_{\phi-1}(\coth \gamma')-Q_{\phi-1}(\coth \gamma')
P_{\phi-1}(\coth \gamma) \right) \\
&& B_{\Gamma,\Gamma'}= { 1 \over {\sinh^2 \gamma}}
\left( P'_{\phi-1}(\coth \gamma)
Q_{\phi-1}(\coth \gamma')-P_{\phi-1}(\coth \gamma')
Q'_{\phi-1}(\coth \gamma) \right) 
\end{eqnarray}
where $\gamma=\delta \Gamma$, $\gamma'=\delta \Gamma'$, 
$\phi=(1+\sqrt 5)/2$ and $Q_{\nu}(y)$ and $P_{\nu}(y)$ are associated Legendre
functions: they are two linearly independent solutions of the equation
(\ref{legendreq}) (with $(1+b) \to \nu (1+\nu)$).

The probability for a $(\pm)$ link to be decimated at $\Gamma$  
given that its last decimation occurred at $\Gamma'$
is therefore
\begin{eqnarray}
&& \rho^{\pm}(\Gamma,\Gamma')=- \partial_\Gamma 
\int_0^{\infty} d\zeta D^{\pm}_{\Gamma,\Gamma'}(\zeta)
 =  \frac { u^{+}_{\Gamma} u^{-}_{\Gamma} }
 {u^{\mp}_{\Gamma'} } u^{\pm}_{\Gamma} A_{\Gamma,\Gamma'} 
\end{eqnarray}
This leads to the equation (\ref{pq}) given in the text in
terms of the the reduced variables $y=\coth \gamma$ and 
$y'=\coth \gamma'$.

\section{dynamics within a well}
\label{app:inwellbiased}

\subsection{Probability that a bond has degenerate minima}
\label{degeneratebiased}

\begin{figure}[thb]  
\centerline{\fig{8cm}{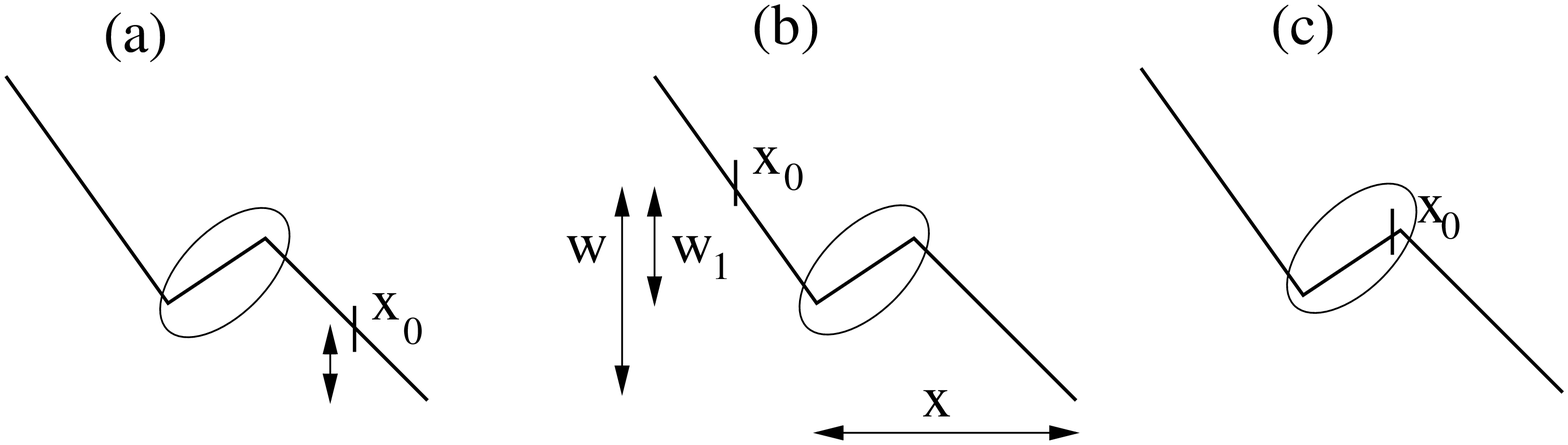}}
\caption{
\label{rgvalley}
\narrowtext Different terms which contribute to
the RG equation for $S$.}
\end{figure}

Let us introduce the  probability 
$S_{\Gamma}(\zeta,l,x,w)$ that: 
a given point (denoted $x_0$ in 
Fig. \ref{rgvalley})
belongs at $\Gamma$ to a bond of barrier
$F=\Gamma + \zeta$, of length $l$,
is at a distance $x$ from the min of the bond and is at a potential $w$
above the potential of the minimum of the bond. 
One has that by definition $0<x<l$ and its normalization
with respect to $x$ and $w$ is
$\int_0^l dx \int_0^{\Gamma + \zeta} dw  S_{\Gamma}(\zeta,l,x,w) =
l P_{\Gamma}(\zeta,l)/\int_l P_{\Gamma}(l)$, which is the
probability that a given point belongs to a bond with $F,l$.
The RG equation for $S_{\Gamma}(\zeta,l,x,w)$ reads:
\begin{eqnarray}
&& (\partial_\Gamma - \partial_\zeta) S_{\Gamma}(\zeta,l,x,w)
=-2 P_\Gamma(0) S_{\Gamma}(\zeta,l,x,w) 
+ P_\Gamma(0,.)*_l P_\Gamma(.,.)*_{\zeta,l} S_{\Gamma}(.,.,x,w) \nonumber \\
&& + \int_{\zeta_1,l_1,x_1,w_1,l_2,\zeta_3,l_3} 
S_{\Gamma}(\zeta_1,l_1,x_1,w_1)  P_{\Gamma}(0,l_2) P_
{\Gamma}(\zeta_3,l_3) \nonumber \\
&& \delta(\zeta-(\zeta_1+\zeta_3)) \delta(l-(l_1+l_2+l_3))
\delta(x-(x_1+l_2+l_3)) \delta(w-(w_1+\zeta_3)) \nonumber \\
&& + \int_{\zeta_1,l_1,x_2,w_2,l_2,\zeta_3,l_3}
P_{\Gamma}(\zeta_1,l_1) S_{\Gamma}(0,l_2,x_2,w_2) P_{\Gamma}(\zeta_3,l_3) 
\nonumber \\
&& \delta(\zeta-(\zeta_1+\zeta_3)) \delta(l-(l_1+l_2+l_3)) 
\delta(x-(l_1+l_2-x_2)) \delta(w-(w_2+\zeta_1)) \label{initialrg}
\end{eqnarray}
where each term is described in 
Fig. (\ref{rgvalley})
We notice that the 
evolution equation for $S_{\Gamma}(\zeta,l,x,w=0)$ decouples
and leads to the form:
\begin{equation}  \label{decomposition}
S_{\Gamma}(\zeta,l,x,w=0)= \frac{2}{\Gamma^2} 
( \delta(x) P_{\Gamma}(\zeta,l) + R_{\Gamma}(\zeta,l,x) )
\end{equation}
where the delta-function part represents the probability that the point
$x_0$ happens to be exactly at 
the bottom of the renormalized bond in which case $w=0$ by definition.
The function $R_\Gamma(\zeta,l,x)$ is the probability that a renormalized bond at scale
$\Gamma$ has $(\zeta,l)$ and a distinct degenerate minimum at a finite distance $x$.
For small $x$ ($x = O(1)$) this function is non-universal. We  compute
this function in the scaling regime $x \sim \Gamma^2$ where it
is universal and of order $1/\Gamma$. We use the rescaled variables
$\eta=\zeta/\Gamma$, $\lambda=l/\Gamma^2$ and $X=x/\Gamma^2$
such that $R_\Gamma(\zeta,l,x) = \Gamma^{-6} R(\eta,\lambda,X)$
and obtain the following fixed point RG equation
for $R$:
\begin{eqnarray} 
&& 0 = \Gamma \partial_\Gamma  R_\Gamma(\eta,\lambda,X) =
((1+ \eta) \partial_\eta + 2 \lambda \partial_\lambda
+ 2 X \partial_{X} + 6 )  
R(\eta,\lambda,X)  \nonumber \\
&&
+ P(0,.)*_{\lambda} P(.,.)*_{\eta,\lambda} R(.,.,X) 
+2 P(\eta,\lambda-X) (P(0,.)*_{\lambda = X} P(0,.)) 
\label{rescrg}
\end{eqnarray}
This equation was obtained by substituting the decomposition
(\ref{decomposition}) into
(\ref{initialrg}) in the spirit of an expansion
in powers of $1/\Gamma$, where the zeroth-order equation is
satisfied by the delta-function part. The
order $O(1/\Gamma)$ equation yields the equation
for $R_\Gamma(\zeta,l,x)$ where the delta-function part acts now
as a source in the last two terms of (\ref{initialrg})
leading to the last term in (\ref{rescrg}). This term
describes the probability that between $\Gamma$ and $\Gamma+d\Gamma$
a new bond with one degenerate
minimum (a distance of order $\Gamma^2$ away from the lowest edge)
is created via the decimation of a bond whose neighbor also has 
$\zeta \sim 0$---cases (b) and (c) in Fig. \ref{rgvalley}
with $w=w_1=0$)---the probability of this is of order $1/\Gamma$.

Before proceeding further, we notice that it is easy to also
keep track of the barrier $\Gamma_0$ between the
two degenerate minima. We define 
$R_\Gamma(\zeta,l,x,\Gamma_0)$ as the 
probability that a renormalized bond at scale
$\Gamma$ has $(\zeta,l)$ and 
a distinct degenerate minimum at a finite distance $x$
separated from the minimum by a barrier $\Gamma_0$. It takes the scaling form
$R_\Gamma(\zeta,l,x,\Gamma_0) = \Gamma^{-7} R_\Gamma(\eta,\lambda,X,u)$
with $u=\Gamma_0/\Gamma$. The normalization is
$R_{\Gamma}(\eta,\lambda,X)
= \int_0^{1} du R_{\Gamma}(\eta,\lambda,X,u)$. The scaling form
satisfies the fixed point RG equation:
\begin{eqnarray}
&& 0 = \Gamma \partial_\Gamma  R(\eta,\lambda,X,u)
= ((1+ \eta) \partial_\eta + 2 \lambda \partial_\lambda
+ 2 X \partial_{X} + u \partial_u + 7 ) 
R(\eta,\lambda,X,u) \nonumber \\
&& + P(0,.)*_{\lambda} P(.,.)*_{\eta,\lambda} R(.,.,X,u) 
+2 P(\eta,\lambda-X) (P(0,.)*_{\lambda = X} P(0,.))
\delta(u-1) 
\end{eqnarray}
where the last term corresponds to barriers $\Gamma_0=\Gamma$
created upon decimation. 

Remarkably, one can find the complete solution
of this equation in a factorized form:
\begin{eqnarray}   
R(\eta,\lambda,X,u) = P(\eta, \lambda - X) r(X,u)
\label{decoupled}
\end{eqnarray}
where $r(X,u)$ satisfies
\begin{eqnarray}
&& 0 = ( 2 X \partial_{X} + u \partial_u + 4 ) 
r(X,u)
+ 2 (P(0,.)*_{\lambda = X} P(0,.))
\delta(u-1)
\end{eqnarray}
whose solution is:
\begin{eqnarray}
&& r(X,u) = \theta(1-u) u^{-4} \hat{r}(X/u^2) \\
&& \hat{r}(X) = 2 (P(0,.)*_{\lambda = X} P(0,.))
\label{soludecoupled2}
\end{eqnarray}
which, using (\ref{solu}),
yields formula (\ref{soludecoupled}) given in the text.

The biased case can be studied similarly. The corresponding
quantities (as usual $\pm$ designate descending and ascending bonds,
respectively) also satisfy 
$S_{\Gamma}^{\pm}(\zeta,l,x,w=0)= \frac{1}{\overline{l}_\Gamma} 
( \delta(x) P_{\Gamma}^{\pm} + R_{\Gamma}^{\pm}(\zeta,l,x) )$
and one finds that the RG equation for 
$R^{\pm}_\Gamma(\zeta,l,x,\Gamma_0)$ is
\begin{eqnarray}
&&  (\partial_\Gamma - \partial_\zeta) R_{\Gamma}^{\pm}(\zeta,l,x,\Gamma_0)
=  (P^{\pm}_\Gamma(0) - P^{\mp}_\Gamma(0) ) 
R_{\Gamma}^{\pm}(F,l,x,\Gamma_0) \nonumber \\
&& + P_{\Gamma}^{\mp}(0,.)*_l P_{\Gamma}^{\pm}(.,.)*_{\zeta,l} R_{\Gamma}^{\pm}(.,.,x,\Gamma_0) 
+ 2 P_\Gamma^{\pm}(\zeta,l-x) 
P_\Gamma^{\mp}(0,.)*_x P_{\Gamma}^{\pm}(0,.) \delta(\Gamma-\Gamma_0) 
\end{eqnarray}
The solution again factorizes into
\begin{eqnarray}
\label{decoupledbiased}
 R_{\Gamma}^{\pm}(\zeta,l,x,\Gamma_0)
= P^{\pm}_{\Gamma}(\zeta,l-x) r_{\Gamma}(x,\Gamma_0)  
\end{eqnarray}
where $r_{\Gamma}(x,\Gamma_0)=2 \theta(\Gamma-\Gamma_0)  
P_{\Gamma_0}^{+}(0,.)*_x P_{\Gamma_0}^{-}(0,.)$ does not depend on the direction
of the bias. Its Laplace transform is simply
\begin{eqnarray}
\int_0^{\infty} dx e^{-px} r_{\Gamma} (x,\Gamma_0)
= \theta(\Gamma-\Gamma_0) 2 U_{\Gamma_0}^{+}(p) U_{\Gamma_0}^{-}(p)
= 2 \theta(\Gamma-\Gamma_0) \frac{ p+\delta^2 }{ \sinh^2(\Gamma_0 \sqrt{p+\delta^2})}
\end{eqnarray}
so that finally
\begin{eqnarray}
 r_{\Gamma} (x,\Gamma_0)
= \theta(\Gamma-\Gamma_0) \frac {1}{\Gamma_0^4} 
\hat{r} \left(\frac{x}{\Gamma_0^2}\right) e^{- x \delta^2}
\label{functionrbiased}
\end{eqnarray}
where $\hat{r}$ is the function for the symmetric case
introduced in (\ref{soludecoupled}).

\subsection{Relationship to  the
associated Schr\"odinger operator Green's function}
\label{greenbiased}

The disorder averaged Green function defined in
Section \ref{greenfunction}
is exactly related to 
the probability $S_{\Gamma}(\zeta,l,x,w)$ introduced 
above. In the symmetric case, one
can restrict to $x>0$, and one has:
\begin{equation}
\overline {{\cal G}(x,0,t)} 
= {1 \over 2} \int_{0}^{\infty} d\zeta \int_0^{\zeta+\Gamma} dw e^{-w/2} 
\int_{x }^{\infty} dl S_{\Gamma}(\zeta,l,x,w)
\end{equation}
The factor ${1 \over 2}$ is simply the probability to be on a descending
bond ($x>0$). This can be expressed using the rescaled variables 
$\zeta=\Gamma \eta$ , $w=\Gamma u$ ,
$l=\Gamma^2 \lambda$ , $x=\Gamma^2 X$ and simplified
using that for large $\Gamma$, we may replace 
$e^{-\Gamma u/2}$ by ${2 \over \Gamma} \delta(u)$.
Using (\ref{decomposition}) one obtains:
\begin{equation}
\overline {{\cal G}(x,0,t)} 
= {2 \over \Gamma^5 } \int_{0}^{\infty} d\eta  
\int_{X}^{\infty} d\lambda ~~ R(\eta,\lambda,X).
\end{equation}
Using (\ref{decoupled}) one finds the result of the text
(\ref{greenf}). 

In the biased case we obtain an expression 
for the averaged Green function 
in terms of the functions
$S_{\Gamma}^{\pm} (\zeta,l,x,0)$

\begin{eqnarray}
 \overline {{\cal G}(x>0,0,t)}  = 2 \int_{0}^{\infty} d\zeta   
\int_{\vert x \vert}^{\infty} dl S_{\Gamma}^+ (\zeta,l, \vert x \vert,0)
\end{eqnarray}

\begin{eqnarray}
 \overline {{\cal G}(x<0,0,t)}= 2 \int_{0}^{\infty} d\zeta  
\int_{\vert x \vert}^{\infty} dl S_{\Gamma}^-(F,l, \vert x \vert,0)
\end{eqnarray}

Using (\ref{decoupledbiased}) we finally get
$\overline {{\cal G}(x>0,0,t)}  =  \overline {{\cal G}(x<0,0,t)} = 
{\cal G}( \vert x \vert ,t)$ with
\begin{eqnarray}
{\cal  G}(|x|,t) = \frac{2}{\overline{l}_{\Gamma}} \int_0^{\Gamma} d \Gamma_0
r_{\Gamma} (|x| ,\Gamma_0)
  = \frac{2}{\overline{l}_{\Gamma}} \frac{1}{\Gamma^3} 
G\left(\frac{|x|}{{\Gamma}^2}\right)  e^{- |x| \delta^2}
\end{eqnarray}
where the function $G$ has been introduced in (\ref{functionG}),
leading to formula (\ref{biasedgreenf}) in the text. 

\section {Solution of the two time RG equations}
\label{app:twotime}
 
In this Appendix we solve explicitly the RG equations (\ref{rgomega})
for the quantities 

$\Omega^{\epsilon \epsilon'}_{\Gamma,\Gamma'} (\zeta,x_L,x_R ; x_L',x_R')$,
$\epsilon = \pm 1$ $\epsilon' = \pm 1$ from which we can obtain
the two-time diffusion front $\overline{{\rm Prob}(xt,x'
t'|00)}$. We consider the general
biased case and discuss the particular limit of the symmetric case.

We first introduce the Laplace transforms:
\begin{eqnarray}
&& {\hat \Omega}^{\epsilon \epsilon'}_{\Gamma,\Gamma'} \left(\zeta, \mu, \nu ; \mu', \nu' \right)
\nonumber \\
&& =\int_{x_L' >0, x_R' >0, x_L>0, x_R >0} 
e^{-\mu'x_R'} e^{-\nu'x_L'} e^{-\mu(x_R-x_R')}
e^{-\nu(x_L-x_L')} 
\Omega^{\epsilon \epsilon'}_{\Gamma,\Gamma'} (\zeta,x_L,x_R ; x_L',x_R') 
\end{eqnarray}
Since we consider large $\Gamma'$ we can use the
fixed point solution (\ref{solu-biased}):
\begin{eqnarray}
P^{\pm}_{\Gamma}(\zeta,\mu)=\int_0^{\infty} dl e^{- \mu l} P^{\pm}_{\Gamma}(\zeta,l)
=U^{\pm}_{\Gamma}(\mu) e^{-\zeta u^{\pm}_{\Gamma}(\mu)}
\end{eqnarray}
The RG equations (\ref{rgomega}) can be then written in Laplace variables as:
\begin{eqnarray}
&& \left(\partial_{\Gamma}-\partial_\zeta \right)
{\hat \Omega}^{\pm \epsilon'}_{\Gamma,\Gamma'} \left(\zeta, \mu, \nu ; \mu', \nu' \right)
= -2 U^{\mp}_{\Gamma}(0) 
\Omega^{\pm \epsilon'}_{\Gamma,\Gamma'} 
\left(\zeta, \mu, \nu ; \mu', \nu' \right) \nonumber \\
&& + U^{+}_{\Gamma}(\mu) U^{-}_{\Gamma}(\mu) \int_0^{\infty} d\zeta_1 
e^{-(\zeta-\zeta_1) u^{\pm}_{\Gamma}(\mu)}
{\hat \Omega}^{\pm \epsilon'}_{\Gamma,\Gamma'} \left(\zeta_1, \mu, \nu ; \mu', \nu' \right)  \nonumber \\
&& + U^{+}_{\Gamma}(\nu) U^{-}_{\Gamma}(\nu) \int_0^{\infty} d\zeta_1 e^{-(\zeta-\zeta_1)
u^{\pm}_{\Gamma}(\nu)}
{\hat \Omega}^{\pm \epsilon'}_{\Gamma,\Gamma'} \left(\zeta_1, \mu, \nu ; \mu', \nu' \right) \nonumber \\
&& + {\hat \Omega}^{\mp \epsilon'}_{\Gamma,\Gamma'} \left(0, \mu, \nu ; \mu', \nu' \right)
U^{\pm}_{\Gamma}(\mu) U^{\pm}_{\Gamma}(\nu) \int_0^{\infty} d\zeta_1 
e^{-(\zeta-\zeta_1) u^{\pm}_{\Gamma}(\mu)}
 e^{-\zeta_1 u^{\pm}_{\Gamma}(\nu)}
\end{eqnarray}

together with the initial conditions at $\Gamma=\Gamma'$ 
given in (\ref{omegainit}) 
which become:
\begin{eqnarray}
  {\hat \Omega}^{\epsilon \epsilon'}_{\Gamma',\Gamma'} \left(\zeta, \mu, \nu ; \mu', \nu' \right)
& = & \delta_{\epsilon \epsilon'}
\int_0^{\infty} dx_L' \int_0^{\infty} dx_R'
 e^{-\mu'x_R'} e^{-\nu'x_L'}  \omega^{\epsilon'}_{\Gamma'} (\zeta,x_R',x_L') \nonumber \\
& = &\delta_{\epsilon \epsilon'} { 1 \over {\overline{l}_{\Gamma'}}}
{1 \over {\mu'-\nu'}}
\left( U^{\epsilon'}_{\Gamma'}(\nu') e^{-\zeta u^{\epsilon'}_{\Gamma'}(\nu') }
-U^{\epsilon'}_{\Gamma'}(\mu') e^{-\zeta 
u^{\epsilon'}_{\Gamma'}(\mu')} \right).
\end{eqnarray}

We  look for solutions of the form
\begin{eqnarray}
 {\hat \Omega}^{\epsilon \epsilon'}_{\Gamma,\Gamma'} \left(\zeta, \mu, \nu ; \mu', \nu' \right)
= && A^{\epsilon \epsilon'}_{\Gamma,\Gamma'} \left( \mu, \nu ; \mu', \nu' \right)
 e^{-\zeta u^{\epsilon}_{\Gamma}(\mu)} + B^{\epsilon \epsilon'}_{\Gamma,\Gamma'} 
\left( \mu, \nu ; \mu', \nu' \right) e^{-\zeta u^{\epsilon}_{\Gamma}(\nu)}  \nonumber \\
&& +C^{\epsilon \epsilon'}_{\Gamma,\Gamma'} \left( \mu, \nu ; \mu', \nu' \right)
 e^{-\zeta u^{\epsilon'}_{\Gamma'}(\mu')}
+D^{\epsilon \epsilon'}_{\Gamma,\Gamma'} \left( \mu, \nu ; \mu', \nu' \right)
 e^{-\zeta u^{\epsilon'}_{\Gamma'}(\nu')}.
\end{eqnarray}
It is  useful to introduce the functions
\begin{eqnarray}
&& \theta^{\epsilon \epsilon'}_{\Gamma,\Gamma'} \left( \mu, \nu ; \mu', \nu' \right)
= \Omega^{\epsilon \epsilon'}_{\Gamma,\Gamma'} \left(\zeta=0, \mu, \nu ; \mu', \nu' \right) \nonumber \\
&& = A^{\epsilon \epsilon'}_{\Gamma,\Gamma'} \left( \mu, \nu ; \mu', \nu' \right)
 + B^{\epsilon \epsilon'}_{\Gamma,\Gamma'} \left( \mu, \nu ; \mu', \nu'\right)
+C^{\epsilon \epsilon'}_{\Gamma,\Gamma'} \left( \mu, \nu ; \mu', \nu' \right)
+D^{\epsilon \epsilon'}_{\Gamma,\Gamma'} \left( \mu, \nu ; \mu', \nu' \right)
\nonumber \\
&& \sigma^{\epsilon \epsilon'}_{\Gamma,\Gamma'} \left( \mu, \nu ; \mu', \nu' \right)
= \int_0^{\infty} d\zeta \Omega^{\epsilon \epsilon'}_{\Gamma,\Gamma'}   
\left(\zeta, \mu, \nu ; \mu', \nu' \right) \nonumber \\
&& = { {A^{\epsilon \epsilon'}_{\Gamma,\Gamma'} \left( \mu, \nu ; \mu', \nu' \right)}
\over {u^{\epsilon}_{\Gamma}(\mu)}}
+{ {B^{\epsilon \epsilon'}_{\Gamma,\Gamma'} \left( \mu, \nu ; \mu', \nu' \right)}
\over {u^{\epsilon}_{\Gamma}(\nu)}}
+{ {C^{\epsilon \epsilon'}_{\Gamma,\Gamma'} \left( \mu, \nu ; \mu', \nu' \right)}
\over {u^{\epsilon'}_{\Gamma'}(\mu')}}
+{ {D^{\epsilon \epsilon'}_{\Gamma,\Gamma'} \left( \mu, \nu ; \mu', \nu' \right)}
\over {u^{\epsilon'}_{\Gamma'}(\nu')}}
\end{eqnarray}
For each initial condition indexed by one of $\epsilon'=\pm 1$
it is most convenient to work with the eight functions
$\theta^{\epsilon \epsilon'}$, $\sigma^{\epsilon \epsilon'}$,  
$C^{\epsilon \epsilon'}$ and $D^{\epsilon \epsilon'}$ with $\epsilon = \pm 1$.

We first consider the equations for $C^{\epsilon \epsilon'}$ 
and $D^{\epsilon \epsilon'}$.
These equations are homogenous and thus easier to solve.
The equations for $C^{\epsilon \epsilon'}$ read
\begin{eqnarray}
\partial_{\Gamma} C^{\epsilon \epsilon'}_{\Gamma,\Gamma'} 
\left( \mu, \nu ; \mu', \nu' \right) = 
&& \left( - 2 U^{-}_{\Gamma}(0) \delta_{\epsilon,+1}
- 2 U^{+}_{\Gamma}(0) \delta_{\epsilon,-1}
- u^{\epsilon'}_{\Gamma'}(\mu')
+ { {U^{-}_{\Gamma}(\mu) U^{+}_{\Gamma}(\mu)} 
\over {u^{\epsilon}_{\Gamma}(\mu)-u^{\epsilon'}_{\Gamma'}(\mu')}} 
\right.  \nonumber \\
&& \left. + { {U^{-}_{\Gamma}(\nu) U^{+}_{\Gamma}(\nu) }
\over {u^{\epsilon}_{\Gamma}(\nu)-u^{\epsilon'}_{\Gamma'}(\mu')}} \right)
 C^{\epsilon \epsilon'}_{\Gamma,\Gamma'} \left( \mu, \nu ; \mu', \nu' \right)
\end{eqnarray}
with initial conditions at $\Gamma=\Gamma'$  indexed by $\epsilon'$:
\begin{eqnarray}
C^{\epsilon \epsilon'}_{\Gamma',\Gamma'} \left( \mu, \nu ; \mu', \nu' \right) 
=  - \delta_{\epsilon \epsilon'} {1 \over {\overline{l}_{\Gamma'} } }
{ U^{\epsilon'}_{\Gamma'}(\mu') \over {\mu'-\nu'}}.  
\end{eqnarray}
Similarly the equations for $D^{\pm}$ read
\begin{eqnarray}
\partial_{\Gamma} 
D^{\epsilon \epsilon'}_{\Gamma,\Gamma'} 
\left( \mu, \nu ; \mu', \nu' \right) = 
&& \left( - 2 U^{-}_{\Gamma}(0) \delta_{\epsilon,+1}
- 2 U^{+}_{\Gamma}(0) \delta_{\epsilon,-1}
- u^{\epsilon'}_{\Gamma'}(\nu')
+ { {U^{-}_{\Gamma}(\mu) U^{+}_{\Gamma}(\mu)} 
\over {u^{\epsilon}_{\Gamma}(\mu)-u^{\epsilon'}_{\Gamma'}(\nu')}} 
\right. \nonumber \\
&& \left. + { {U^{-}_{\Gamma}(\nu) U^{+}_{\Gamma}(\nu)}
\over {u^{\epsilon}_{\Gamma}(\nu)-u^{\epsilon'}_{\Gamma'}(\nu')}}\right)
D^{\epsilon \epsilon'}_{\Gamma,\Gamma'} \left( \mu, \nu ; \mu', \nu' \right)
\end{eqnarray}
with initial conditions at $\Gamma=\Gamma'$ are indexed by $\epsilon'$:
\begin{eqnarray}
D^{\epsilon \epsilon'}_{\Gamma',\Gamma'} \left( \mu, \nu ; \mu', \nu' \right) 
=  \delta_{\epsilon \epsilon'} {1 \over {\overline{l}_{\Gamma'}} } 
{ U^{\epsilon'}_{\Gamma'}(\nu') \over {\mu'-\nu'}}.  
\end{eqnarray}

To find the solution one notices that each matrix element 
$C^{\epsilon \epsilon'}$ and $D^{\epsilon \epsilon'}$
satisfies its own differential equation.
Thus, since the initial condition is diagonal in $\epsilon \epsilon'$,
the solution is also diagonal. It is found to be:

\begin{eqnarray}
&& C^{\epsilon \epsilon'}_{\Gamma,\Gamma'} \left( \mu, \nu ; \mu', \nu' \right) 
= \nonumber \\
&&
{- \delta_{\epsilon \epsilon'} \over {\overline{l}_{\Gamma'}} }
\frac{U^{\epsilon'}_\Gamma(0)^2}{U^{\epsilon'}_{\Gamma'}(0)^2}
{{U^{\epsilon'}_{\Gamma'}(\mu')} \over {(\mu'-\nu')}}
{ {(u^{\epsilon'}_{\Gamma'}(\mu')-u^{\epsilon'}_{\Gamma'}(\mu)) 
(u^{\epsilon'}_{\Gamma'}(\mu')-u^{\epsilon'}_{\Gamma'}(\nu))} 
\over {(u^{\epsilon'}_{\Gamma'}(\mu')-u^{\epsilon'}_{\Gamma}(\mu)) 
(u^{\epsilon'}_{\Gamma'}(\mu')-u^{\epsilon'}_{\Gamma}(\nu))} } 
e^{-(\Gamma-\Gamma') u^{\epsilon'}_{\Gamma'}(\mu') }
\nonumber \\
&& D^{\epsilon \epsilon'}_{\Gamma,\Gamma'} \left( \mu, \nu ; \mu', \nu' \right) 
= \nonumber \\
&& {\delta_{\epsilon \epsilon'} \over {\overline{l}_{\Gamma'}} }
\frac{U^{\epsilon'}_\Gamma(0)^2}{U^{\epsilon'}_{\Gamma'}(0)^2}
{{U^{\epsilon'}_{\Gamma'}(\nu')} \over {(\mu'-\nu')}}
{ {(u^{\epsilon'}_{\Gamma'}(\nu')-u^{\epsilon'}_{\Gamma'}(\mu)) (u^{\epsilon'}_{\Gamma'}(\nu')-u^{\epsilon'}_{\Gamma'}(\nu))} \over {(u^{\epsilon'}_{\Gamma'}(\nu')-u^{\epsilon'}_{\Gamma}(\mu)) 
(u^{\epsilon'}_{\Gamma'}(\nu')-u^{\epsilon'}_{\Gamma}(\nu))} } 
e^{-(\Gamma-\Gamma') u^{\epsilon'}_{\Gamma'}(\nu') }
\end{eqnarray}

For each $\epsilon'$ the four remaining functions
$\theta^{\pm ,\epsilon'}$ and $\sigma^{\pm ,\epsilon'}$
satisfy the following system of four differential equations:

\begin{eqnarray}
&& \partial_\Gamma \theta^{+ \epsilon'} = - ( 2 u_\Gamma^{-}(0) + u_\Gamma^{+}(\mu)
+ u_\Gamma^{+}(\nu) ) \theta^{+ \epsilon'} + u_\Gamma^{+}(\mu) u_\Gamma^{+}(\nu)
\sigma^{+ \epsilon'}
+ p^{+}_{\Gamma,\Gamma'} \delta_{\epsilon',+} \\
&& \partial_\Gamma \theta^{- \epsilon'} = - ( 2 u_\Gamma^{+}(0) + u_\Gamma^{-}(\mu)
+ u_\Gamma^{-}(\nu) ) \theta^{- \epsilon'} + u_\Gamma^{-}(\mu) u_\Gamma^{-}(\nu)
\sigma^{- \epsilon'}
+ p^{-}_{\Gamma,\Gamma'} \delta_{\epsilon',-} \\
&& \partial_\Gamma \sigma^{+ \epsilon'} = \left( - 2 u_\Gamma^{-}(0) + 
\frac{U_\Gamma^{+}(\mu) U_\Gamma^{-}(\mu)}{u_\Gamma^{+}(\mu)}
+ 
\frac{U_\Gamma^{+}(\nu) U_\Gamma^{-}(\nu)}{u_\Gamma^{+}(\nu)} \right) 
\sigma^{+ \epsilon'}
- \theta^{+ \epsilon'}
+ 
\frac{U_\Gamma^{+}(\mu) U_\Gamma^{+}(\nu)}{u_\Gamma^{+}(\mu) u_\Gamma^{+}(\nu)}
\theta^{- \epsilon'} \\
&& \partial_\Gamma \sigma^{- \epsilon'} = \left( - 2 u_\Gamma^{+}(0) + 
\frac{U_\Gamma^{+}(\mu) U_\Gamma^{-}(\mu)}{u_\Gamma^{-}(\mu)}
+ 
\frac{U_\Gamma^{+}(\nu) U_\Gamma^{-}(\nu)}{u_\Gamma^{-}(\nu)} \right)
 \sigma^{- \epsilon'}
- \theta^{- \epsilon'}
+ 
\frac{U_\Gamma^{-}(\mu) U_\Gamma^{-}(\nu)}{u_\Gamma^{-}(\mu) u_\Gamma^{-}(\nu)}
\theta^{+ \epsilon'}
\label{systemaging}
\end{eqnarray}

We note that the system for $\epsilon'=+1$ and the system for
$\epsilon'=-1$ are identical except for the inhomogeneous
part, that we have defined as:

\begin{eqnarray}
p^{\epsilon'}_{\Gamma ,\Gamma'}  = && - 
\frac{1}{u^{\epsilon'}_{\Gamma'}(\mu')} (
u^{\epsilon'}_{\Gamma}(\mu) - u^{\epsilon'}_{\Gamma'}(\mu') )
( u^{\epsilon'}_{\Gamma}(\nu) - u^{\epsilon'}_{\Gamma'}(\mu') )
C^{\epsilon' \epsilon'}_{\Gamma,\Gamma'} \nonumber \\
&& - \frac{1}{u^{\epsilon'}_{\Gamma'}(\nu')} (
u^{\epsilon'}_{\Gamma}(\mu) - u^{\epsilon'}_{\Gamma'}(\nu') )
( u^{\epsilon'}_{\Gamma}(\nu) - u^{\epsilon'}_{\Gamma'}(\nu') )
D^{\epsilon' \epsilon'}_{\Gamma,\Gamma'}.
\end{eqnarray}
To exhibit explicitly the $\Gamma$ dependence, it is useful to rewrite
\begin{eqnarray}
 p^{+}_{\Gamma,\Gamma'} \left( \mu, \nu ; \mu', \nu' \right)
 = \frac{1}{\overline{l}_{\Gamma} (\mu'-\nu')}
( f^{+}_{\Gamma'}( \mu, \nu ; \mu' ) 
e^{-(\Gamma-\Gamma') u^{-}_{\Gamma'}(\mu') }
- f^{+}_{\Gamma'}( \mu, \nu ; \nu' ) 
e^{-(\Gamma-\Gamma') u^{-}_{\Gamma'}(\nu') } )
\end{eqnarray}
\begin{eqnarray}
 p^{-}_{\Gamma,\Gamma'} \left( \mu, \nu ; \mu', \nu' \right)
 = \frac{1}{\overline{l}_{\Gamma} (\mu'-\nu')}
( f^{-}_{\Gamma'}( \mu, \nu ; \mu' ) 
e^{-(\Gamma-\Gamma') u^{+}_{\Gamma'}(\mu') }
- f^{-}_{\Gamma'}( \mu, \nu ; \nu' ) 
e^{-(\Gamma-\Gamma') u^{+}_{\Gamma'}(\nu') } )
\end{eqnarray}
where we have introduced the two $\Gamma$-independent functions 
\begin{eqnarray}
 f^{\epsilon'}_{\Gamma'} \left( \mu, \nu ; \mu' \right)
=  {{U^{\epsilon'}_{\Gamma'}(\mu')} \over {u^{\epsilon'}_{\Gamma'}(\mu')}}
 {(u^{\epsilon'}_{\Gamma'}(\mu')-u^{\epsilon'}_{\Gamma'}(\mu)) 
(u^{\epsilon'}_{\Gamma'}(\mu')-u^{\epsilon'}_{\Gamma'}(\nu))} 
\end{eqnarray}

The first step is to obtain the solutions of the 
above two (identical) homogeneous systems (\ref{systemaging}). 
Remarkably, these
can be constructed from the functions $U^{\pm}_\Gamma(p)$ and
$u^{\pm}_\Gamma(p)$ using the differential equations (\ref{equpm}).
We  find the four independent solutions of
the homogeneous system to be:

\begin{eqnarray}
&&
\left\{ 
\theta_1^{+} = n_{\Gamma} U^{+}_\Gamma(\mu), ~~
\theta_1^{-} = - n_{\Gamma} U^{-}_\Gamma(\nu), ~~
\sigma_1^{+} = n_{\Gamma} \frac{U^{+}_\Gamma(\mu)}{u^{+}_\Gamma(\mu)}, ~~
\sigma_1^{-} = - n_{\Gamma} \frac{U^{-}_\Gamma(\nu)}{u^{-}_\Gamma(\nu)} 
\right\}  \\
&&
\left\{ 
\theta_2^{+} = n_{\Gamma} U^{+}_\Gamma(\mu) u^{+}_\Gamma(\nu), ~~
\theta_2^{-} = n_{\Gamma} U^{-}_\Gamma(\nu) u^{-}_\Gamma(\mu), ~~
\sigma_2^{+} = n_{\Gamma} \frac{ U^{+}_\Gamma(\mu)} 
{u^{+}_\Gamma(\mu) u^{+}_\Gamma(\nu)} \right. \nonumber \\  
&& \times \quad (u^{+}_\Gamma(\nu)^2 - U^{+}_\Gamma(\nu) U^{-}_\Gamma(\nu)), 
\left. ~~ \sigma_2^{-} 
= n_{\Gamma} \frac{ U^{-}_\Gamma(\nu)}{ u^{-}_\Gamma(\mu) u^{-}_\Gamma(\nu) }
(u^{-}_\Gamma(\mu)^2 - U^{+}_\Gamma(\mu) U^{-}_\Gamma(\mu))\right\} \\
&&
\left\{ \theta_3^{+} = n_{\Gamma} U^{+}_\Gamma(\nu), ~~
\theta_3^{-} = - n_{\Gamma} U^{-}_\Gamma(\mu), ~~
\sigma_3^{+} = n_{\Gamma} \frac{U^{+}_\Gamma(\nu)}{u^{+}_\Gamma(\nu)}, ~~
\sigma_3^{-} = - n_{\Gamma} \frac{U^{-}_\Gamma(\mu)}{u^{-}_\Gamma(\mu)} 
\right\} \\
&& 
\left\{ \theta_4^{+} = n_{\Gamma} U^{+}_\Gamma(\nu) u^{+}_\Gamma(\mu), ~~
\theta_4^{-} = n_{\Gamma} U^{-}_\Gamma(\mu) u^{-}_\Gamma(\nu), ~~
\sigma_4^{+} = n_{\Gamma} \frac{ U^{+}_\Gamma(\nu)}
{ u^{+}_\Gamma(\mu) u^{+}_\Gamma(\nu)}\right. \nonumber \\
&& \times (u^{+}_\Gamma(\mu)^2 - U^{+}_\Gamma(\mu) U^{-}_\Gamma(\mu)),
\left. \sigma_4^{-} = n_{\Gamma} 
\frac{ U^{-}_\Gamma(\mu)}{ u^{-}_\Gamma(\mu) u^{-}_\Gamma(\nu) }
(u^{-}_\Gamma(\nu)^2 - U^{+}_\Gamma(\nu) U^{-}_\Gamma(\nu)) \right\}  
\label{systemsolu}
\end{eqnarray}
where $n_{\Gamma}=1/{\overline{l_{\Gamma}}}=
{{\delta^2} \over {\sinh^2(\Gamma \delta)}}$ and with
$\partial_\Gamma n_\Gamma = - (u^{+}_\Gamma + u^{-}_\Gamma) n_\Gamma$.

It will be useful to consider the matrix formed by these solutions

\begin{eqnarray*}
N_\Gamma =  \pmatrix{ 
& \theta^{+}_1 & \theta^{+}_2 & \theta^{+}_3 & \theta^{+}_4 \\ 
& \theta^{-}_1 & \theta^{-}_2 & \theta^{-}_3 & \theta^{-}_4 \\
& \sigma^{+}_1 & \sigma^{+}_2 & \sigma^{+}_3 & \sigma^{+}_4 \\ 
& \sigma^{-}_1 & \sigma^{-}_2 & \sigma^{-}_3 & \sigma^{-}_4 }
\end{eqnarray*}
From the usual properties of systems of linear equations the
Wronskian $W_\Gamma= \det[N_\Gamma]$ satisfies the simple equation,
$\partial_\Gamma W_\Gamma = Tr [M_\Gamma]  W_\Gamma$ where 
$M$ is the matrix formed by the coefficients of the homogeneous part of the
linear system. One can easily integrate this equation,
or one can  compute directly the determinant, and use
the definitions (\ref{systemsolu}) to simplify the result (after a tedious
calculation). Both give the same, remarkably simple,  result:
\begin{eqnarray}
W_\Gamma= - n_{\Gamma}^4 (\mu - \nu)^2 \frac{ 
U^{+}_\Gamma(\mu) U^{-}_\Gamma(\mu) U^{+}_\Gamma(\nu) U^{-}_\Gamma(\nu)
}{ u^{+}_\Gamma(\mu) u^{-}_\Gamma(\mu) u^{+}_\Gamma(\nu) u^{-}_\Gamma(\nu) }.
\end{eqnarray}
Since this is not zero, this shows that the four solutions given above
are linearly independent. Thus we are now in a position to write 
the solutions of the two linear differential systems (\ref{systemaging})
with the inhomogeneous terms. It is found, as usual, as a linear combination of
the four independent solutions (\ref{systemsolu}) of the homogeneous system:

\begin{eqnarray}
\pmatrix{ 
& \theta^{+ \epsilon'} \nonumber \\
& \theta^{- \epsilon'} \nonumber \\
& \sigma^{+ \epsilon' } \nonumber \\
& \sigma^{- \epsilon'} } = \sum_{i=1}^4 \lambda^{i \epsilon'}_\Gamma
\pmatrix{ 
& \theta_{i,\Gamma}^{+}(\mu,\nu) \nonumber \\
& \theta_{i,\Gamma}^{-}(\mu,\nu) \nonumber \\
& \sigma_{i,\Gamma}^{+}(\mu,\nu) \nonumber \\
& \sigma_{i,\Gamma}^{-}(\mu,\nu) }  \equiv
N_\Gamma \cdot \pmatrix{ 
& \lambda_\Gamma^{1 \epsilon'} \nonumber \\
& \lambda_\Gamma^{2 \epsilon'} \nonumber \\
& \lambda_\Gamma^{3 \epsilon'} \nonumber \\
& \lambda_\Gamma^{4 \epsilon'} }
\end{eqnarray}
where $\lambda^{i \epsilon'}_\Gamma \equiv 
\lambda^{i \epsilon'}_{\Gamma \Gamma'}(\mu,\nu,\mu',\nu')$
are the coefficients of the linear combinations.
Using the standard method 
one finds the following equations for the coefficients:

\begin{eqnarray}
N_\Gamma \cdot \pmatrix{ 
& \partial_\Gamma \lambda^{1 +} \nonumber \\
& \partial_\Gamma \lambda^{2 +} \nonumber \\
& \partial_\Gamma \lambda^{3 +} \nonumber \\
& \partial_\Gamma \lambda^{4 +} }  = 
\pmatrix{ & p^{+}_{\Gamma,\Gamma'} \nonumber \\
& 0 \nonumber \\
& 0 \nonumber \\
& 0 }
\end{eqnarray}
\begin{eqnarray*}
N_\Gamma \cdot \pmatrix{ 
& \partial_\Gamma \lambda^{1 -} \\
& \partial_\Gamma \lambda^{2 -} \\
& \partial_\Gamma \lambda^{3 -} \\
& \partial_\Gamma \lambda^{4 -} } = 
\pmatrix{ & 0 \\
& p^{-}_{\Gamma,\Gamma'} \\
& 0 \\
& 0 } 
\end{eqnarray*}
The initial condition for the $\lambda^{i \epsilon'}_\Gamma$ at 
$\Gamma=\Gamma'$ are fixed by the initial conditions
\begin{eqnarray}
\theta^{\epsilon \epsilon'}_{\Gamma',\Gamma'} \left(\mu, \nu ; \mu', \nu' \right)
=  && \sum_{i=1}^4  \lambda^{(i),\epsilon'}_{\Gamma',\Gamma'} \left( \mu, \nu ; \mu', \nu' \right) \theta^{\epsilon }_i(\Gamma',\mu,\nu)
={\hat \Omega}^{\epsilon \epsilon'}_{\Gamma',\Gamma'} 
\left(\zeta=0, \mu, \nu ; \mu', \nu' \right) \nonumber \\
= && \delta_{\epsilon \epsilon'} { 1 \over {\overline{l}_{\Gamma'}}}
{1 \over {\mu'-\nu'}}
\left( U^{\epsilon'}_{\Gamma'}(\nu') 
-U^{\epsilon'}_{\Gamma'}(\mu') \right) \\
\sigma^{\epsilon \epsilon'}_{\Gamma',\Gamma'} \left(\mu, \nu ; \mu', \nu' \right) 
=  && \sum_{i=1}^4  \lambda^{(i),\epsilon'}_{\Gamma',\Gamma'} \left( \mu, \nu ; \mu', \nu' \right) \sigma^{\epsilon }_i(\Gamma',\mu,\nu)
=\int_0^{\infty} d\zeta
 {\hat \Omega}^{\epsilon \epsilon'}_{\Gamma',\Gamma'} 
\left(\zeta, \mu, \nu ; \mu', \nu' \right) \nonumber \\
= && \delta_{\epsilon \epsilon'} { 1 \over {\overline{l}_{\Gamma'}}}
{1 \over {\mu'-\nu'}}
\left( { {U^{\epsilon'}_{\Gamma'}(\nu') }
\over { u^{\epsilon'}_{\Gamma'}(\nu') } }
- { {U^{\epsilon'}_{\Gamma'}(\mu') } \over { u^{\epsilon'}_{\Gamma'}(\mu')} } \right)
\end{eqnarray}
Finally, we find, for $\epsilon'=+1$, the solution of (\ref{systemaging}) 
with the above initial conditions:

\begin{eqnarray}
&& \theta^{\pm +}_{\Gamma,\Gamma'} =
\sum_{i=1}^{4} \theta_{i,\Gamma}^{\pm}
\left[ \int_{\Gamma'}^{\Gamma} d\tilde{\Gamma} 
(N^{-1}_{\tilde{\Gamma}})_{i,1} p^{+}_{\tilde{\Gamma},\Gamma'}
+ (N^{-1}_{\Gamma'})_{i,1} \theta^{++}_{\Gamma',\Gamma'}
+ (N^{-1}_{\Gamma'})_{i,3} \sigma^{++}_{\Gamma',\Gamma'} \right]  \\
&& \sigma^{\pm +}_{\Gamma,\Gamma'} =
\sum_{i=1}^{4} \sigma_{i,\Gamma}^{\pm}
\left[ \int_{\Gamma'}^{\Gamma} d\tilde{\Gamma} 
(N^{-1}_{\tilde{\Gamma}})_{i,1} p^{+}_{\tilde{\Gamma},\Gamma'}
+ (N^{-1}_{\Gamma'})_{i,1} \theta^{++}_{\Gamma',\Gamma'}
+ (N^{-1}_{\Gamma'})_{i,3} \sigma^{++}_{\Gamma',\Gamma'} \right]
\label{soluplus}
\end{eqnarray} 
and, for $\epsilon'=-1$:
\begin{eqnarray}
&& \theta^{\pm -}_{\Gamma,\Gamma'} =
\sum_{i=1}^{4} \theta_{i,\Gamma}^{\pm}
\left[ \int_{\Gamma'}^{\Gamma} d\tilde{\Gamma} 
(N^{-1}_{\tilde{\Gamma}})_{i,2} p^{-}_{\tilde{\Gamma},\Gamma'}
+ (N^{-1}_{\Gamma'})_{i,2} \theta^{--}_{\Gamma',\Gamma'}
+ (N^{-1}_{\Gamma'})_{i,4} \sigma^{--}_{\Gamma',\Gamma'} \right]  \\
&& \sigma^{\pm -}_{\Gamma,\Gamma'} =
\sum_{i=1}^{4} \sigma_{i,\Gamma}^{\pm}
\left[ \int_{\Gamma'}^{\Gamma} d\tilde{\Gamma} 
(N^{-1}_{\tilde{\Gamma}})_{i,2} p^{-}_{\tilde{\Gamma},\Gamma'}
+ (N^{-1}_{\Gamma'})_{i,2} \theta^{--}_{\Gamma',\Gamma'}
+ (N^{-1}_{\Gamma'})_{i,4} \sigma^{--}_{\Gamma',\Gamma'} \right]
\label{soluminus}
\end{eqnarray}

The next step is to evaluate $N_\Gamma^{-1}$, the inverse 
of the matrix $N_\Gamma$. Remarkably,the inverse admits a simple explicit
form in terms of the functions $u_\Gamma^{\pm}$ and $U_\Gamma^{\pm}$, which can
be found after some tedious calculations using the form
(\ref{solu-biased-u}). It reads:

\begin{eqnarray}
N^{-1}_\Gamma =  \frac{1}{n_\Gamma (\nu - \mu) }
\pmatrix{ 
&  - \frac{\mu}{U_\Gamma^{+}(\mu)}
& - \frac{\nu}{U_\Gamma^{-}(\nu)}
& \frac{u_\Gamma^{+}(\nu) u_\Gamma^{+}(\mu) u_\Gamma^{-}(\mu) }{U_\Gamma^{+}(\mu)} 
& \frac{u_\Gamma^{-}(\mu) u_\Gamma^{+}(\nu) u_\Gamma^{-}(\nu) }{U_\Gamma^{-}(\nu)} 
\nonumber \\ 
&  - \frac{u_\Gamma^{+}(\mu)}{U_\Gamma^{+}(\mu)} 
& \frac{u_\Gamma^{-}(\nu)}{U_\Gamma^{-}(\nu)} 
& \frac{u_\Gamma^{+}(\mu) u_\Gamma^{+}(\nu)}{U_\Gamma^{+}(\mu)}
& - \frac{u_\Gamma^{-}(\mu) u_\Gamma^{-}(\nu)}{U_\Gamma^{-}(\nu)}
\nonumber \\
& \frac{\nu}{U_\Gamma^{+}(\nu)}
& \frac{\mu}{U_\Gamma^{-}(\mu)}
& - \frac{u_\Gamma^{+}(\mu) u_\Gamma^{+}(\nu) u_\Gamma^{-}(\nu) }{U_\Gamma^{+}(\nu)}
& - \frac{u_\Gamma^{-}(\nu) u_\Gamma^{+}(\mu) u_\Gamma^{-}(\mu) }{U_\Gamma^{-}(\mu)} 
\nonumber \\ 
& \frac{u_\Gamma^{+}(\nu)}{U_\Gamma^{+}(\nu)} 
& - \frac{u_\Gamma^{-}(\mu)}{U_\Gamma^{-}(\mu)} 
& - \frac{u_\Gamma^{+}(\mu) u_\Gamma^{+}(\nu)}{U_\Gamma^{+}(\nu)}
& \frac{u_\Gamma^{-}(\mu) u_\Gamma^{-}(\nu)}{U_\Gamma^{-}(\mu)}
} 
\end{eqnarray}

We have thus obtained the quantities of interest, namely
the $\sigma^{\epsilon \epsilon'}_{\Gamma,\Gamma'}(\mu,\nu;\mu'\nu')
= \int_\zeta \Omega^{\epsilon \epsilon'}_{\Gamma,\Gamma'}(\mu,\nu;\mu'\nu')$.
The two-time probability can then be obtained from the $\sigma$ as follows.

The Laplace transforms in the different sectors ($x>0$, $x'>0$), 
($x<0$, $x'<0$),($x<0$ and $x'>0$) and ($x>0$ and $x'<0$)
are respectively:

\begin{eqnarray}
{\hat P}^{++}_{\Gamma,\Gamma'} (p,p') 
& = & \int_0^{\infty} dx \int_0^{\infty} dx' e^{-p'x'} e^{-p(x-x')}
\overline{\mbox{Prob}(xt,x't'|00)} \nonumber \\
& = & 
{\sigma}^{++}_{\Gamma,\Gamma'} ( \mu=p , \nu=0 ; \mu'=p' , \nu'=0) \\
{\hat P}^{--}_{\Gamma,\Gamma'} (p,p') 
& = & \int_{-\infty}^{0} dx \int_{-\infty}^{0} dx' e^{p'x'} e^{p(x-x')}
\overline{\mbox{Prob}(xt,x't'|00)} \nonumber \\
& = & 
{\sigma}^{--}_{\Gamma,\Gamma'} ( \mu=0 , \nu=p ; \mu'=0 , \nu'=p') \\
{\hat P}^{-+}_{\Gamma,\Gamma'} (p,p') 
& = & \int_{-\infty}^{0} dx \int_0^{\infty} dx' e^{-p'x'} e^{p(x-x')}
\overline{\mbox{Prob}(xt,x't'|00)} \nonumber \\
& = & 
{\sigma}^{-+}_{\Gamma,\Gamma'} (  \mu=0 , \nu=p ; \mu'=p+p' , \nu'=p)\\
{\hat P}^{+-}_{\Gamma,\Gamma'} (p,p') 
& = & \int_{-\infty}^{0} dx' \int_0^{\infty} dx e^{p'x'} e^{-p(x-x')}
\overline{\mbox{Prob}(xt,x't'|00)} \nonumber \\
& = & 
{\sigma}^{+-}_{\Gamma,\Gamma'} (   \mu=p , \nu=0 ; \mu'=p, \nu'=p+p') 
\label{sectors}
\end{eqnarray}

From these one can compute the distribution
$Q(y,t,t')$ $=\int_{-\infty}^{+\infty} dx' \overline{\mbox{Prob}((x'+y),t,x',t'|0,0)}$
of the relative deplacement $y=x(t)-x(t')$.
Its Laplace transform in the sectors $y>0$ and
$y<0$ are respectively:

\begin{eqnarray}
Q^{+}_{\Gamma,\Gamma'} (p) 
& = & \int_0^{\infty} dy e^{-py} Q(y,t,t') \nonumber \\
& = & \int_{0}^{\infty} dx' \int_{x'}^{\infty} dx  e^{-p(x-x')}
\overline{\mbox{Prob}(xt,x't'|00)} \nonumber \\ 
&& \quad + \int_{-\infty}^{0} dx' \int_{0}^{\infty} dx e^{-p(x-x')}
\overline{\mbox{Prob}(xt,x't'|00)} \nonumber \\
& = & {\sigma}^{++}_{\Gamma,\Gamma'} (  \mu=p , \nu=0 ; \mu'=0 , \nu'=0) 
+{\sigma}^{+-}_{\Gamma,\Gamma'} (  \mu=0 , \nu=p ; \mu'=p , \nu'=p) \\
Q^{-}_{\Gamma,\Gamma'} (p) 
& = & \int_{-\infty}^{0} dy e^{py} Q(y,t,t') \nonumber \\
& = & \int_{-\infty}^{0} dx' 
\int_{-\infty}^{0} dx  
e^{p(x-x')}
\overline{\mbox{Prob}(xt,x't'|00)} \nonumber \\ 
&& \quad + \int_{-\infty}^{0} dx \int_{0}^{\infty} dx' e^{p(x-x')}
\overline{\mbox{Prob}(xt,x't'|00)} \nonumber \\
& = & {\sigma}^{--}_{\Gamma,\Gamma'} (  \mu=0 , \nu=p ; \mu'=0 , \nu'=0) 
+{\sigma}^{-+}_{\Gamma,\Gamma'} (  \mu=p , \nu=0 ; \mu'=p , \nu'=p)
\end{eqnarray}

In order to compute these distributions,
we have explicitly evaluated the sums and integrals in (\ref{soluplus},
\ref{soluminus}). As an example let us examine:
\begin{eqnarray}
\sigma^{+ +}_{\Gamma,\Gamma'} (\mu,\nu;\mu',\nu')= && 
\sum_{i=1}^{4} \sigma_{i,\Gamma}^{+} (\mu,\nu)
[ \int_{\Gamma'}^{\Gamma} d\tilde{\Gamma} 
(N^{-1}_{\tilde{\Gamma}})_{i,1}(\mu,\nu) 
p^{+}_{\tilde{\Gamma},\Gamma'}(\mu,\nu;\mu',\nu') \nonumber \\
&& + (N^{-1}_{\Gamma'})_{i,1} (\mu,\nu) 
\theta^{++}_{\Gamma',\Gamma'} (\mu',\nu')
+ (N^{-1}_{\Gamma'})_{i,3} (\mu,\nu) 
\sigma^{++}_{\Gamma',\Gamma'} (\mu',\nu')]
\end{eqnarray}
It turns out that all integrals that appear in this expression
are simple exponentials. This remarkable property remains true 
for all other elements
and is the reason why the calculation, though tedious, can be carried out
explicitly for this problem.

{\it Explicit results for the symmetric case $\delta=0$}

Here we give the explicit expression for the Laplace transform
of $\overline{{\rm Prob}(x t, x' t' \vert 0 0)}$ in the different sectors
defined in (\ref{sectors}). For the sector where the product $x x'>0$, we find:

\begin{eqnarray}
&& {\hat P}^{++}_{\Gamma=\alpha \Gamma',\Gamma'} 
\left( p=\frac{s^2}{\Gamma'^2},p'=\frac{r^2}{\Gamma'^2} \right)
= {\hat P}^{--}_{\Gamma=\alpha \Gamma',\Gamma'} 
\left( p=\frac{s^2}{\Gamma'^2},p'=\frac{r^2}{\Gamma'^2} \right) \nonumber \\
&& =\frac{\coth s}{\alpha^2 r^2 s} \left( 1-\frac{1}{\cosh r} \right)
+\frac{1}{\alpha^2 r^2 s \cosh(\alpha s)}
\left[ -  \left( 1-\frac{1}{\cosh r } \right) \right. \nonumber \\
&& \quad\quad \cosh s 
\coth s +\left( 1-\frac{r}{\sinh r } \right) \sinh s  \nonumber  \\
&&  + \frac{(r \coth r -1)(r\coth r -s \coth s )}
{ \cosh r  (s^2-r^2 \coth^2 r )}
\{ -s \cosh s -r \coth r \sinh s  \nonumber \\
&& \left. +e^{-(\alpha-1) r \coth r} 
(s \cosh( \alpha s )+r \coth r \sinh( \alpha s ) ) \} \right] \nonumber  \\
&&+\frac{\tanh( \alpha s)}{\alpha^2 r^2 s}
\left[ \frac{r}{\sinh r }-1+s \coth s \left(1-\frac{1}{\cosh r}\right)
\right.  \nonumber \\
&&\left. -\frac{(r \coth r-1) (r \coth r-s \coth s) \tanh r}{ r\cosh r} 
(1-e^{-(\alpha-1) r \coth r} )  \right] 
\label{resultsymm1}
\end{eqnarray}
For the sector where $x x'< 0$, we find:
\begin{eqnarray}
&& {\hat P}^{+-}_{\Gamma=\alpha \Gamma',\Gamma'} 
\left( p =\frac{s^2}{\Gamma'^2},p'=\frac{r^2-s^2}{\Gamma'^2} \right)
=\int_{-\infty}^{0} dx' e^{r^2 \frac{x'}{\Gamma'^2} }
\int_{0}^{+\infty} dx e^{-s^2 \frac{x}{\Gamma'^2} } 
\overline{{\rm Prob}(x t, x' t' \vert 0 0)} \nonumber \\
&& = \frac{1}{\alpha^2 (s^2-r^2) s} 
\left( \left( \frac{\cosh s}{\cosh r}-1\right) \coth 
s+s-r\frac{\sinh s}{\sinh r} \right) \nonumber \nonumber \\ 
&& \quad\quad + \frac{1}{\alpha^2 (s^2-r^2) s \cosh( \alpha s ) }
\left[ \frac{1}{ \sinh s} \left(1-\frac{\cosh s}{\cosh r}\right)
\right.  \nonumber \\
&& +\frac{(r \coth r-1)(r\coth r-s \coth s)}
{ \cosh r (s^2-r^2 \coth^2 r)}
 \left\{ -s \cosh s-r \coth r \sinh s \right. \nonumber  \\
&&\left. \left. 
\phantom{\frac{}{}} +e^{-(\alpha-1) r \coth r} (s \cosh( \alpha s )+
r \coth r \sinh( \alpha s )) \right\} \right] \nonumber  \\
&&+\frac{\tanh( \alpha s )}{\alpha^2 (s^2-r^2) s} 
\left[ \cosh s \left( \frac{r}{\sinh r}-\frac{s}{\sinh s} \right)
+1-\frac{\cosh s}{\cosh r} \right. \nonumber \\
&&+ \frac{(r \coth r-1)(r\coth r-s \coth s)}
{ \cosh r (s^2-r^2 \coth^2 r)}
 \{ r \coth r \cosh s+s \sinh s  \nonumber \\
&& \left. \phantom{\frac{}{}} - e^{-(\alpha-1) r \coth r}
 (r \coth r \cosh( \alpha s )+s \sinh(\alpha s)) \} \right]
\label{resultsymm2}
\end{eqnarray}
and the same expression for ${\hat P}^{-+}$.

We have performed a similar calculation in the biased 
case, but the corresponding full expression for the 
Laplace transform is too lengthy to give here.
Some particular limits are discussed in the text.
We give here the explicit expression of the two time
correlator:
\begin{eqnarray}
\label{corrb}
&&  \overline{\langle x(t) x(t')\rangle} = 
\frac{1}{
32 \delta^4 \sinh^2 \gamma' \sinh^2 \gamma } 
\left(A(\gamma,\gamma')+ B(\gamma,\gamma') 
e^{-(\gamma-\gamma') {\rm coth}(\gamma')}  \right)\nonumber
\end{eqnarray}
with:
\begin{eqnarray}
&& A(\gamma,\gamma')=
\cosh(2 \gamma) (\sinh(2 \gamma)-\gamma)
 [ \sinh(4\gamma')+\sinh (2 \gamma')-6 \gamma' \cosh(2 \gamma')] \nonumber  \\
&& + \cosh (2 \gamma) \left[- \gamma' \sinh (4 \gamma') + 
\frac{\cosh (4 \gamma')}{2} 
+ 4 \gamma'^2 \cosh (2 \gamma')- 
2 \gamma' \sinh (2 \gamma')-\frac{1}{2}\right] \nonumber \\
&&+\sinh (2 \gamma) [ - \gamma' \cosh (4 \gamma') + \sinh (4 \gamma') 
- \gamma' \cosh (2 \gamma')  + \sinh (2 \gamma')+2 \gamma' 
  - 6 \gamma'^2 \coth \gamma' ]  \nonumber \\
&& +\gamma [-2 \gamma' (\cosh (4 \gamma')+4 \cosh (2 \gamma')-2)
+3(\sinh(4 \gamma')+\sinh(2 \gamma'))-12 \gamma'^2 \coth \gamma' ] \nonumber  \\
&& +2 \gamma' \sinh (6 \gamma') - 2 \cosh (6 \gamma') 
 - \left(6 \gamma'^2+\frac{5}{2}\right) \cosh(4 \gamma')  
+ 9 \gamma'\sinh(4 \gamma') \nonumber \\
&&+(6 \gamma'^2-2) \cosh (2 \gamma') + 4 \gamma' \sinh(2 \gamma') 
 - 4 \gamma'^3 \coth(\gamma') + 8 \gamma'^2+ \frac{13}{2}   \\ 
&& \hbox{ } \nonumber \nonumber \\
&& B(\gamma,\gamma')=
\cosh (3\gamma+\gamma' )( \sinh(2 \gamma')-2 \gamma')^2 \nonumber  \\
&& +\cosh (\gamma+\gamma') ( \sinh (2 \gamma')-2 \gamma')
[-8 \gamma' \cosh (2 \gamma') 
 +5 \sinh (2\gamma')+4(\gamma-\gamma') 
\frac{\gamma'^2}{\sinh (\gamma')^2} \nonumber \\
&& -4 \gamma \gamma' \coth \gamma' -2 \gamma'] 
+\sinh(\gamma+\gamma') 4 \gamma'^2(\sinh(2 \gamma')-2 \gamma') \nonumber \\
&&+\cosh \gamma (\sinh (2 \gamma')-2 \gamma') \left[\sinh (3 \gamma')
+4 \gamma \cosh(\gamma')+\sinh (\gamma')
-4 \gamma \frac{\gamma'}{\sinh \gamma'}\right]
\end{eqnarray}

\section{  Disorder averaged probability distribution
for a finite size system }

\label{app:fullfinitesize}

In this Appendix we consider a finite size system $0<x<L$ using results of
Sections \ref{finitesizerg} and \ref{finitesolu}.
We start with a RR system, i.e.,  reflecting boundaries
on each end.

We will denote by $b=2 k + 2$, $k=0,1,2,..+ \infty$, the number
of renormalized bonds in the system. The disorder averaged
distribution can be written as a sum:
\begin{eqnarray}
\overline{P_{0 L}(x,t \vert x_0,0)}= 
\sum_{k=0}^{k=+\infty}
\sum_{N=1}^{2 k + 2} P_{2 k + 2}^{N,L}(x,t \vert x_0,0)
\label{completesum}
 \end{eqnarray}
where $P_{2 k + 2}^{N,L}(x,t \vert x_0,0)$ 
corresponds to the contribution
of the case where the starting point 
$x_0$ is on the $N^{th}$ bond (see Figure \ref{figfinitegen}).
One must distinguish
between $N=2n + 1$ odd, when the particle starts on a descending
bond and $x>x_0$ and $N=2 n + 2$ even when it starts on an ascending
bond and $x<x_0$. Thus, in addition to (\ref{completesum}) above
we will also be interested in the explicit decomposition:
\begin{eqnarray}
&& P_{0 L}(x,t \vert x_0,0)= \theta(x-x_0) P^{+}_{0 L}(x,t \vert x_0,0)
+ \theta(x_0 - x) P^{-}_{0 L}(x,t \vert x_0,0)  \\
&& \theta(x-x_0) P^{+}_{0 L}(x,t \vert x_0,0) = 
\sum_{k=0}^{k=+\infty}
\sum_{n=0}^{k} P_{2 k + 2}^{2 n + 1,L}(x,t \vert x_0,0)
 \end{eqnarray}

One has (see Figure \ref{figfinitegen}), for $n=0,..k$:
\begin{eqnarray}
P_{2 k + 2}^{2 n + 1,L}(x,t \vert x_0,0) =
\left\langle 
\delta\left(x - \sum_{i=1}^{2 n + 1} l_i \right) 
\theta \left(\sum_{i=1}^{2 n } l_i < x_0 < x\right) 
\right\rangle_{2 k + 2}
 \end{eqnarray}
where $\langle ..\rangle_{2k+2}$ denotes the average over the $2k+2$ 
bond sector of the finite size measure RR 
in (\ref{decoupledrr}). There is a similar 
formula for an even number of initial bonds. Throughout we will
define Laplace transforms $x \to p, x_0 \to p_0, L \to q$
as follows:
\begin{eqnarray}
P_{2 k + 2}^{N}(p,p_0,q) = 
\int_0^{+\infty} dL \int_0^{L} dx \int_0^{L} dx_0 
e^{- (p x + p_0 x_0 + q L)}
P_{2 k + 2}^{N,L}(x,t \vert x_0,0).
\end{eqnarray}
One finds:
\begin{eqnarray}
&& P_{2 k + 2}^{2 n + 1}(p,p_0,q) =
\frac{\overline{l}_\Gamma}{p_0} 
E^{+}_{p+p_0+q} (P^{-}_{p+p_0+q} P^{+}_{p+p_0+q})^{n-1}
P^{-}_{p+p_0+q} ( P^{+}_{q+p} - P^{+}_{p+p_0+q} )
(P^{-}_{q} P^{+}_{q})^{k-n} E^{-}_q  \\
&& P_{2 k + 2}^{2 n + 2}(p,p_0,q) = \frac{\overline{l}_\Gamma}{p_0}
E^{+}_{p+p_0+q} (P^{-}_{p+p_0+q} P^{+}_{p+p_0+q})^{n}
(P^{-}_{q} - P^{-}_{p_0+q} ) P^{+}_q
(P^{-}_{q} P^{+}_{q})^{k-n-1} E^{-}_q
\end{eqnarray}
the first formula being valid for $1 \le n \le k$ (and $k \ge 1$)
and the second for $0 \le n \le k-1$ (and $k \ge 1$).
Finally for the two edge bonds one has:
\begin{eqnarray}
&& P_{2 k + 2}^{1}(p,p_0,q) = \frac{\overline{l}_\Gamma}{p_0}
( E^{+}_{q+p} - E^{+}_{p+p_0+q} )
(P^{-}_{q} P^{+}_{q})^{k} E^{-}_q \\
&& P_{2 k + 2}^{2 k + 2}(p,p_0,q) = \frac{\overline{l}_\Gamma}{p_0}
E^{+}_{p+p_0+q} (P^{-}_{p+p_0+q} P^{+}_{p+p_0+q})^{k}
( E^{-}_q - E^{-}_{p_0+q} ) 
\end{eqnarray}
for any $k$. Resumming and using the identities (\ref{normcheck}) yields:
\begin{eqnarray}
P^{+}(p,p_0,q) & = & \frac{1}{p_0 q} \frac{E^+_{q+p}}{E^{+}_q}
+\frac{1}{\overline{l}_\Gamma p_0 q (p+p_0+q)}
\frac{P^{-}_{p+p_0+q} P^{+}_{p+q} -1}{
E^{-}_{p+p_0+q} E^{+}_q } \\
P(p,p_0,q) & = & 
\frac{1}{\overline{l}_\Gamma p_0 q (p+p_0+q)}
\frac{P^{-}_{p+p_0+q} P^{+}_{p+q} -
P^{-}_{p_0+q} P^{+}_{q}}{
E^{-}_{p+p_0+q} E^{+}_q }
+
\frac{1}{p_0 q} \frac{E^+_{q+p}}{E^{+}_q} \nonumber \\
&& \quad  - 
\frac{1}{p_0 (p + p_0+q)}
\frac{E^-_{p_0+q}}{E^{-}_{p+p_0+q}} 
\end{eqnarray}
A simpler expression holds at coinciding points:
\begin{eqnarray}
\int_0^{+\infty} dL \int_0^{+\infty} dx_0 
e^{-p_0 x_0 -q L } \overline{ P_{0 L}(x_0,t \vert x_0,0)}
= \frac{1}
{\overline{l}_\Gamma  q (p_0+q) E^{-}_{p_0+q} E^{+}_q }
\end{eqnarray}

A similar calculation in the case of absorbing boundaries
(AA case) gives:
\begin{eqnarray}
&& P(p,p_0,q) = \frac{( P^{+}_{q+p} P^{-}_{q} 
- P^{+}_{p+p_0+q} P^{-}_{p_0+q} ) }
{\overline{l}_\Gamma p_0 q (p+p_0+q)E^{+}_{p+p_0+q} E^{-}_q }
\end{eqnarray}
as well as the semi-infinite limit $L=\infty$ with an absorbing boundary 
at $x=0$ 
\begin{eqnarray}
&& P_{0 \infty}(p,p_0) = \frac{( P^{+}_{p}  - P^{+}_{p+p_0} P^{-}_{p_0} ) }
{\overline{l}_\Gamma p_0  (p+p_0)E^{+}_{p+p_0} }
\end{eqnarray}
At coinciding points in semi-infinite system
this becomes simply 
\[
\int_0^{+\infty} dx_0 
e^{-p_0 x_0  } \overline{ P(x_0,t \vert x_0,0)}
= \frac{ P^{+}_{p_0}  }
{\overline{l}_\Gamma   p_0 E^{+}_{p_0}  }.
\] 
In the RA case the result is
\begin{eqnarray}
P(p,p_0,q) =\frac{( E^{+}_{q+p} ) P^{-}_{q}} 
{p_0 q E^{-}_q}
+  \frac{ ( P^{-}_{q} P^{-}_{p+p_0+q}  P^{+}_{q+p} - P^{-}_{q+p_0} )  }
{\overline{l}_\Gamma p_0 q (p+q+p_0) E^{-}_{p+p_0+q} E^{-}_q } 
\end{eqnarray}
The AR case being obtained by the global exchange of
$+$ and $-$ as well as $x \to L -x$ and $x_0 \to L-x_0$
(i.e $p \to - p$, $p_0 \to - p_0$ and $q \to q + p + p_0$).


\begin{thebibliography}{10}
\bibitem{footnote2}
see formulae (43-50) in \cite{laloux_pld_sinai}) given that in the small $\mu$
  limit $(t-t')^\mu \sim t^{\mu} (1 - \exp(\frac{1}{\mu}(\gamma-\gamma')))^\mu$
  can be replaced by $t^{\mu}$ in the exponential of (\ref{diffbias}).
\end{thebibliography}

\end{document}